\documentclass{emulateapj}
\usepackage{amsmath, natbib, color}

\received{}
\revised{}
\accepted{}

\shorttitle{}
\shortauthors{}

\begin{document}

\title{Reverberation Mapping of Luminous Quasars at High-$z$}

\author{
Paulina Lira\altaffilmark{1},
Shai Kaspi\altaffilmark{2},
Hagai Netzer\altaffilmark{3},
Ismael Botti\altaffilmark{4},
Nidia Morrell\altaffilmark{5},
Juli\'an Mej\'ia-Restrepo\altaffilmark{1},
Paula S\'anchez\altaffilmark{1},
Jorge Mart\'inez\altaffilmark{1},
Paula L\'opez\altaffilmark{1}
}

\altaffiltext{1}{Departamento de Astronom\'ia, Universidad de Chile, Casilla 36D, Santiago, Chile.}
\altaffiltext{2}{Wise Observatory, School of Physics and Astronomy, Tel Aviv University, Tel Aviv 69978, Israel}
\altaffiltext{3}{School of Physics and Astronomy, Tel Aviv University, Tel Aviv 69978, Israel}
\altaffiltext{4}{Facultad de Ingenier\'ia, Universidad del Desarrollo, Av.~Plaza 680, Las Condes, Santiago, Chile}
\altaffiltext{5}{Las Campanas Observatory, Carnegie Observatories, Casilla 601, La Serena, Chile}

\begin{abstract}

We present Reverberation Mapping (RM) results for 17 high-redshift, high-luminosity quasars with good quality R-band and emission line light curves. We are able to measure statistically significant lags for Ly$\alpha$ (11 objects), SiIV (5 objects), CIV (11 objects), and CIII] (2 objects). Using our results and previous lag determinations taken from the literature, we present an updated CIV radius--luminosity relation and provide for the first time radius--luminosity relations for Ly$\alpha$, SiIV and CIII]. While in all cases the slope of the correlations are statistically significant, the zero points are poorly constrained because of the lack of data at the low luminosity end. We find that the emissivity weighted distance from the central source of the Ly$\alpha$, SiIV and CIII] line emitting regions are all similar, which corresponds to about half that of the H$\beta$ region. We also find that 3/17 of our sources show an unexpected behavior in some emission lines, two in the Ly$\alpha$ light curve and one in the SiIV light curve, in that they do not seem to follow the variability of the UV continuum. Finally, we compute RM black hole masses for those quasars with highly significant lag measurements and compare them with CIV single--epoch (SE) mass determinations. We find that the RM-based black hole mass determinations seem smaller than those found using SE calibrations.

\end{abstract}


\section{Introduction}

These days it is widely accepted that all massive galaxies harbor a
massive Black Hole (BH) in their centers (Kormendy \& Ho, 2013). To
determine the properties of these BHs is therefore crucial for the
understanding of galaxy formation and evolution. The ability to
measure BH masses and accretion rates in AGN using reverberation
mapping techniques has enabled real physical comparison between active
and dormant BHs in the local universe, but BH mass estimates from AGN
become even more crucial at high-z since stellar dynamical estimates
from the study of normal galaxies are clearly not feasible at
redshifts of cosmological interest ($z>1$).

Reverberation Mapping (RM) uses the lag between variations in the
central ionizing source and the response of the Broad Line Region
(BLR) emission lines to directly measure the BLR size (Blandford \&
McKee, 1982). Assuming a gravitationally bound system and measuring
the BLR line widths, it is possible to infer the mass of the central
BH. This assumption has proven to be correct for those objects with
measurements from several lines: the observed anti-correlation between
the line Doppler widths and their distance from the central black
hole, are consistent with virialized motion of the BLR gas in the deep
potential of the central BH (Onken \& Peterson 2002).

\begin{table*}
\renewcommand{\thetable}{\arabic{table}}
\centering
\caption{Sample summary}
\begin{tabular}{lccccccccc}
\tablewidth{0pt}
\hline
\hline
Quasar & z & RA DEC (J2000) & R     & $\lambda L_{\lambda}(1350$\AA) & $\lambda L_{\lambda}(5100$\AA) & $P_{843\ {\rm MHz}}$ & $P_{1.4\ {\rm GHz}}$ & $R_{radio}$ & M$_{\rm BH}^{\rm SE}$(CIV)\\
       &   &                & (mag) & ($10^{46}$ ergs s$^{-1}$) & ($10^{46}$ ergs s$^{-1}$) & (mJy)           & (mJy)            &     & ($10^9$ M$_{\odot}$)\\
\hline
CT1061         & 3.373 & 10 48 56  -16 37 09 &  $16.20 \pm 0.12$ & $33.88 \pm 3.15$ & $10.28 \pm 0.12$& N/A            & $<0.5$      & $<1$   & 3.42 \\
CT250          & 2.407 & 04 11 45  -42 54 44 &  $17.69 \pm 0.14$ & $4.97 \pm 0.99$  & $2.88 \pm 0.07$ & $< 6$          & N/A         & $<21$ & 4.65 \\
CT252          & 1.890 & 04 18 10  -45 32 17 &  $16.40 \pm 0.10$ & $0.00 \pm 0.00$  & $3.03 \pm 0.06$ & $< 6$          & N/A         & $<6$ & 1.26 \\
CT286          & 2.556 & 10 17 23  -20 46 58 &  $16.89 \pm 0.13$ & $11.16 \pm 1.83$ & $6.28 \pm 0.13$ & N/A            & $17.7 \pm 0.7$ & $48 \pm 10$  & 2.27 \\
CT320          & 2.956 & 13 17 44  -31 47 13 &  $17.82 \pm 0.11$ & $6.35 \pm 0.91$  & $3.86 \pm 0.07$ & $< 6$          & $<0.5$      & $<2$ & 6.23 \\
CT367          & 2.601 & 22 00 36  -35 02 17 &  $17.14 \pm 0.14$ & $5.89 \pm 1.80$  & $4.66 \pm 0.18$ & $< 6$          & $<0.5$      & $<1$ & 5.01 \\
CT406          & 3.183 & 10 39 09  -23 13 25 &  $17.66 \pm 0.13$ & $8.13 \pm 0.75$  & $6.18 \pm 0.07$ & N/A            & $3.0 \pm 0.5$ & $8.4 \pm 3.4$  & 4.80 \\
CT564          & 2.659 & 21 50 15  -44 11 23 &  $17.05 \pm 0.12$ & $9.95 \pm 1.51$  & $3.65 \pm 0.07$ & $< 6$          & N/A         & $<10$ & 1.99 \\
CT650          & 2.662 & 04 55 22  -42 16 17 &  $17.28 \pm 0.11$ & $7.59 \pm 1.83$  & $4.77 \pm 0.14$ & $< 6$          & N/A         & $<16$ & 1.88 \\
CT803          & 2.741 & 00 04 48  -41 57 28 &  $17.02 \pm 0.12$ & $10.11 \pm 1.20$ & $8.20 \pm 0.12$ & $< 6$          & N/A         & $<5$ & 4.09 \\
CT953          & 2.535 & 21 59 54  -40 05 50 &  $17.00 \pm 0.11$ & $9.99 \pm 1.97$  & $6.12 \pm 0.15$ & $< 6$          & N/A         & $<7$ & 5.98 \\
CT975          & 2.866 & 22 38 13  -32 48 24 &  $17.46 \pm 0.17$ & $8.78 \pm 1.57$  & $4.70 \pm 0.11$ & $< 6$          & $<0.5$      & $<3$ & 5.35 \\
HB890329-385   & 2.433 & 03 31 06  -38 24 05 &  $17.54 \pm 0.12$ & $5.80 \pm 0.93$  & $5.40 \pm 0.11$ & $24.3 \pm 1.3$ & $29.8 \pm 1.0$  & $50 \pm 9$   & 6.74 \\
2QZJ002830     & 2.403 & 00 28 30  -28 17 06 &  $17.05 \pm 0.14$ & $8.98 \pm 1.74$  & $3.76 \pm 0.09$ & N/A            & $<0.5$      & $<1$  & 7.64 \\
2QZJ214355     & 2.620 & 21 43 55  -29 51 59 &  $17.17 \pm 0.11$ & $9.17 \pm 1.02$  & $2.82 \pm 0.04$ & N/A            & $<0.5$      & $<2$  & 5.63 \\
2QZJ221516     & 2.706 & 22 15 16  -29 44 23 &  $16.71 \pm 0.14$ & $14.29 \pm 1.86$ & $10.18 \pm 0.17$& N/A            & $467 \pm 14$ & $576 \pm 100$ & 1.56 \\
2QZJ224743     & 2.590 & 22 47 43  -31 03 07 &  $16.65 \pm 0.10$ & $13.00 \pm 1.05$ & $4.29 \pm 0.04$ & $< 6$          & $2.7 \pm 0.6$ & $5.9 \pm 2.8$ & 15.55\\
\hline
\hline
\multicolumn{10}{l}{Radio fluxes at 843 MHz and 1.4 GHz were taken from the SUMSS and NVSS catalogs, respectively.}\\ 
\multicolumn{10}{l}{N/A implies that the sources were not covered by the footprint of the survey.}\\  
\multicolumn{10}{l}{$R_{radio}$ was obtained using the $P_{1.4\ {\rm GHz}}$ measurements, except for unavailable objects (`N/A'), where $P_{843\ {\rm MHz}}$ was used instead.}\\ 
\multicolumn{10}{l}{$\lambda L_{\lambda}(1350$\AA) measurements were obtained from the spectroscopic data.}\\
\multicolumn{10}{l}{$\lambda L_{\lambda}(5100$\AA) luminosities were obtained either from 2MASS photometry or by extrapolating our own R-band magnitudes (see text).}\\
\multicolumn{10}{l}{Uncertainties in $\lambda L_{\lambda}(1350$\AA) and $\lambda L_{\lambda}(5100$\AA) were assumed to correspond to the observed R-band variability.}\\
\multicolumn{10}{l}{Single--epoch (SE) BH masses (M$_{\rm BH}^{\rm SE}$) have been obtained using the measured CIV FWHM (Mej\'ia-Restrepo et al.~2016).}\\
\end{tabular}
\end{table*}

To date reverberation mapping results cover almost 5 orders of
magnitude in luminosity but are still limited to luminosities $\lambda
L_{\lambda} (5100$\AA) $<10^{46}$~erg/sec, with the bulk of sources
found well below $<10^{45}$~erg/sec (e.g., Kaspi et al.~2000, 2005,
2007, Peterson et al.~2004, Bentz et al.~2006, 2009, 2013). Hence,
such results cannot be directly applied to high-$z$, high-L sources,
that contain the most massive BHs, since measuring their BLR size
requires an extrapolation by up to {\it two orders of magnitude} in
luminosity. This hampers the calibration of the so called
radius--luminosity relations, which enable the determination of BH
masses from a single spectroscopic observation without requiring
source monitoring. `Single--epoch' black hole mass determinations are
readily obtained from large spectroscopic surveys, but require the
extrapolation of the radius--luminosity relation when high-$z$,
high-luminosity quasars are studied. It is therefore clear that to
have statistically significant results for BH demographics and their
mass growth with cosmic time, it is first necessary to determine well
calibrated radius--luminosity relations that are representative of the
full span of AGN luminosity.

So far the few attempts to carry out reverberation mapping of very
high-luminosity quasars have seldom proved successful (e.g., Welsh et
al.~2000, Trevese et al.~2006, Kaspi et al.~2007, Trevese et al.~2014,
Saturni et al.~2016). There are mainly two reasons for this: first,
most high-luminosity sources show very low amplitude variations ($<
20$\%) on short time scales and require very extended (many years)
monitoring to observe significant flux variations and to overcome the
$(1+z)$ time delay. Second, as it is usually observed in monitoring
campaigns, amplitudes for emission line light curves are smaller than
that of the continuum emission, as the emission line response is
averaged over the very large ($\sim 1$ pc) quasar BLR geometry. As
monitoring of high-$z$, high-luminosity quasars often only samples a
few, low amplitude continuum flux variation `events', the line
response can become extremely weak. This requires the ability to
measure emission line fluxes to an accuracy of about a few percent
(e.g., Kaspi et al., 2007), and the implementation of tailored
observing strategies.

In this article we present results from a $\ga 10$ year monitoring of
high-$z$, high-luminosity quasars. In section 2 we describe the sample
selection and data acquisition and treatment. Section 3 deals with
time series analysis, while section 4 presents the specifics on the
cross-correlation analysis. Section 5 presents the results for the
radius--luminosity relations. Finally, section 6 and 7 discusses and
summarizes the findings. A concordance cosmology with $\Omega_m = 0.3$
and $\Omega_{\Lambda} = 0.7$ is adopted throughout this paper.

\begin{figure}
\begin{center}
\includegraphics[scale=0.8,trim=40 0 0 40]{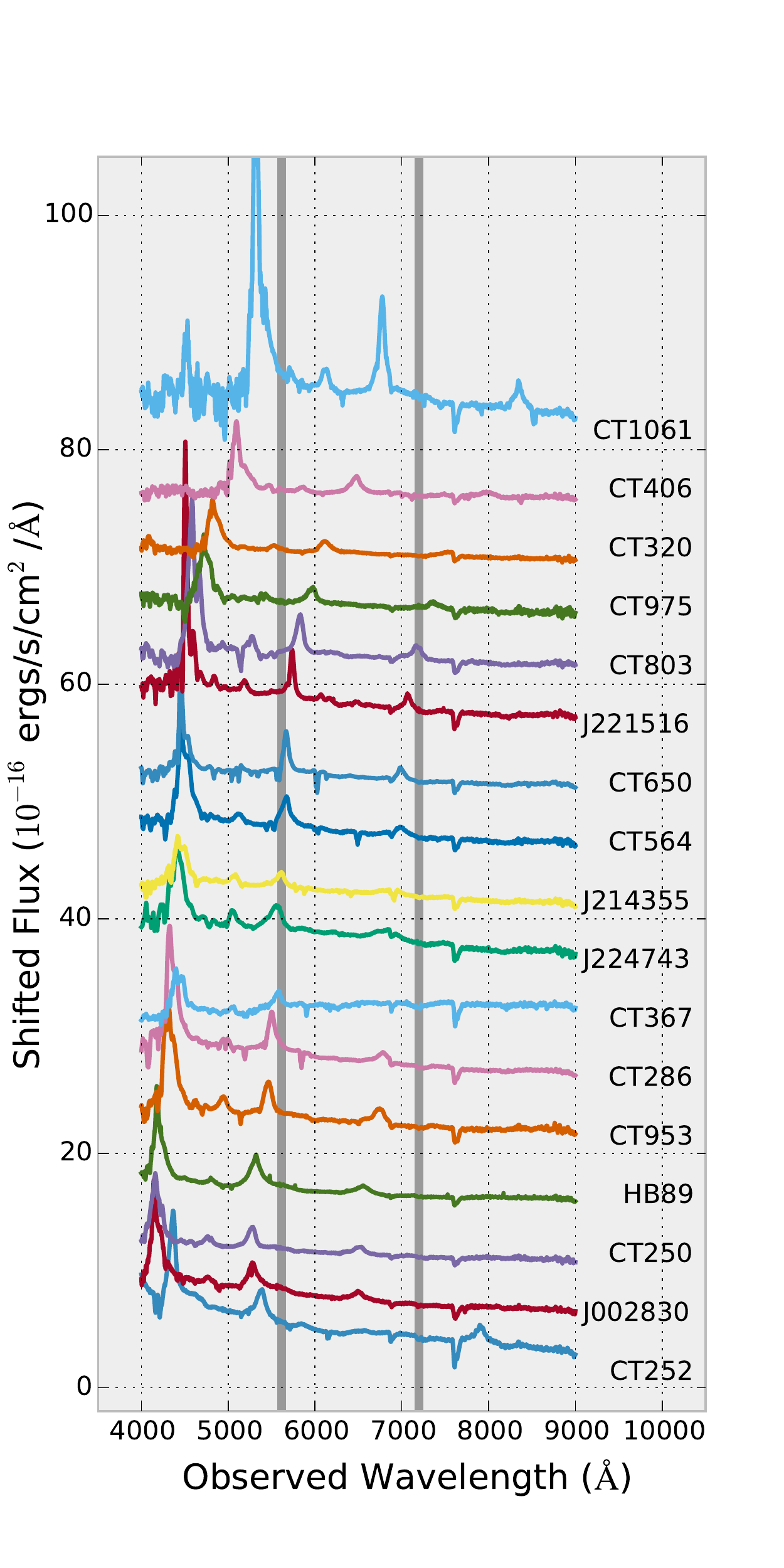}
\caption{Flux calibrated, Galactic extinction corrected, mean
  spectra of the 17 quasars reported in this work. The spectra have
  been shifted in the y-axis for display purposes, starting with a
  zero shift for CT252, 5 units of flux for J002830, 10 units of flux
  for CT250, and so on. The gray lines approximately demark the
  wavelength region corresponding to the R-band filter.}
\end{center}
\end{figure}

\begin{figure*}
\begin{center}
\includegraphics[scale=1.2,trim=0 0 0 50]{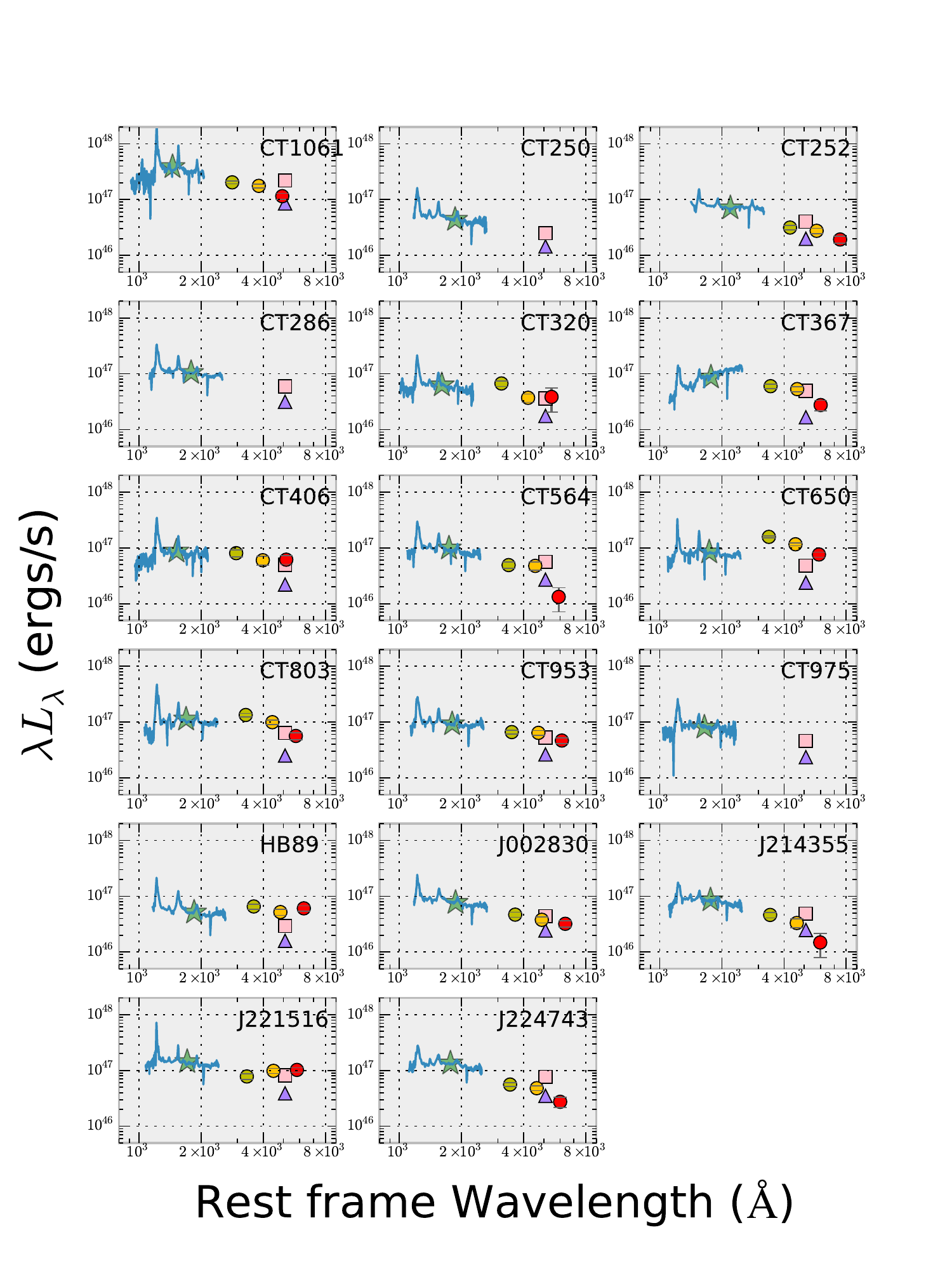}
\caption{Rest-frame Spectral Energy Distributions (SEDs) of the
  quasars in our sample. Mean spectra for each source are shown
  together with our $R$-band (stars) and JHK 2MASS (circles)
  photometry. Two extrapolations to 5100\AA\ are also included: from
  1350\AA\ (triangles) using the correlation determined in
  Mej\'ia-Restrepo et al.~(2016), and from the $R$-band photometry
  (squares) using the UV power-law index determined by Vanden Berk et
  al.~(2001).}
\end{center}
\end{figure*}

\section{Sample and data}

\subsection{Sample selection}

The targets were drawn from the Cal\'an-Tololo survey (Maza et al.,
1996 and references therein), the 2dF QSO Redshift Survey (Croom et
al., 2004), and the Hewitt-Burbidge QSO compilation (Hewitt \&
Burbidge, 1989). They correspond to quasars of very high luminosity,
typically $M_B \sim -29$ magnitudes, located at the high luminosity
end of the quasar luminosity function (Bongiorno et al., 2007; Ross et
al., 2013). A first spectroscopic run carried out in March 2006 with
the du Pont telescope at Las Campanas Observatory (LCO) allowed us to
corroborate their quasar nature and the presence of intense emission
lines suitable for reverberation mapping. At this redshift range, four
lines are readily seen in the spectra of all our quasars: Ly$\alpha$,
SiIV, CIV and CIII].

Towards the end of the monitoring campaign, when confirming the
redshift of CT252, we realized that the published value for this
source (see Maza et al.~(1993), where a redshift of 2.5 was provided)
is much lower than the rest of the sample, at $z = 1.818$. This was
probably because its very strong and non-symmetric CIV line was
mistakingly identified as Ly$\alpha$. Hence for this source Ly$\alpha$
is not visible in our spectra, but instead we can observe the MgII
emission line.

The redshifts for the quasars were obtained from our own data from the
CIII] line, except for CT320 where the line fit quality was poor. For
  this object the CIV redshift is reported instead. For CT252 the MgII
  redshift is given. Redshifts are found in Table 1 together with some
  basic information for each quasar such as the R-band magnitude, its
  standard deviation (see Section 3) and radio flux measurements (see
  below). Single--epoch virial masses are also reported using the
  calibration for the CIV emission line presented in Mej\'ia-Restrepo
  et al.~(2016). These were determined using our FWHM measurements of
  the CIV emission line. A virial factor $f$ of one was assumed, which
  is appropriate for velocities estimated from FWHMs (e.g., Grier et
  al.~2017).

\subsubsection{Spectral properties}

The mean spectra of each quasar can be seen in Figure 1. These have
been flux calibrated and corrected for Galactic extinction assuming
the extinction law of Cardelli, Clayton \& Mathis (1989) and $R_V=3.1$
in the observed frame. $E(B-V)$ values were in the range of 0.01 to
0.04 magnitudes. For display purposes the spectra have been shifted in
the vertical direction ordered by redshift (see caption for
details). The absorption feature seen at the constant wavelength of
$\sim$ 7600\AA\ corresponds to a telluric O$_2$ absorption. Rest-frame
line luminosities, widths and equivalent widths (EWs) are reported in
Table 2.

Figure 1 shows rather broad range of spectral properties, from objects
with very broad lines, such as CT953 and J224743, and others with much
narrower features, such as CT650 and J221516. Also, the lines can be
very prominent (i.e., with large EW), like in the case of CT1061 and
CT803, or rather weak (i.e., small EW), such as those in J002839 and
J214355. The spectral shapes are rather uniform, with the exception of
CT252, for which we are observing a very different spectral range in
the rest frame, and CT367, which clearly shows a red spectral shape.

In between gray lines we highlight the 5620-7200\AA\ wavelength range,
which roughly corresponds to the width of the R-band filter. Depending
on the redshift of the source, the observed R-band
5620-7200\AA\ wavelength range corresponds to a mean rest-frame
wavelength of 1465\AA\ for our highest redshift source (CT1061),
1530\AA\ for our second highest (CT406), and 2271\AA\ for the lowest
redshift source (CT252). The remaining sources are found in the range
1620-1880\AA. It can be seen that this region of the spectra contains
the CIII] emission line, which as we will see, does not show strong
  variability. However, for CT1061, CT320, CT406, CT564, CT650, CT803,
  CT975 and J221516 the CIV line is redshifted into the R-band
  coverage. This could affect the analysis of the
  variability. However, the very broad nature of the R-band and the
  small amplitude observed in the line variations secures a negligible
  interference: using the EW values presented in Table 2 and assuming
  a width of the R-band filter of 2200 \AA, it can be seen that the
  total CIV line flux would contribute at most 10\%\ to the
  observed-frame R-band photometry.

\subsubsection{Radio-loudness and SEDs}

It is of interest to determine the radio-loudness of the quasars in
our sample. Usually, a radio-to-optical flux-ratio threshold of
$R_{radio} = f(6{\rm cm})/f(4400{\rm \AA}) = 10$ is adopted to
separate radio-loud (RL) from radio-quiet (RQ) systems, while values
between 10 and 100 are sometimes referred to as radio-intermediate. We
searched two surveys for radio sources consistent with the positions
of our quasars. First, the Sydney University Molonglo Sky Survey
(SUMSS) Source Catalog (Mauch et al.~2003), which covers the southern
sky for declinations $-50 < \delta < -30$ degrees at 843 MHz, reaches
a depth of 6 mJy/beam, and has a spatial resolution of $45\times45\ /
\cos(|\delta|)$ square arcseconds. We also searched the National Radio
Astronomy Observatory Very Large Array Sky Survey (NVSS) catalog
(Condon et al.~1998), which covers the sky north of -40 degrees at 1.4
GHz, reaching a depth of 0.45 mJy/beam, and with a spatial resolution
of 45 arcseconds. The presence of counterparts was confirmed by eye
inspection of the radio maps. We K-corrected the radio measurements
assuming a power law spectral energy distribution of the form $S_{\nu}
\propto \nu^{-\alpha}$, with index $\alpha=0.75$ (Wang et al.~2007;
Momjian et al.~2014).

To determine the rest-frame optical fluxes we obtained J,H and K
magnitudes from the 2MASS All-Sky Catalog of Point Sources (Cutri et
al.~2003). No 2MASS photometry was available for CT250, CT286 or
CT975. We also obtained fluxes at 5100\AA\ applying the
correlation between continuum emission at 1350\AA\ and 5100\AA\ found
in Mej\'ia-Restrepo et al.~(2016), and extrapolating from our $R$-band
photometry using the quasar rest-frame UV power-law index
($\alpha=0.44$) obtained by Vanden Berk et al.~(2001). 

The Spectral Energy Distributions for our sample are presented in
Figure 2. It can be seen that in several cases there is good agreement
between the spectra, their extrapolations to 5100\AA\ and the 2MASS
photometry. However, it is also seen that the extrapolation based on
the work by Vanden Berk et al.~(2001) is a factor $\sim 2$ higher than
that obtained applying the correlation found in Mej\'ia-Restrepo et
al.~(2016), which in most cases falls below the 2MASS
observations. The relation found in Trakhtenbrot \& Netzer (2012)
predicts fluxes about half way between the two previous
extrapolations. An anomalous case is CT650 (and perhaps CT803 and
HB89) which was clearly brighter at the time of the 2MASS
observations. CT367 has a spectral shape that is clearly poorly
represented by the extrapolation to 5100\AA, while for J221516 the
2MASS photometry suggests an up-turn of the 2MASS fluxes towards
longer wavelengths. In summary, only for those objects without 2MASS
photometry and CT650 we will use the spectral 5100\AA\ extrapolation
based on Vanden Berk et al.~(2001) to estimate the rest-frame optical
flux. For all other objects, the 2MASS photometry will be adopted. The
5100\AA\ luminosities are reported in Table 1. To K-correct the
observations to the rest frame 4400\AA\ needed to determine
$R_{radio}$, Vanden Berk et al.~(2001) spectral index was again used.

The results on the radio-loudness are reported in Table 1. Three
sources were found to be radio-loud quasars: CT286, HB89 and J221516
($R_{radio} = 48 \pm 10$, $50 \pm 9$, and $576 \pm 100$,
respectively), the last two already noticed as radio-loud systems in
the literature (Shemmer et al.~2004; Chhetri et al.~2013). The
remaining objects are split into 12 secure radio-quiet systems and two
with upper limits above $R_{radio} = 10$ (CT250 and CT650). Leaving
these two last sources aside, a fraction of 3/15 radio-loud quasars is
found, which is in good agreement with the general quasar population
at the high-end of the luminosity range (Cirasuolo et al., 2003).

\begin{table*}
\renewcommand{\thetable}{\arabic{table}}
\centering
\caption{Line measurements} 
\begin{tabular}{l|ccc|ccc|ccc|ccc}
\tablewidth{0pt}
\hline
\hline
       & \multicolumn{3}{c|}{Ly$\alpha$} & \multicolumn{3}{c|}{SiIV} & \multicolumn{3}{c|}{CIV} & \multicolumn{3}{c}{CIII]}\\
Quasar & L & EW & FWHM                & L & EW & FWHM         & L & EW & FWHM        & L & EW & FWHM           \\
       & ergs s$^{-1}$ & \AA\ & km s$^{-1}$ & ergs s$^{-1}$ & \AA\ & km s$^{-1}$ & ergs s$^{-1}$ & \AA\ & km s$^{-1}$ & ergs s$^{-1}$ & \AA\ & km s$^{-1}$ \\
\hline
CT1061         & 2.1e+46 & 64 & 3355 & 2.4e+45 & 10 & 4843 & 8.9e+45 & 44 & 3218 & 1.4e+45 &  9 & 2452 \\
CT250          & 2.1e+46 & 64 & 3355 & 2.4e+45 & 10 & 4843 & 8.9e+45 & 44 & 3218 & 1.4e+45 &  9 & 2452 \\
CT286          & 2.5e+45 & 57 & 6817 & 3.3e+44 & 10 & 6676 & 1.0e+45 & 35 & 6256 & 3.6e+44 & 17 & 5900 \\
CT320          & 1.3e+45 & 13 & 2934 & 2.8e+44 &  4 & 5312 & 7.8e+44 & 11 & 3493 & 2.3e+44 &  5 & 3796 \\
CT367          & 2.0e+45 & 33 & 5368 & 3.9e+44 &  9 & 6453 & 1.2e+45 & 33 & 6844 & 1.4e+44 &  5 & 2085 \\
CT406          & 1.2e+45 & 23 & 4869 & 1.9e+44 &  5 & 3922 & 1.7e+45 & 40 & 6236 & 2.1e+44 &  4 & 8828 \\
CT564          & 3.8e+45 & 46 & 4943 & 3.8e+44 &  7 & 4814 & 1.8e+45 & 35 & 5623 & 5.9e+44 & 17 & 6361 \\
CT650          & 4.1e+45 & 52 & 7007 & 4.3e+44 &  6 & 6053 & 6.1e+44 & 10 & 3419 & 4.9e+44 & 12 & 8593 \\
CT803          & 1.4e+45 & 20 & 2266 & 1.4e+44 &  3 & 3029 & 1.1e+45 & 21 & 3437 & 6.7e+44 & 18 & 4009 \\
CT953          & 7.3e+45 & 107 & 5506& 8.0e+44 & 13 & 5044 & 3.1e+45 & 54 & 5005 & 3.4e+44 &  7 & 7929 \\
CT975          & 4.2e+45 & 47 & 8199 & 6.6e+44 & 10 & 5875 & 2.3e+45 & 40 & 5970 & 3.1e+44 &  7 & 4468 \\
HB89 0329-385  & 9.5e+44 & 14 & 6275 & 2.9e+44 &  5 & 5231 & 1.1e+45 & 22 & 5831 & 6.6e+44 & 19 & 4070 \\
2QZJ002830     & 2.4e+45 & 46 & 3975 & 1.7e+44 &  4 & 5603 & 1.9e+45 & 57 & 7326 & 6.7e+44 & 28 & 8336 \\
2QZJ214355     & 1.6e+45 & 19 & 4397 & 2.8e+44 &  5 & 6342 & 7.9e+44 & 15 & 6895 & 3.5e+44 &  9 & 4360 \\
2QZJ221516     & 1.1e+45 & 15 & 5857 & 3.4e+44 &  6 & 5239 & 9.4e+44 & 17 & 5888 & 1.4e+44 &  4 & 2893 \\
2QZJ224743     & 5.0e+45 & 41 & 2410 & 4.8e+44 &  5 & 4419 & 2.2e+45 & 25 & 2728 & 6.5e+44 & 10 & 2966 \\
               & 1.0e+45 & 10 & 5206 & 5.1e+44 &  6 & 4685 & 1.9e+45 & 23 & 8861 & 1.1e+44 &  2 & 2010 \\
\hline
       & \multicolumn{3}{c|}{CIV} & \multicolumn{3}{c|}{CIII]} & \multicolumn{3}{c|}{MgII} &&&\\
Quasar & L & EW & FWHM                & L & EW & FWHM          & L & EW & FWHM &&& \\
       & ergs s$^{-1}$ & \AA\ & km s$^{-1}$ & ergs s$^{-1}$ & \AA\ & km s$^{-1}$ & ergs s$^{-1}$ & \AA\ & km s$^{-1}$ &&& \\
CT252   & 1.4e+45 &  28 & 5198    & 2.7e+44 & 7 & 4970 & 4.9e+44 & 16 & 3800 \\
\hline
\hline
\multicolumn{13}{l}{Measurements were obtained from the mean spectrum of each source in the rest frame.}
\end{tabular}
\end{table*}

\subsection{Broad-band imaging}

Photometric monitoring of $\sim$50 high-z AGN started in February
2005. After a few years the less variable systems were dropped and we
continue to monitor $\sim$60\%\ of the original sample. The data were
obtained with the 1.3m and 0.9m SMARTS telescopes using broad-band R
imaging. Several observations per year were acquired in queue
mode. The light curves reported in this work extend until January
2017.

Bias subtraction and flat correction was done in the usual way using
IRAF tasks. PSF differential photometry of the quasars was obtained
using typically 10-12 local stars. The stars where in turn calibrated
against their R-band USNO magnitudes so that the light curves are
finally expressed in flux units. Formal photometric errors for the
stars were generally small. Therefore a 0.015 magnitude error, as
obtained from the median standard deviation of the observed stellar
fluxes, was adopted as a more representative photometric error.

\subsection{Spectroscopy}

To secure an accurate relative spectrophotometric calibration we
followed Maoz et al.~(1990) and Kaspi et al.~(2000) and rotated the
spectrograph so that the quasar and a comparison star are observed
through the same slit. The WFCCD at the du Pont telescope at LCO
proved to be a very reliable instrument allowing to position the
quasar and comparison star within the slit with sub-pixel precision,
for a pixel size of 0.484 arcseconds. The slit width was 8.1 arcseconds
wide throughout the observations.

The spectroscopic monitoring started in April 2007 and the latest data
were obtained in January 2017. We aimed at having at least one
observation per year for each quasar, although weather and
instrumental problems sometimes did not allow us to fulfill this
goal. Typically each observation consisted of three repeated spectra
of 900 to 1200 seconds of exposure time. The spectra were reduced in
the standard way using IRAF tasks: bias subtraction and flat
calibration were applied using bias frames and internal lamps observed
each night. Flats were obtained using the same wide slit as the
science frames. One dimensional spectra of the quasar and comparison
star were extracted using the same fixed aperture along the spectral
direction. Wavelength calibration was determined using helium, neon,
argon calibration `arcs' obtained using a narrow slit.

To achieve the relative spectrophotometry, the spectrum of each quasar
was divided by the heavily smoothed spectrum of the simultaneously
observed comparison star and then the separate `normalized-quasar
spectra' were combined. To secure a homogeneous wavelength calibration
around the Ly$\alpha$, SiIV, CIV and CIII] emission lines (and MgII in
  the case of CT252), sections of the spectra around each line were
  used to perform a cross-correlation analysis. The spectra were then
  shifted according to the cross-correlation results. This was
  particularly important at the blue end of the spectra, where only a
  few lines from the comparison arcs were available, thus making the
  wavelength solution rather unreliable.

Line flux measurements were obtained by adopting two small
pseudo-continuum windows located at each side of the corresponding
emission line. The continuum level under the line in consideration was
then assumed to correspond to the interpolation of a straight line
joining the mean flux obtained from the pseudo-continuum windows. We
checked that the pseudo-continuum windows corresponded to regions with
small values in the rms spectrum of each quasar to avoid the presence
of weak emission or absorption lines. Likewise, to avoid introducing a
spurious line variability signal due to variations of strong
self-absorbing features, we limited the line flux measurements to
regions devoided of absorption lines, which were readily seen in the
rms spectra. As an example, Figure 3 shows the mean and rms spectra
for the emission lines in the quasar CT650. The placement of the
continuum windows and the region adopted for the line flux
measurements are shown. As can be seen, SiIV is heavily absorbed
throughout the profile, while the remaining lines show strong and
variable absorption in their blue wings. Hence the line fluxes were
determined to the right of Ly$\alpha$, CIV and CIII], while the light
  curve of the SiIV was obtained from a small spectral window. Notice,
  however, that if non-variable absorption profiles are present in the
  line profiles, we do not make any attempt to isolate them from the
  line flux measurements.

\begin{figure*}
\begin{center}
\includegraphics[scale=0.8]{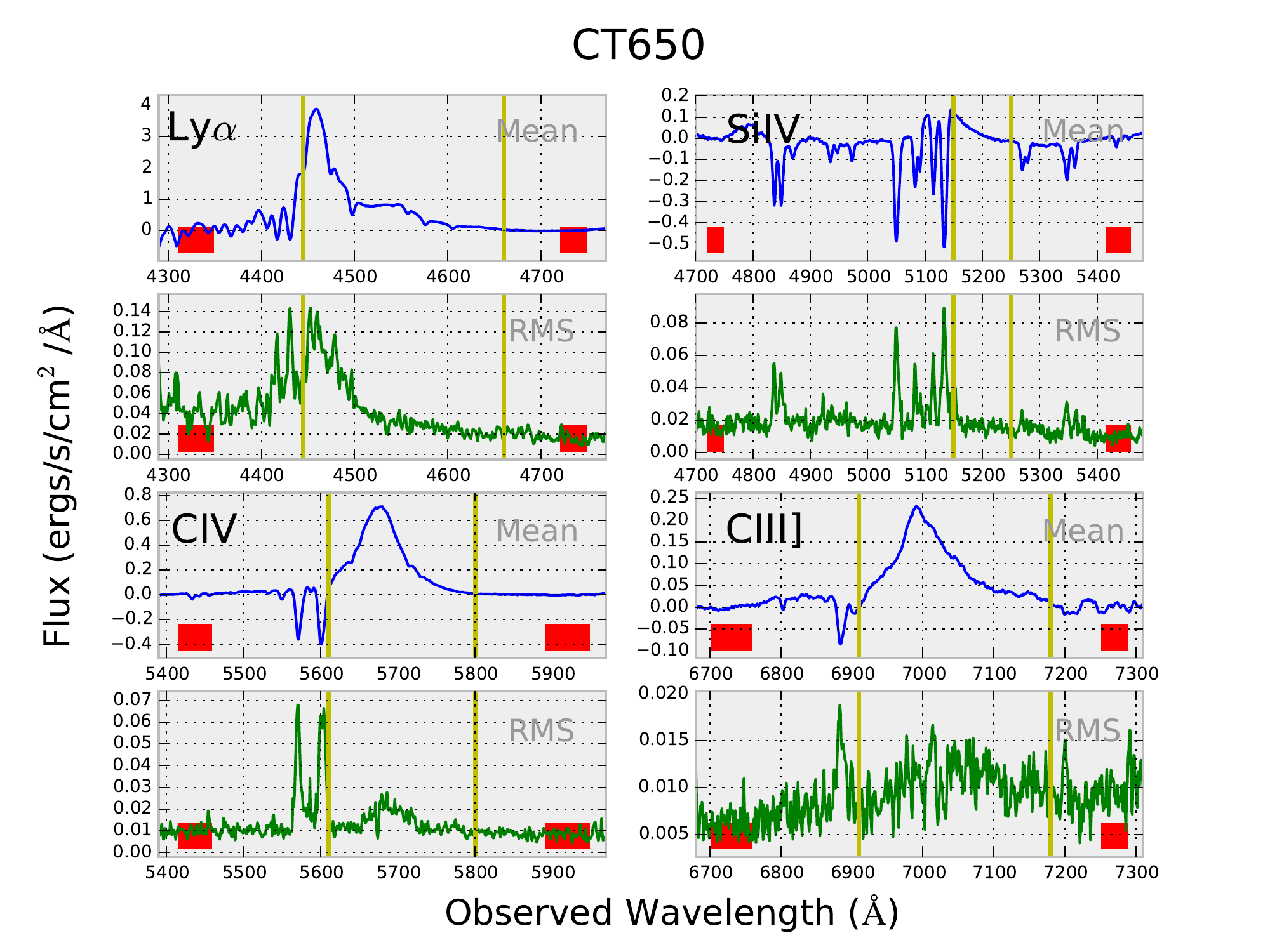}
\caption{Mean (top panels -- in blue) and rms (bottom panels -- in
  green) spectra of CT650 around the Ly$\alpha$, SiIV, CIV and CIII]
  emission lines. The mean spectra have been continuum subtracted
  using the continuum measurements obtained at the positions marked
  by red boxes. The line measurement is obtained as the summation of
  all flux between the vertical yellow lines.}
\end{center}
\end{figure*}

\begin{figure}
\begin{center}
\includegraphics[scale=0.4]{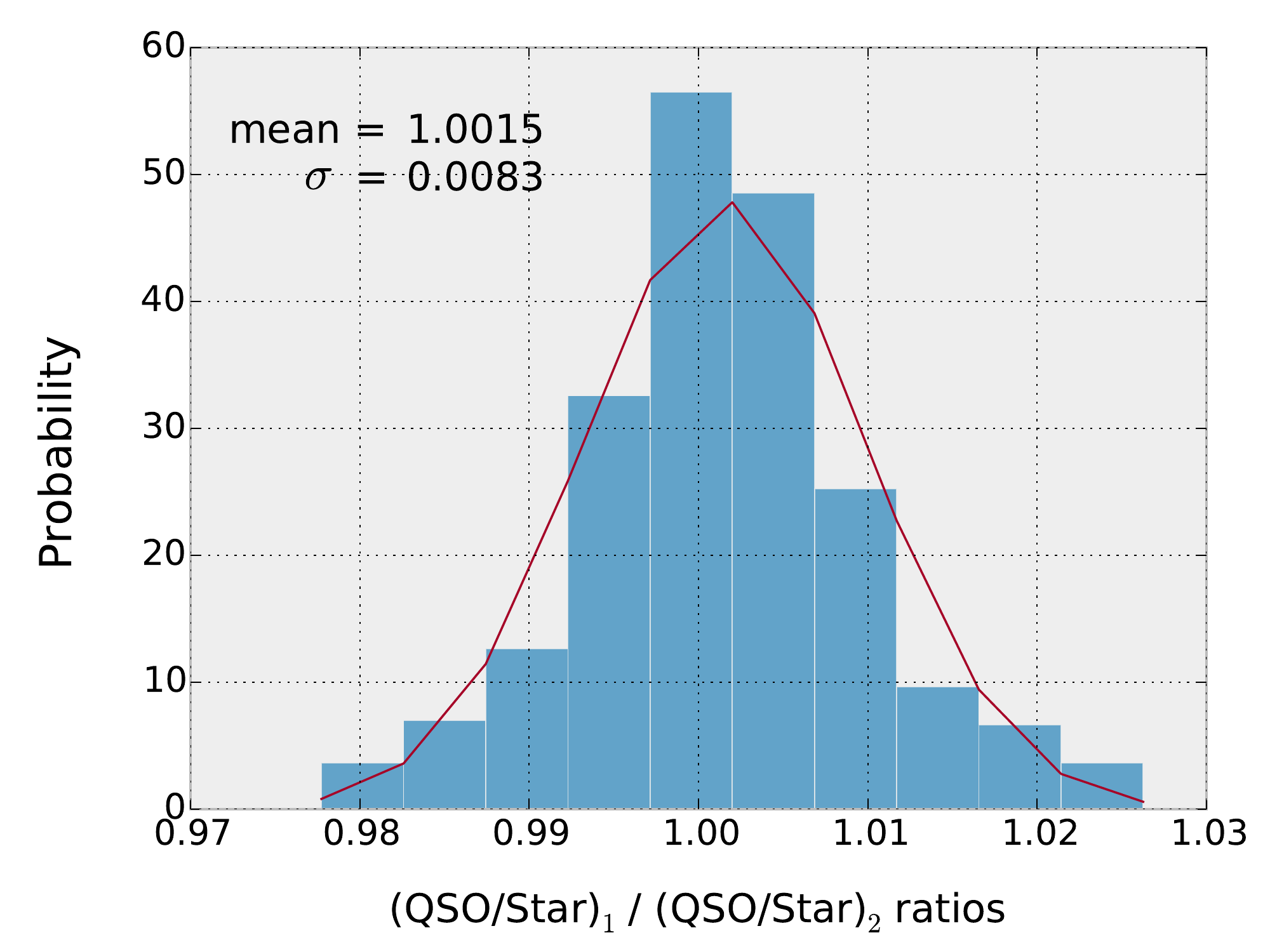}
\caption{Distribution of Quasar/Star ratios.}
\end{center}
\end{figure}

Errors in emission line measurements were estimated assuming that the
line flux $L$ can be expressed as $L = \alpha\, (F-C)$, where $\alpha$
is the scaling by the division of the comparison star, $F$ is the
total flux measured in the regions defined for line flux measurement
(i.e., between the yellow vertical lines shown in Figure 3) and $C$ is
the interpolated continuum as defined above. Hence, the variance for
each line measurement can be written as $\sigma_{\alpha}^2(F-C)^2 +
\alpha^2(\sigma_F^2 + \sigma_C^2)$, where $\sigma$ denotes uncertainty
in the quantity given by the subscript.

To estimate $\sigma_{\alpha}$, which encompasses effects such as poor
centering of the quasar-star pair on the slit or guiding problems, we
determined the ratio of two normalized-quasar spectra obtained during
a single observing run. Since typically three observations were
obtained per night, two such ratios could be constructed per quasar
observation. The ratio distribution using all available data for all
quasars can be seen in Figure 4, where the mean and standard deviation
of the distribution are also given.

To estimate $\sigma_{F}$ we used the error spectrum of each quasar
observation (which is obtained assuming photon Poisson statistics and
the specific gain and read-out noise of the detector) and determined
the total variance as the quadratic sum of the errors from each pixel
within the line window. During this step we did not take into account
the division by the comparison star as their spectra were heavily
smoothed before the division and therefore introduced no further
noise.

Finally, to estimate $\sigma_{C}$ we resorted to Monte Carlo (MC)
simulations because of the rather complex uncertainties that the
determination of the continuum level might introduce. For example,
continuum window placement can be affected by small mismatches in the
wavelength solution, which, in turn, will impact the flux measurement
in regions with strong flux fluctuations, like towards the blue end of
the spectra. This is particularly true for the Ly$\alpha$
pseudo-continuum windows, since the normalized quasar continuum flux
can increase sharply because of the division by a comparison star with
a spectral energy distribution that falls quickly towards the
blue. The Ly$\alpha$ blue pseudo-continuum window is also affected by
the shape of the continuum due to intergalactic absorption.

We obtained 10000 MC realizations for each line measurement where the
flux level in each pseudo-continuum window was determined using fluxes
drawn from a Gaussian distribution around the observed window flux
values. For the standard deviation of the distribution we adopted the
largest value between the flux rms within the pseudo-continuum window
and the photon noise derived from the error spectrum within the same
window. As expected, for Ly$\alpha$ the flux rms was consistently
larger than the photon noise. The determined flux distributions were
then normalized and integrated from the ends to a cumulative value of
0.159 (i.e., corresponding to a $1\sigma$ confidence limit). 
$\sigma_{C}$ was adopted as half the range given by these two limits.

We found that all three terms of the variance,
$\sigma_{\alpha}^2\,(F-C)^2$, $\alpha^2\, \sigma_F^2$ and
$\alpha^2\,\sigma_C^2$, were comparable and necessary to have a full
description of the emission line flux errors.

To increase the number of continuum measurements available for the
variability analysis, we measured the mean value of the continuum in
the 5620-7200\AA\ wavelength range from each spectroscopic
observation, as a proxy for R-band photometric values. These
`spectroscopic' points were later scaled to the broad-band R-band
photometry using a simple $\chi^2$ minimization to bring the mean
`spectroscopic' light-curve in line with the photometric values.

R-band, Ly$\alpha$, SiIV, CIV and CIII] light curves are presented in
  Figure 5 for all objects except for CT252, for which CIV, CIII] and
    MgII light curves are presented. The full database is found in the
    Appendix.

\section{Variability Analysis}

\subsection{General continuum and emission line variability}

\begin{table*}
\renewcommand{\thetable}{\arabic{table}}
\centering
\caption{Light curve variability statistics}
\begin{tabular}{rrrrrrrrrrrrrrrr}
\tablewidth{0pt}
\hline
\hline
Quasar & l.c. & N & $R_{max}$ & $f_{var}$ & $\chi^2$ & $P_{\chi}$ & cc & Quasar & l.c. & N & $R_{max}$ & $f_{var}$ & $\chi^2$ & $P_{\chi}$ & cc\\
\hline
CT1061& Lya&    13&     1.05&   0.00&   0.8&    0.30& y & CT803 & Lya&    21&     1.15&   0.02&   2.0&    1.00& n\\  
&       SiIV&   13&     1.12&   0.00&   0.1&    0.00& n & &       SiIV&   21&     1.31&   0.05&   3.0&    1.00& n\\  
&       CIV&    13&     1.10&   0.01&   1.4&    0.83& y & &       CIV&    21&     1.15&   0.00&   0.5&    0.04& n\\  
&       CIII]&  13&     1.20&   0.00&   0.6&    0.14& n & &       CIII]&  21&     1.26&   0.00&   0.6&    0.06& n\\  
&       R&      65&     1.32&   0.08&   56.8&   1.00& --  &       R&      79&     1.47&   0.08&   50.3&   1.00& --\\  
CT250 & Lya&    17&     1.22&   0.05&   11.9&   1.00& y & CT953 & Lya&    23&     1.35&   0.07&   11.4&   1.00& y\\  
&       SiIV&   17&     1.54&   0.00&   1.0&    0.58& y & &       SiIV&   23&     1.71&   0.08&   3.9&    1.00& y\\  
&       CIV&    17&     1.39&   0.06&   4.7&    1.00& y & &       CIV&    23&     1.27&   0.05&   5.4&    1.00& y\\  
&       CIII]&  17&     1.40&   0.03&   2.0&    0.99& n & &       CIII]&  23&     1.21&   0.00&   1.3&    0.82& n\\  
&       R&      67&     1.47&   0.11&   81.1&   1.00& --  &       R&      74&     1.61&   0.12&   592.3&  1.00& --\\  
CT252 & CIV&    17&     1.25&   0.07&   34.4&   1.00& n & CT975 & Lya&    17&     1.20&   0.03&   1.6&    0.94& n\\  
&       CIII&   17&     1.10&   0.01&   1.3&    0.78& n & &       SiIV&   17&     2.73&   0.00&   0.9&    0.49& n\\  
&       MgII&   17&     1.49&   0.09&   15.6&   1.00& n & &       CIV&    17&     1.33&   0.05&   2.4&    1.00& n\\  
&       R&      62&     1.41&   0.07&   962.7&  1.00& --  &       CIII]&  17&     2.70&   0.12&   2.8&    1.00& n\\  
&       &       &       &       &       &           &     &       R&      63&     1.94&   0.12&   86.5&   1.00& --\\  
CT286 & Lya&    23&     1.20&   0.04&   8.8&    1.00& y & HB89 0329-385&Lya&22&   1.31&   0.05&   7.1&    1.00& y\\  
&       SiIV&   23&     1.36&   0.00&   0.5&    0.04& y & &       SiIV&   22&     1.91&   0.00&   0.9&    0.43& y\\  
&       CIV&    23&     1.25&   0.04&   2.5&    1.00& y & &       CIV&    22&     1.09&   0.00&   1.4&    0.88& n\\  
&       CIII]&  23&     1.16&   0.00&   0.4&    0.01& n & &       CIII]&  22&     1.21&   0.00&   0.9&    0.40& y\\  
&       R&      85&     1.46&   0.10&   572.7&  1.00& --  &       R&      68&     1.38&   0.09&   288.0&  1.00& --\\  
CT320 & Lya&    26&     1.19&   0.04&   14.9&   1.00& y & 2QZJ002830&Lya&  9&     1.22&   0.06&   27.9&   1.00& n\\  
&       SiIV&   26&     1.57&   0.10&   5.6&    1.00& y & &       SiIV&    9&     1.53&   0.11&   5.9&    1.00& n\\  
&       CIV&    26&     1.21&   0.04&   8.6&    1.00& y & &       CIV&     9&     1.27&   0.07&   21.0&   1.00& n\\  
&       CIII]&  26&     1.32&   0.05&   5.8&    1.00& y & &       CIII]&   9&     1.14&   0.03&   2.0&    0.96& n\\  
&       R&      74&     1.53&   0.10&   167.3&  1.00& --  &       R&      54&     1.64&   0.11&   98.7&   1.00& --\\  
CT367 & Lya&    12&     1.19&   0.04&   18.3&   1.00& n & 2QZJ214355&Lya& 16&     1.21&   0.00&   0.9&    0.41& y\\  
&       SiIV&   12&     2.64&   0.24&   94.1&   1.00& n & &       SiIV&   16&     1.33&   0.00&   0.5&    0.08& n\\  
&       CIV&    12&     1.12&   0.03&   5.6&    1.00& n & &       CIV&    16&     1.33&   0.03&   2.5&    1.00& n\\  
&       CIII]&  12&     1.25&   0.05&   4.5&    1.00& n & &       CIII]&  16&     1.60&   0.00&   0.3&    0.00& n\\  
&       R&      66&     1.85&   0.19&   224.7&  1.00& --  &       R&      64&     1.35&   0.07&   439.0&  1.00& --\\  
CT406 & Lya&    15&     1.10&   0.03&   8.3&    1.00& y & 2QZJ221516&Lya& 21&     1.14&   0.03&   19.9&   1.00& y\\  
&       SiIV&   15&     1.52&   0.10&   3.6&    1.00& n & &       SiIV&   21&     1.56&   0.05&   1.2&    0.78& n\\  
&       CIV&    15&     1.25&   0.04&   4.4&    1.00& y & &       CIV&    21&     1.16&   0.03&   2.9&    1.00& y\\  
&       CIII]&  15&     2.52&   0.12&   6.5&    1.00& n & &       CIII]&  21&     1.36&   0.00&   0.9&    0.43& n\\  
&       R&      62&     1.39&   0.07&   59.5&   1.00& --  &       R&      78&     1.34&   0.08&   110.1&  1.00& --\\  
CT564 & Lya&    12&     1.15&   0.04&   11.5&   1.00& y & 2QZJ224743&Lya& 17&     1.46&   0.11&   76.3&   1.00& n\\  
&       SiIV&   12&     4.21&   0.11&   1.2&    0.70& n & &       SiIV&   17&     1.52&   0.11&   12.5&   1.00& n\\  
&       CIV&    12&     1.19&   0.02&   1.7&    0.93& y & &       CIV&    17&     1.11&   0.00&   1.7&    0.96& n\\  
&       CIII]&  12&     1.46&   0.00&   0.7&    0.25& n & &       CIII]&  17&     1.20&   0.01&   2.1&    0.99& n\\  
&       R&      66&     1.45&   0.10&   74.9&   1.00& --  &       R&      60&     1.29&   0.05&   55.2&   1.00& --\\  
CT650 & Lya&    25&     1.17&   0.03&   70.5&   1.00& y &&&&&&&&\\
&       SiIV&   25&     7.66&   0.29&   13.8&   1.00& n &&&&&&&&\\
&       CIV&    25&     1.14&   0.03&   41.4&   1.00& y &&&&&&&&\\
&       CIII]&  25&     1.27&   0.06&   39.8&   1.00& y &&&&&&&&\\
&       R&      76&     2.03&   0.15&   176.7&  1.00& --&&&&&&&&\\
\hline
\hline
\end{tabular}
\end{table*}

\setlength{\unitlength}{1cm}

\begin{figure*}
\includegraphics[scale=0.6,trim=0 0 0 0]{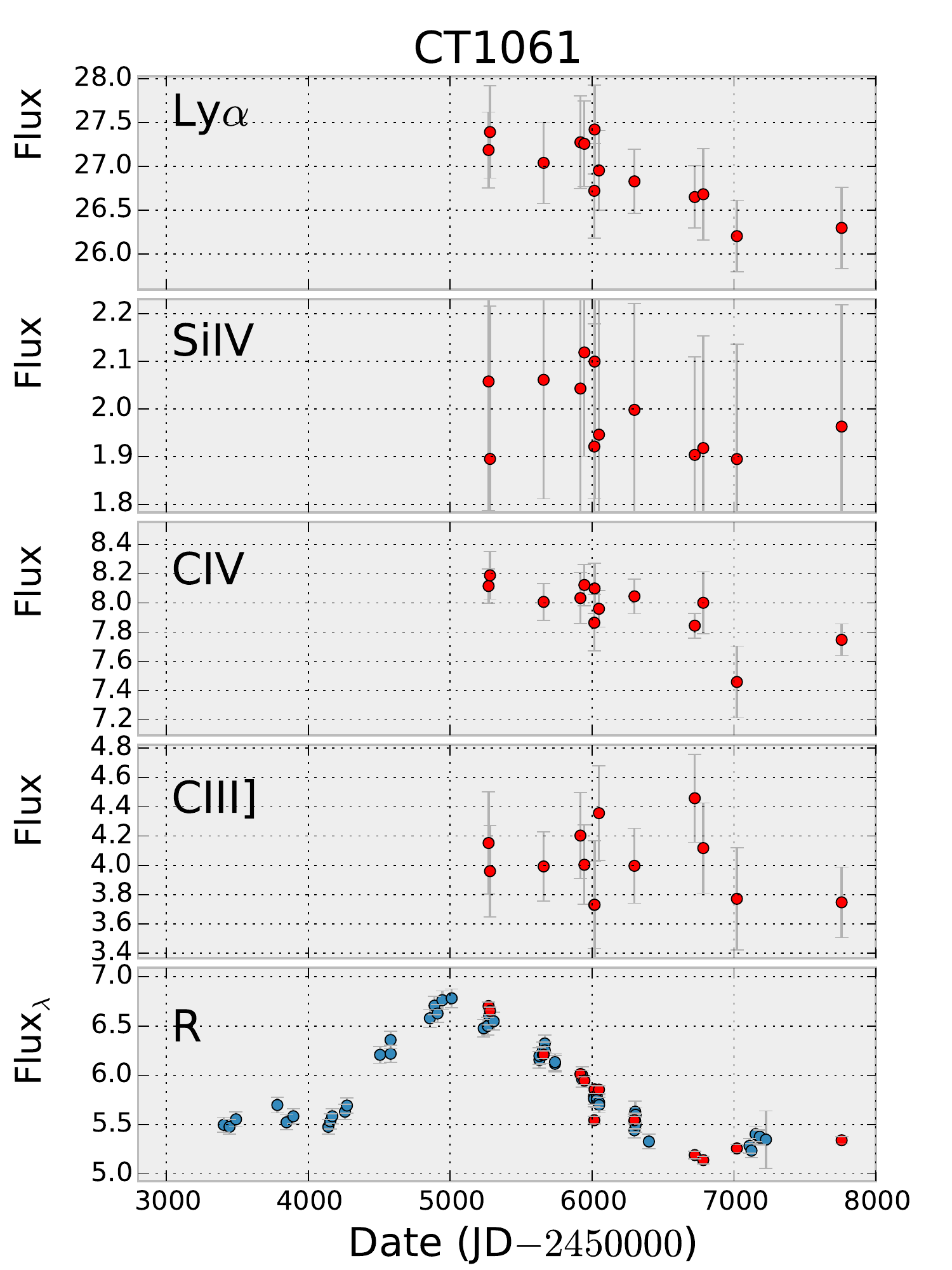}%
\begin{picture}(10,12){
\put(1,6.2){\includegraphics[scale=0.36,trim=0 0 0 0]{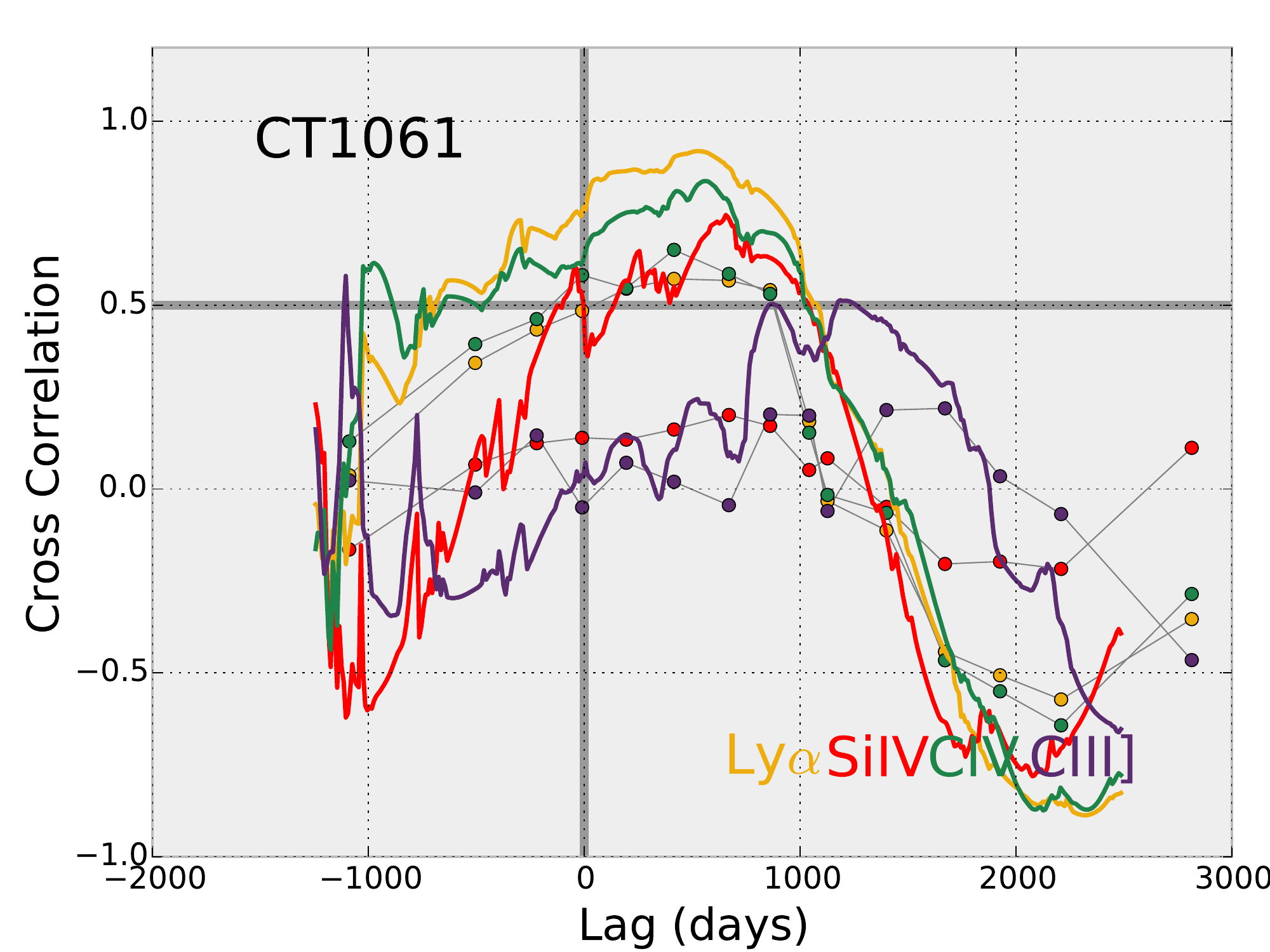}}
\put(0.7,0){\includegraphics[scale=0.40,trim=0 0 0 0]{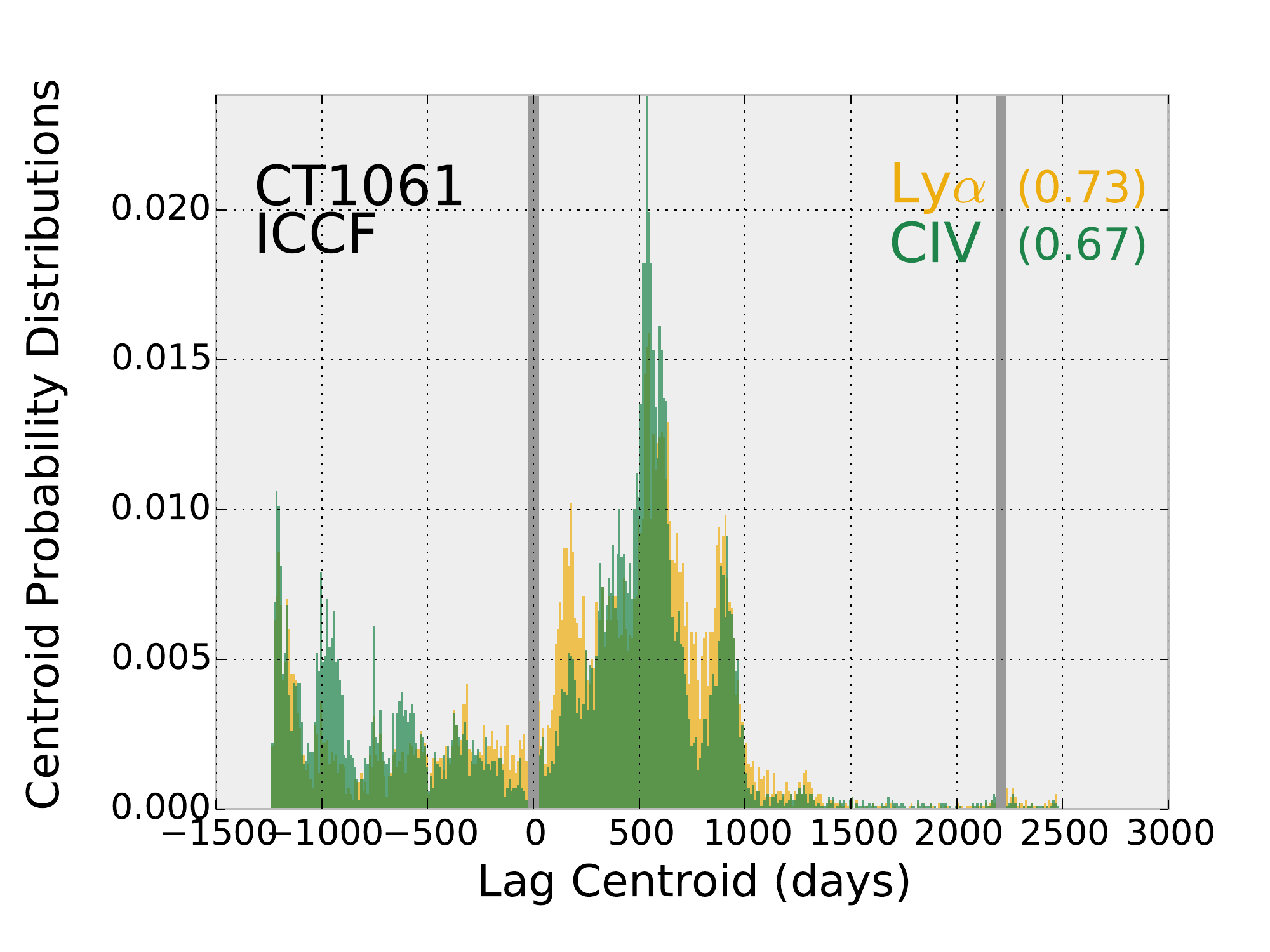}}}
\end{picture}
\caption{{\bf Left:} Line and continuum light curves for all quasars
  with spectroscopic follow-up. Red points correspond to measurements
  taken from the spectroscopic data, while blue points correspond to
  broad-band R photometry. Units are $10^{-14}$ erg s$^{-1}$ cm$^{-2}$
  for the line measurements and $10^{-16}$ erg s$^{-1}$ cm$^{-2}$
  \AA$^{-1}$ for the R-band light curves. {\bf Right, top panels:}
  correlation functions obtained using the ICCF (continuous line) and
  ZDCF (points) methods for the Ly$\alpha$, SiIV, CIV, and CIII]
emission lines. {\bf Right, bottom panels:} Cross Correlation Centroid
Distributions (CCCDs) obtained from the ICCF FR/RSS analysis for those
lines with significant peaks in their ZDCF and ICCF correlation
functions. Thick gray lines show the limits of the distributions used
for lag error determinations. The fraction of the CCCD contained
within the thresholds is shown in parenthesis for each line. For
further details, see the text.}
\end{figure*}

\begin{figure*}
\includegraphics[scale=0.6,trim=0 0 0 0]{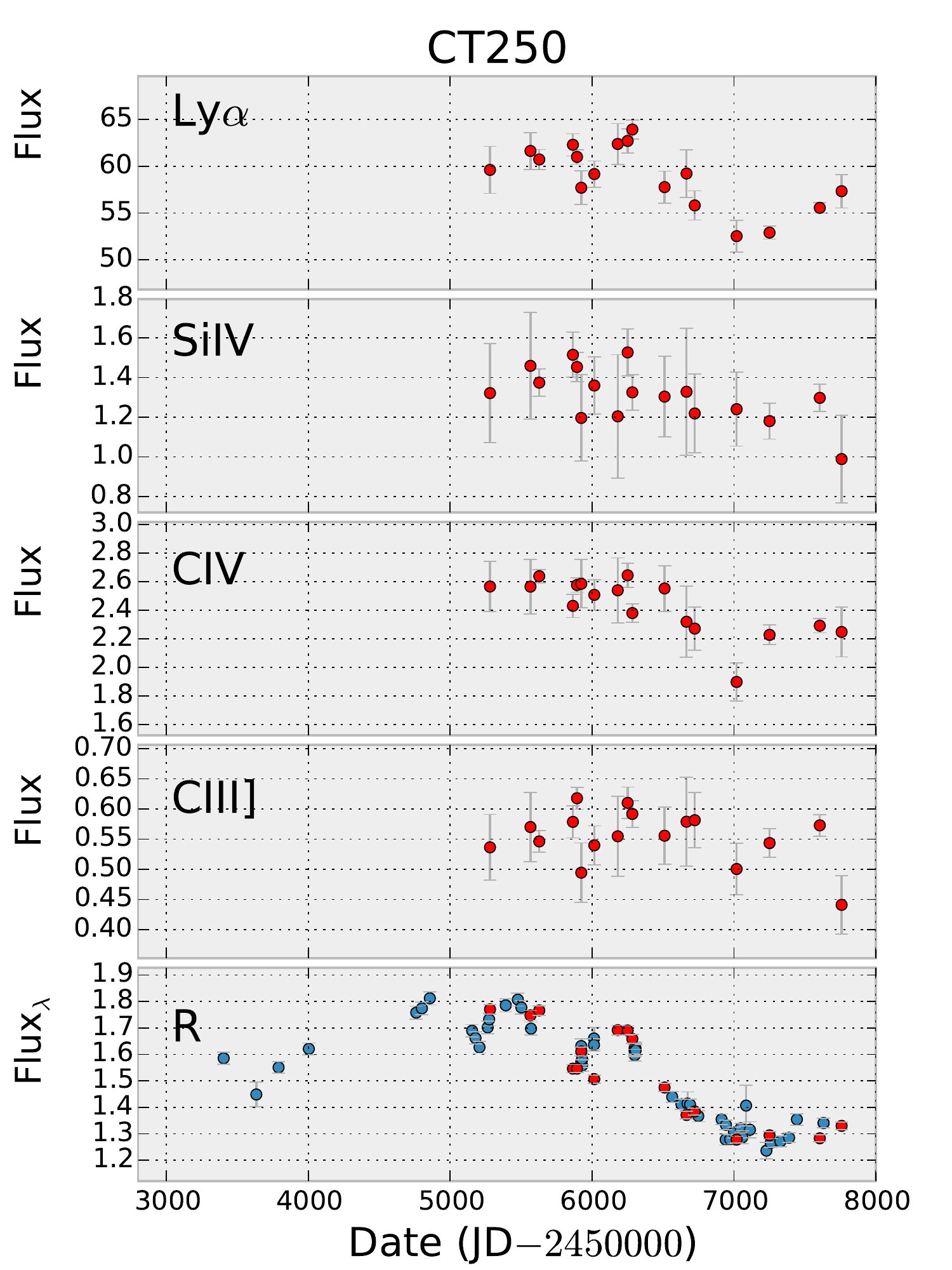}%
\begin{picture}(10,12){
\put(1,6.2){\includegraphics[scale=0.36,trim=0 0 0 0]{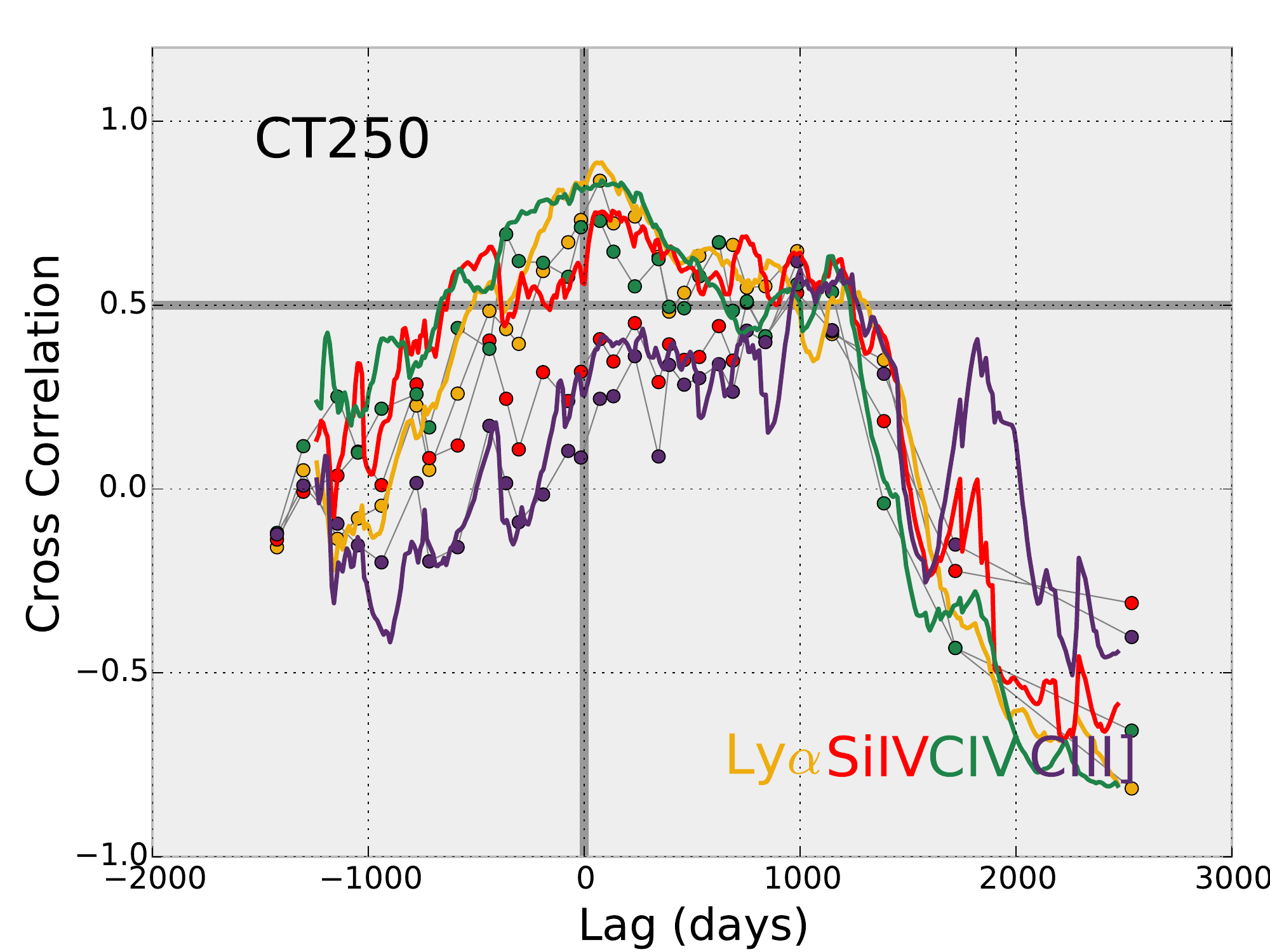}}
\put(0.7,0){\includegraphics[scale=0.40,trim=0 0 0 0]{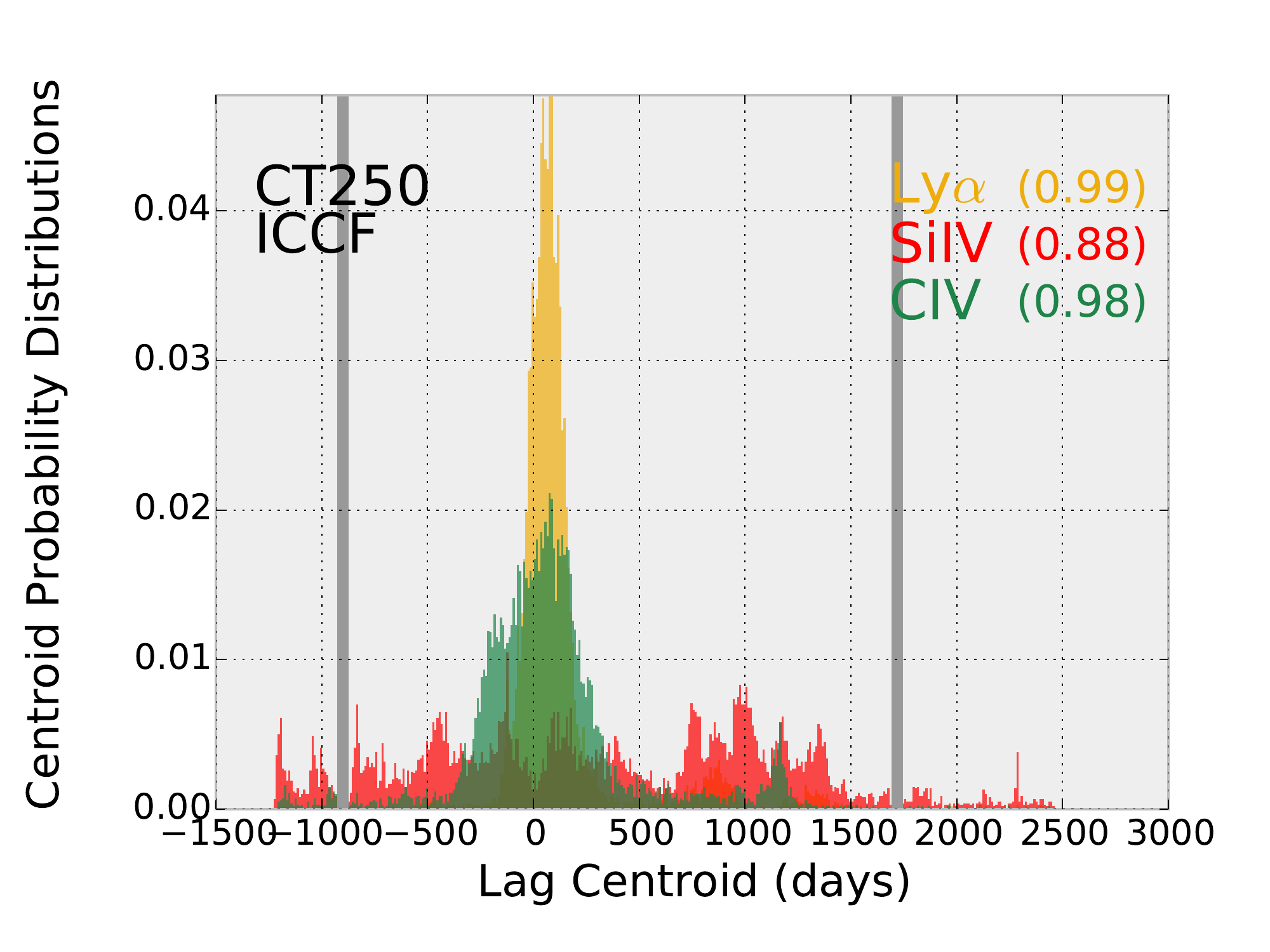}}}
\end{picture}
\includegraphics[scale=0.6,trim=0 0 0 0]{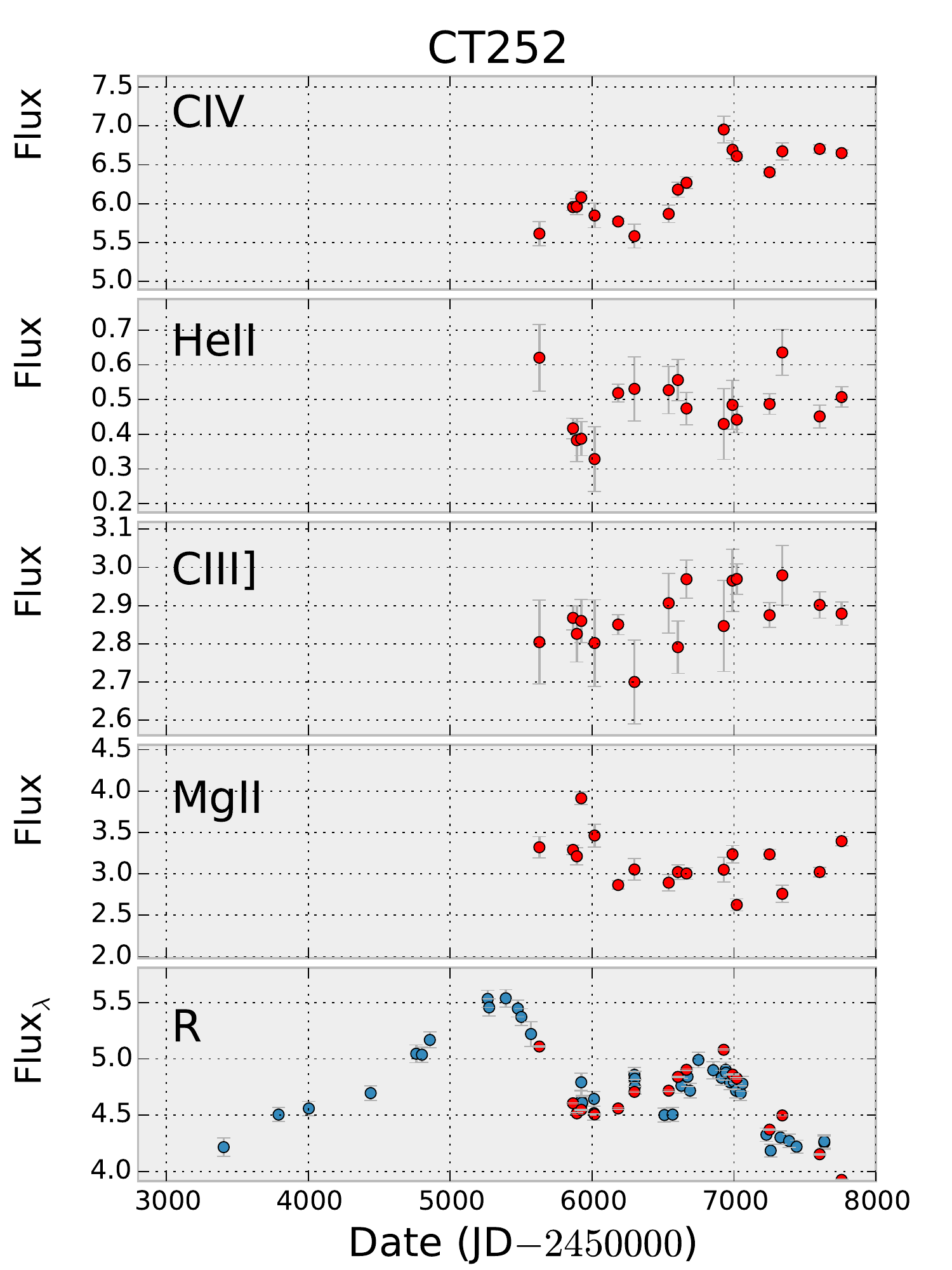}%
\begin{picture}(10,12){
\put(1,6.2){\includegraphics[scale=0.36,trim=0 0 0 0]{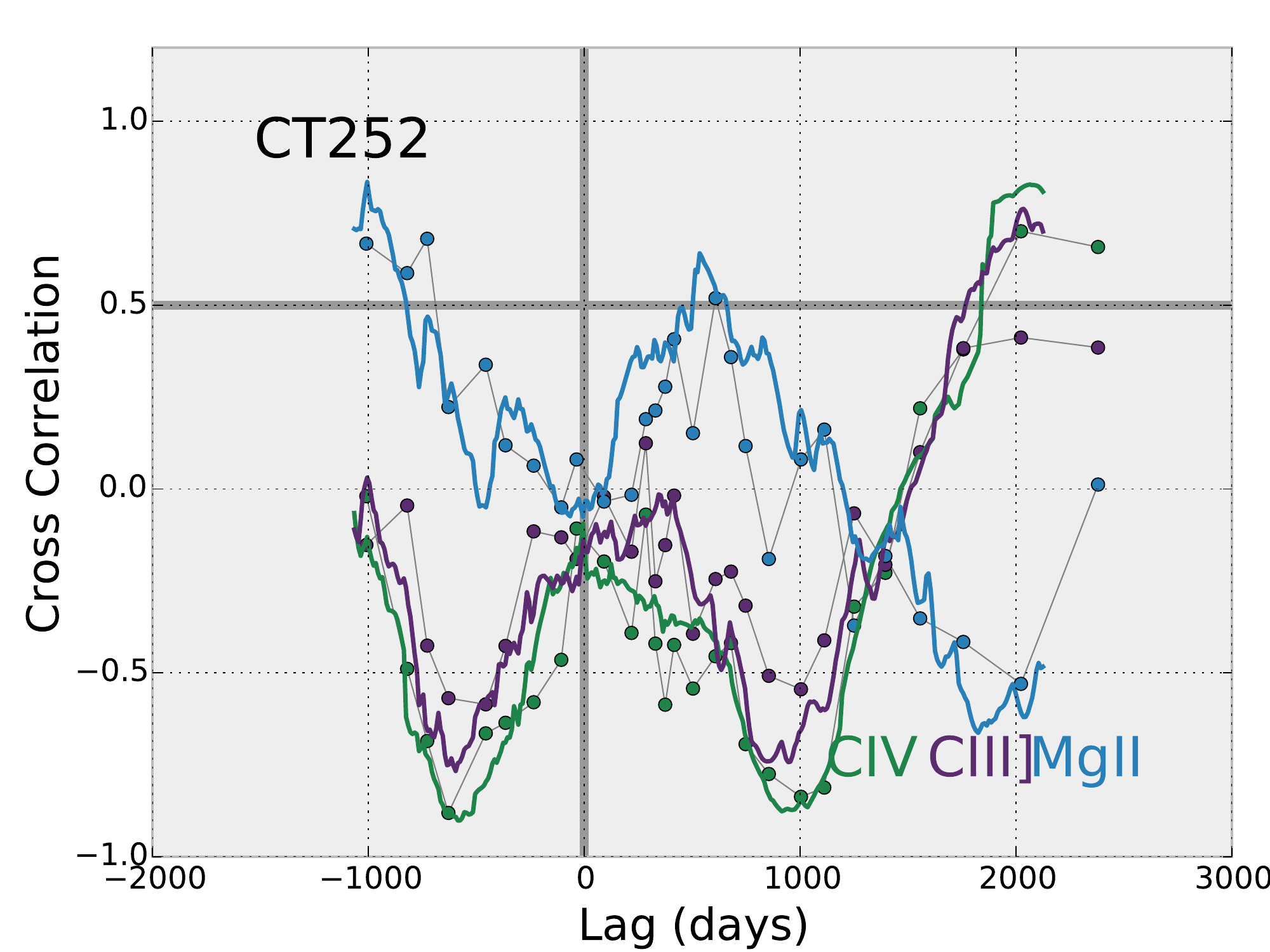}}
\put(0.7,0){\includegraphics[scale=0.40,trim=0 0 0 0]{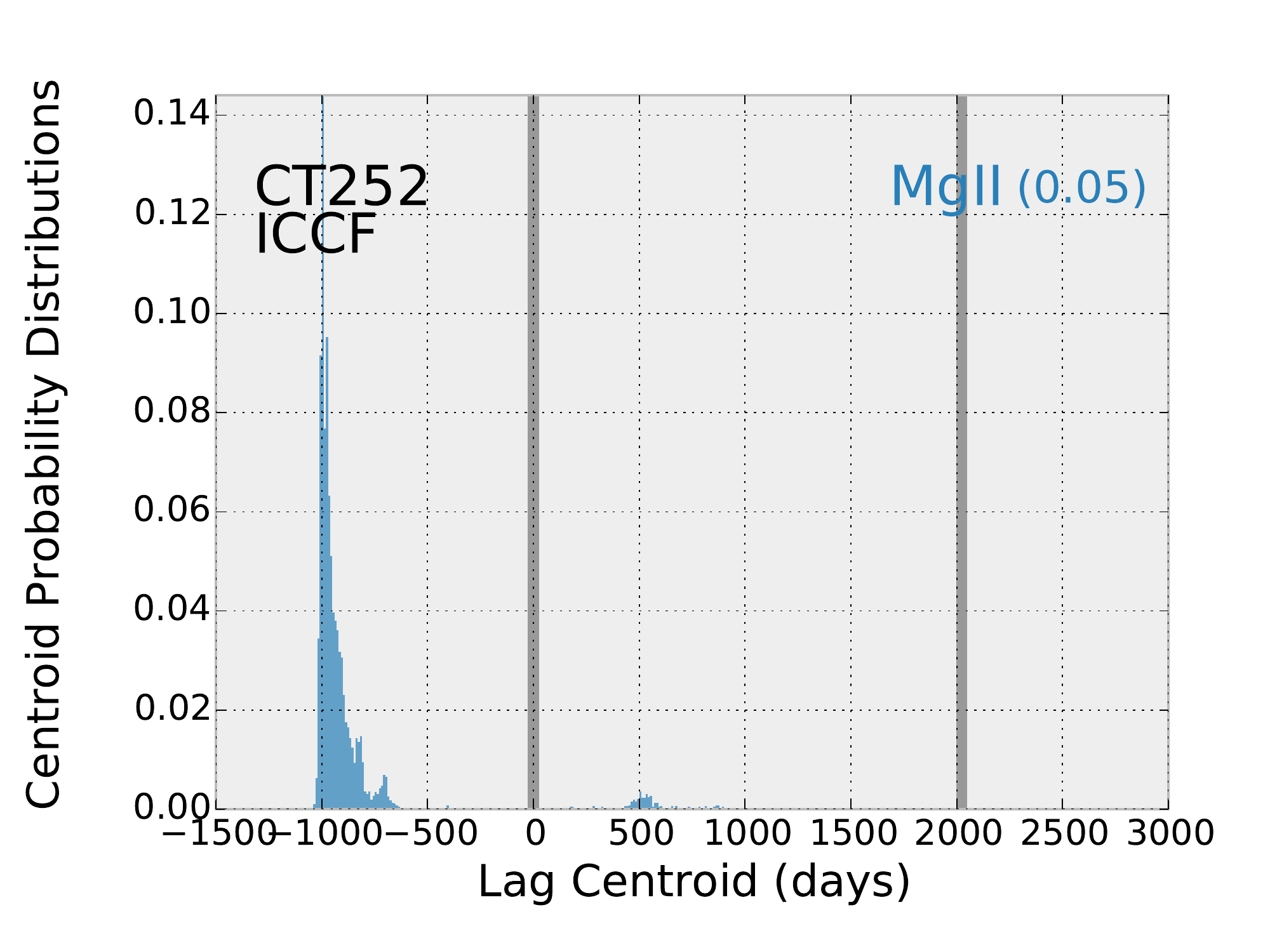}}}
\end{picture}
\end{figure*}

\begin{figure*}
\includegraphics[scale=0.6,trim=0 0 0 0]{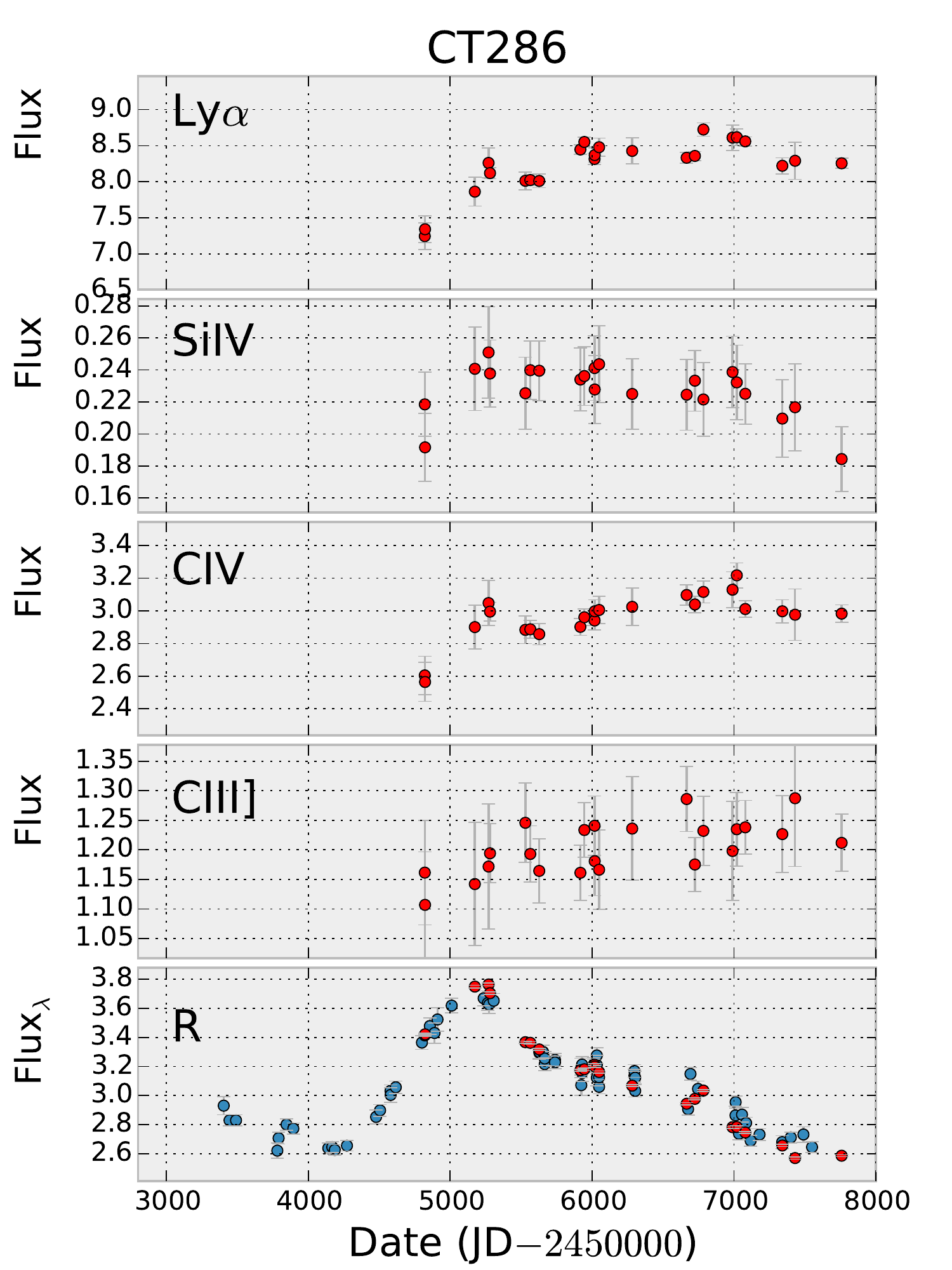}%
\begin{picture}(10,12){
\put(1,6.2){\includegraphics[scale=0.36,trim=0 0 0 0]{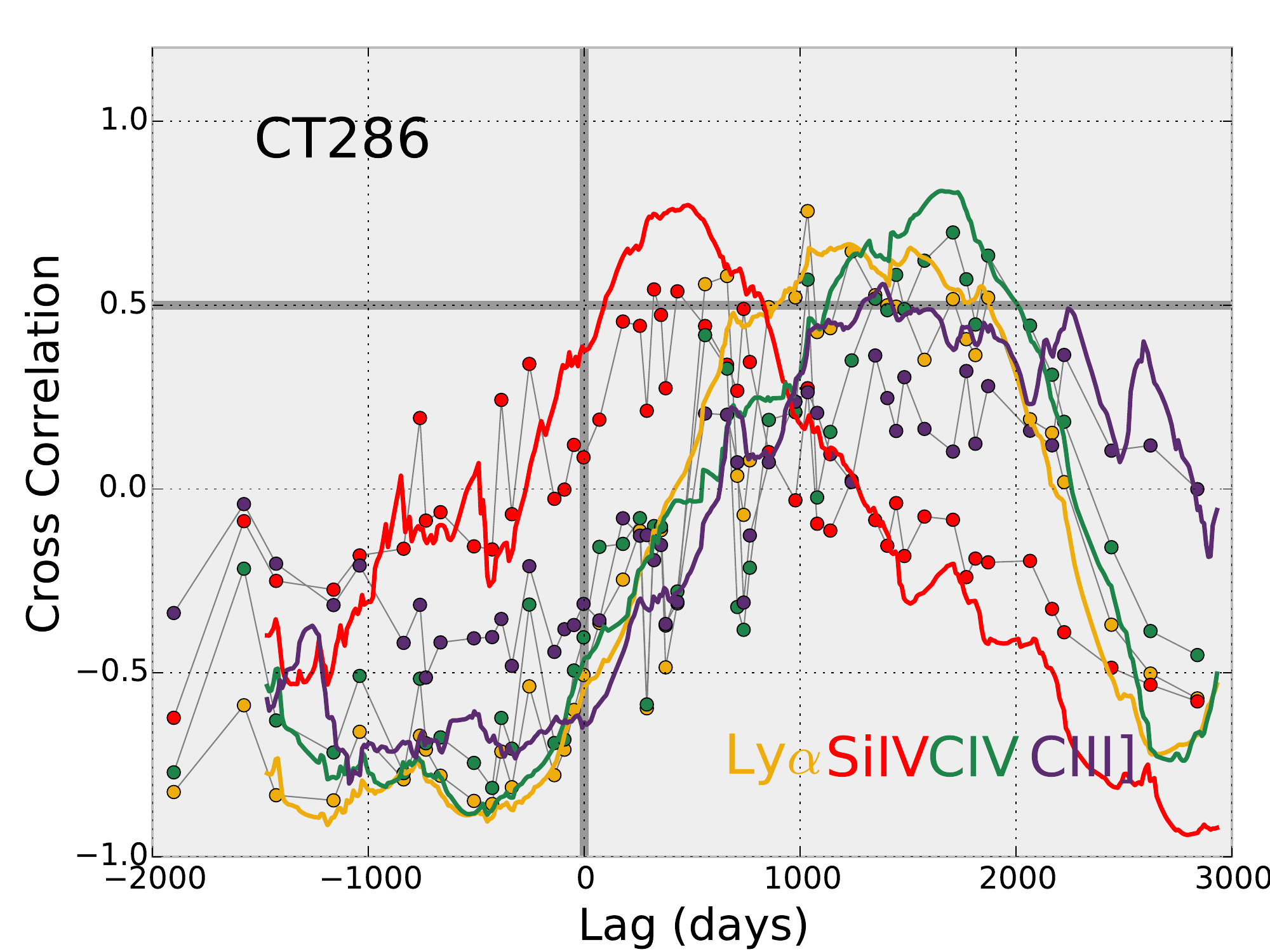}}
\put(0.7,0){\includegraphics[scale=0.40,trim=0 0 0 0]{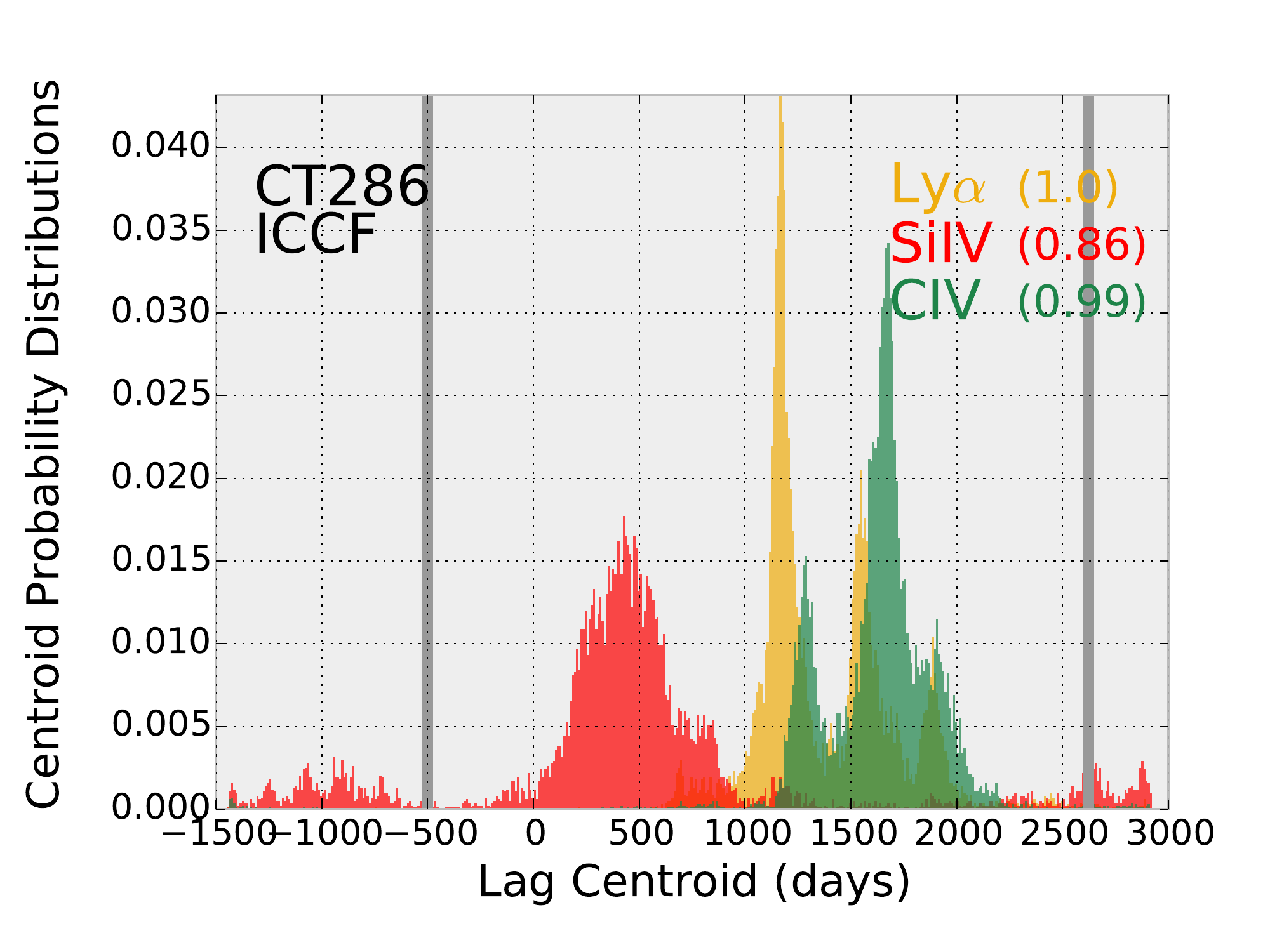}}}
\end{picture}
\includegraphics[scale=0.6,trim=0 0 0 0]{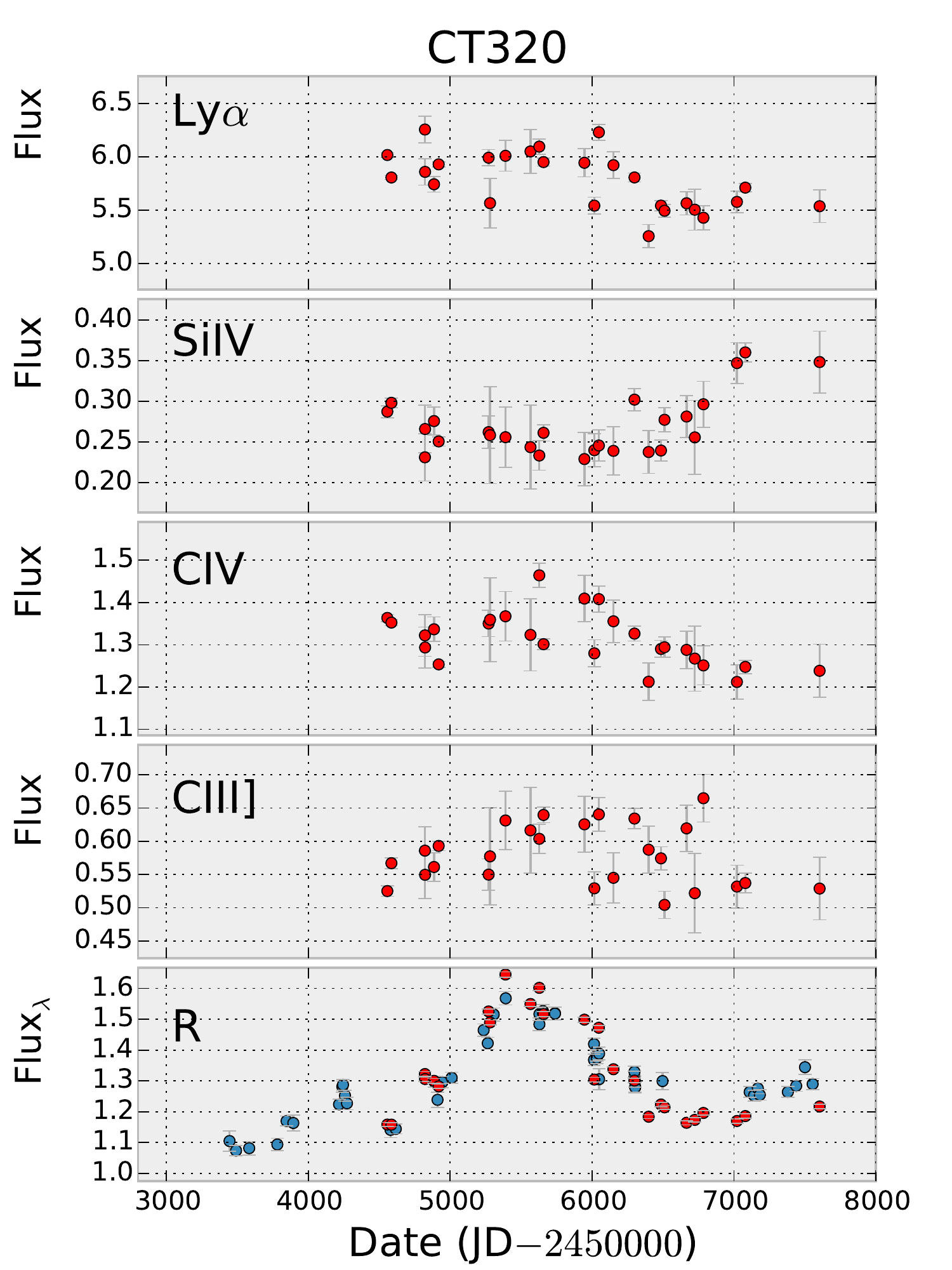}%
\begin{picture}(10,12){
\put(1,6.2){\includegraphics[scale=0.36,trim=0 0 0 0]{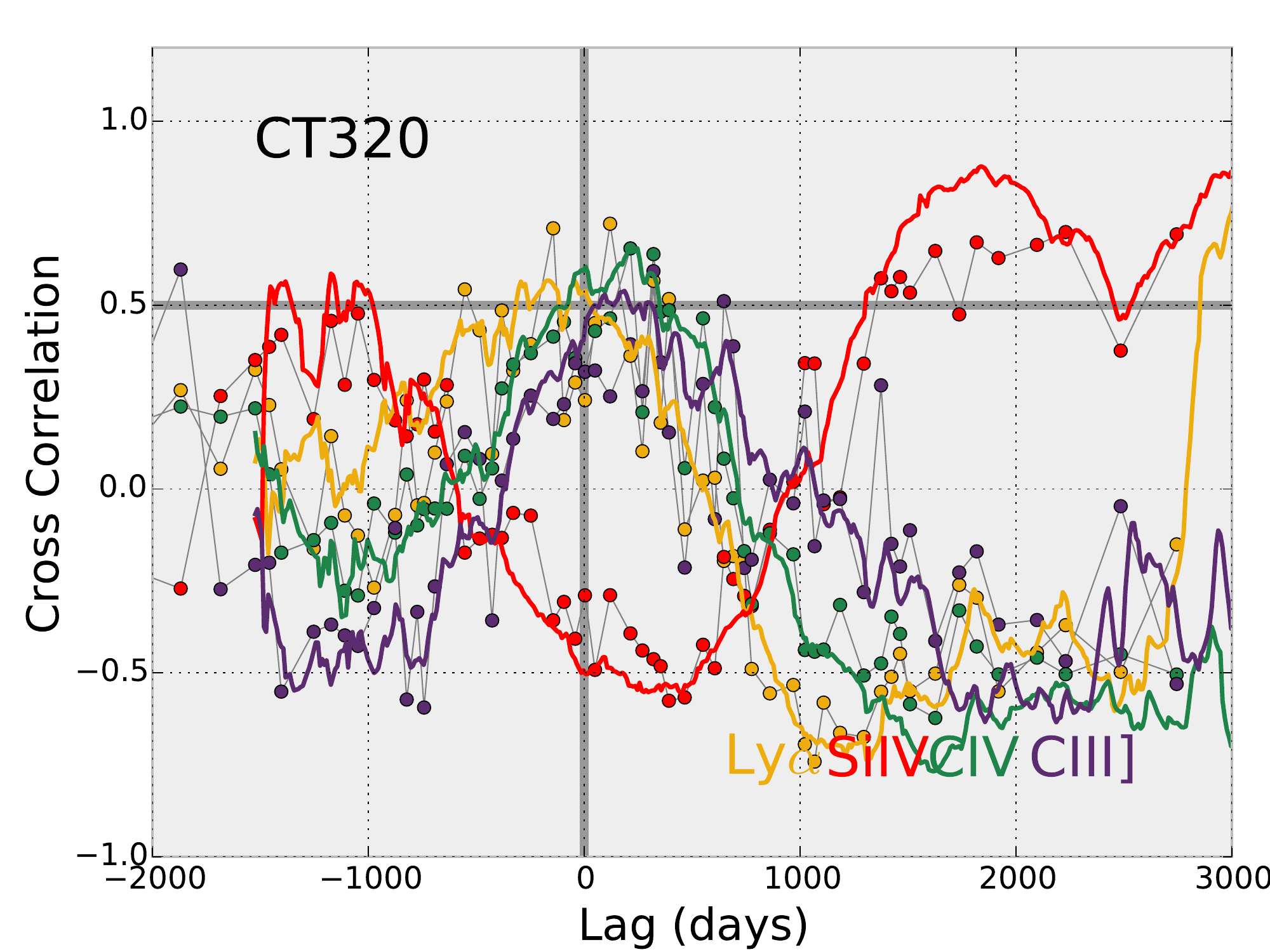}}
\put(0.7,0){\includegraphics[scale=0.40,trim=0 0 0 0]{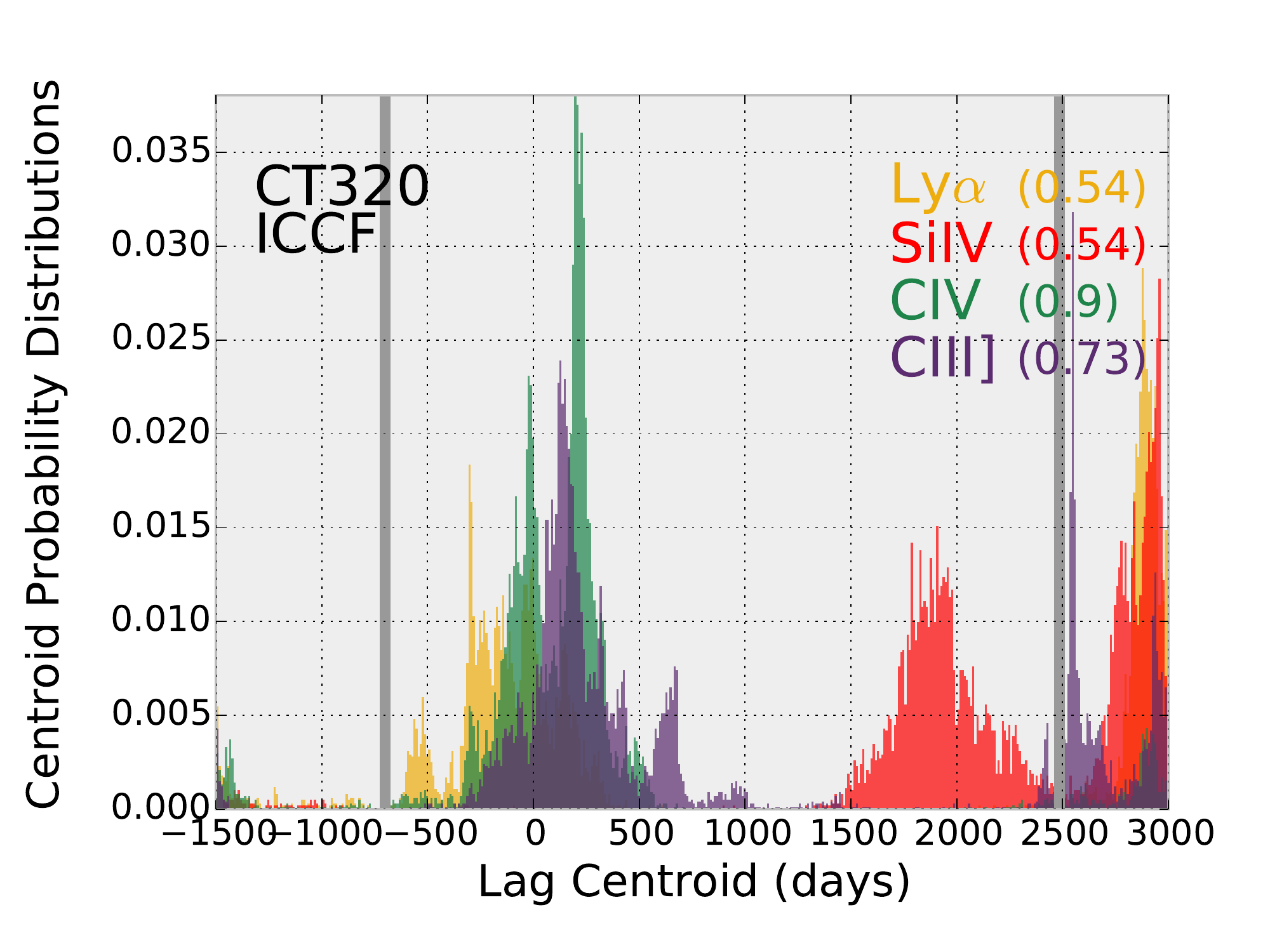}}}
\end{picture}
\end{figure*}

\begin{figure*}
\includegraphics[scale=0.6,trim=0 0 0 0]{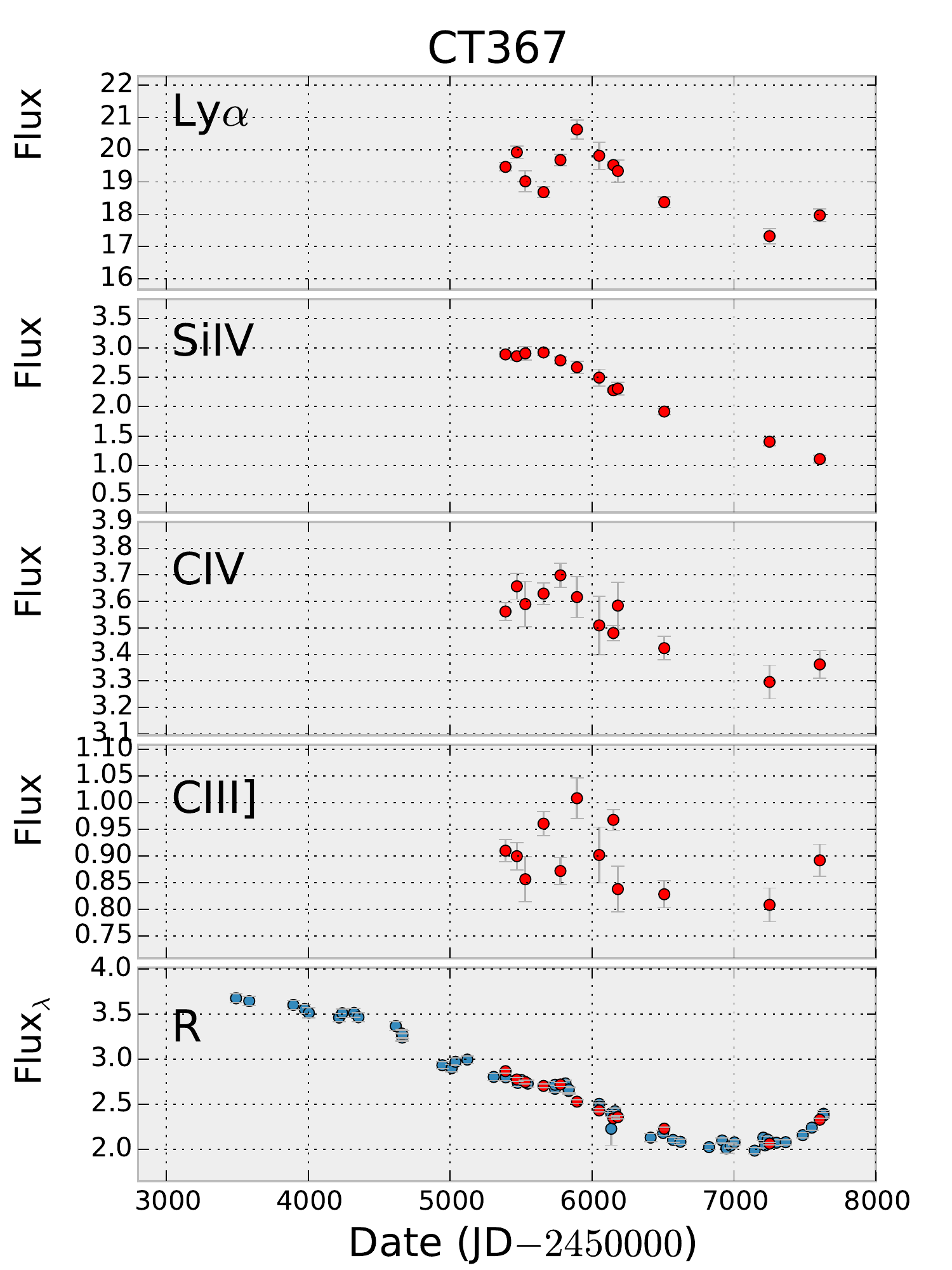}%
\begin{picture}(10,12){
\put(1,3){\includegraphics[scale=0.36,trim=0 0 0 0]{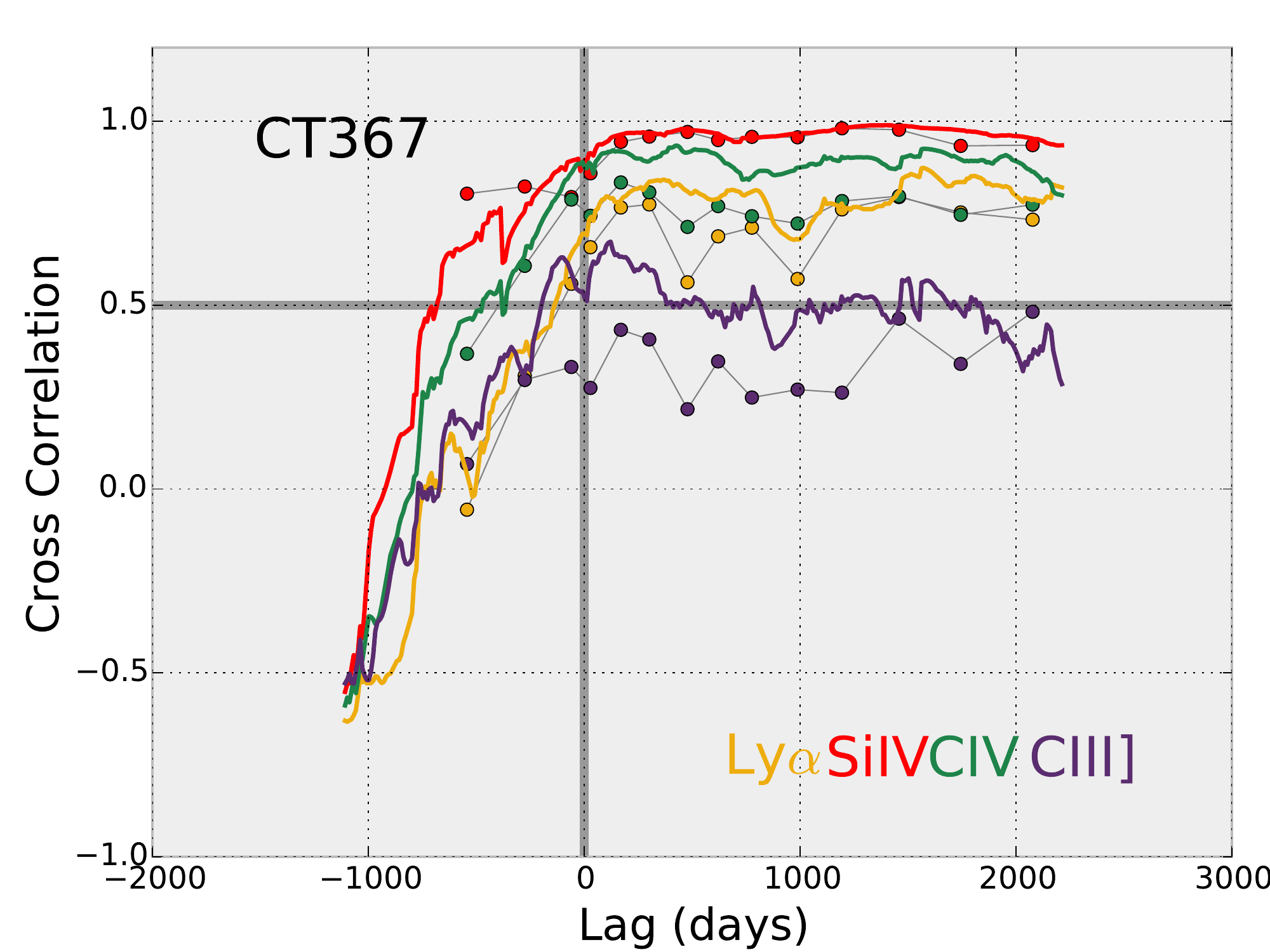}}
\put(0.7,0){}}
\end{picture}
\includegraphics[scale=0.6,trim=0 0 0 0]{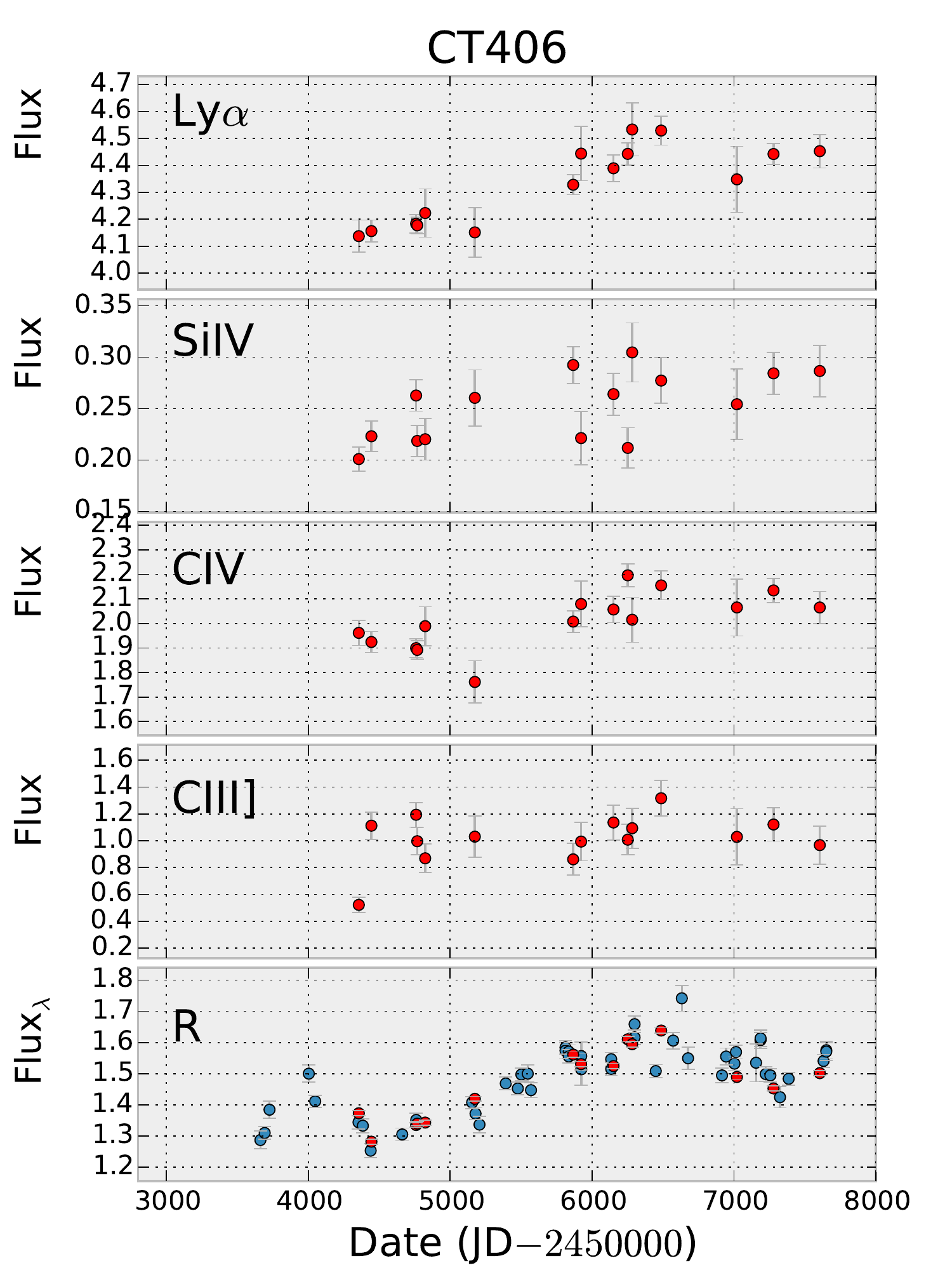}%
\begin{picture}(10,12){
\put(1,6.2){\includegraphics[scale=0.36,trim=0 0 0 0]{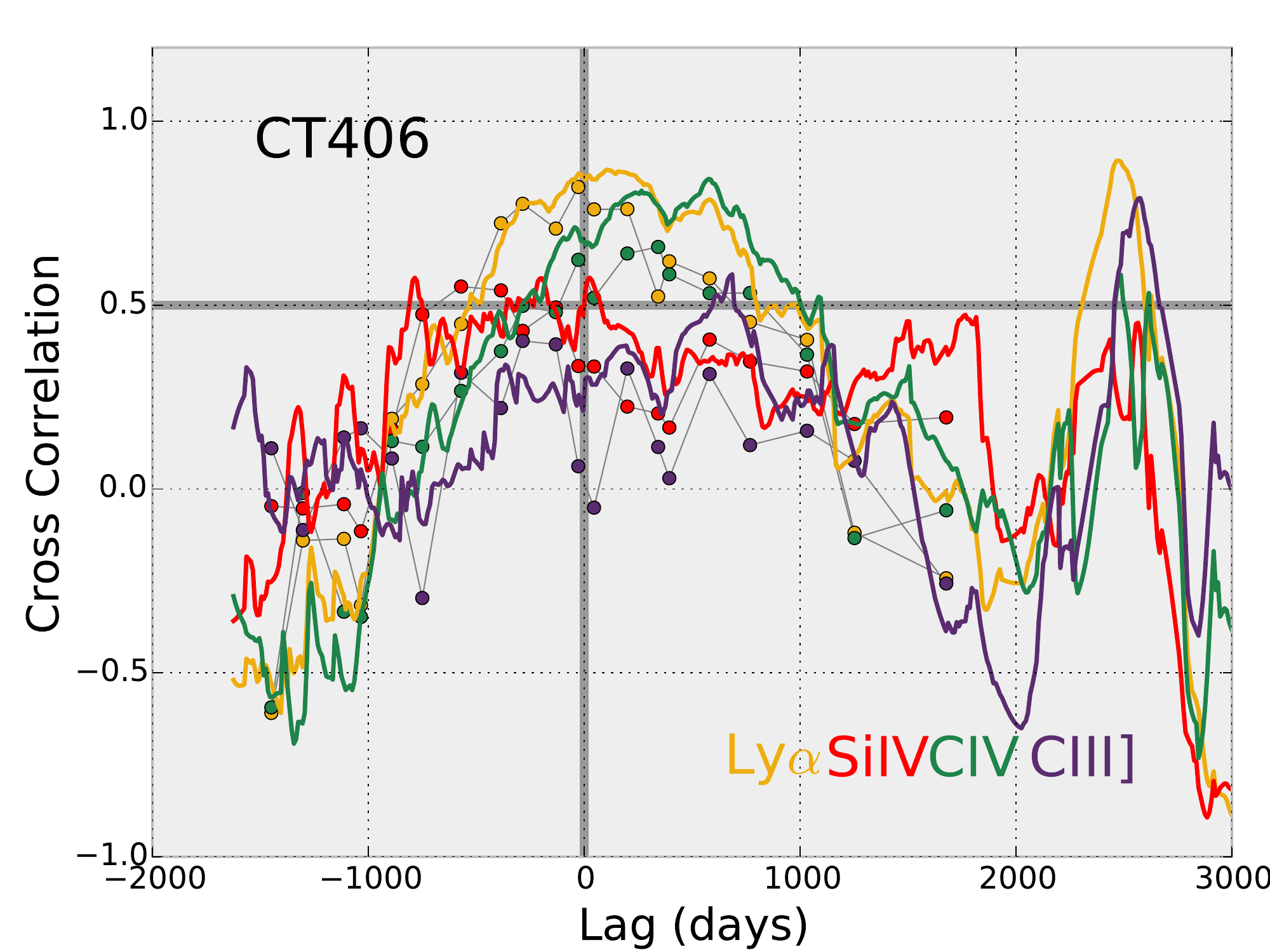}}
\put(0.7,0){\includegraphics[scale=0.40,trim=0 0 0 0]{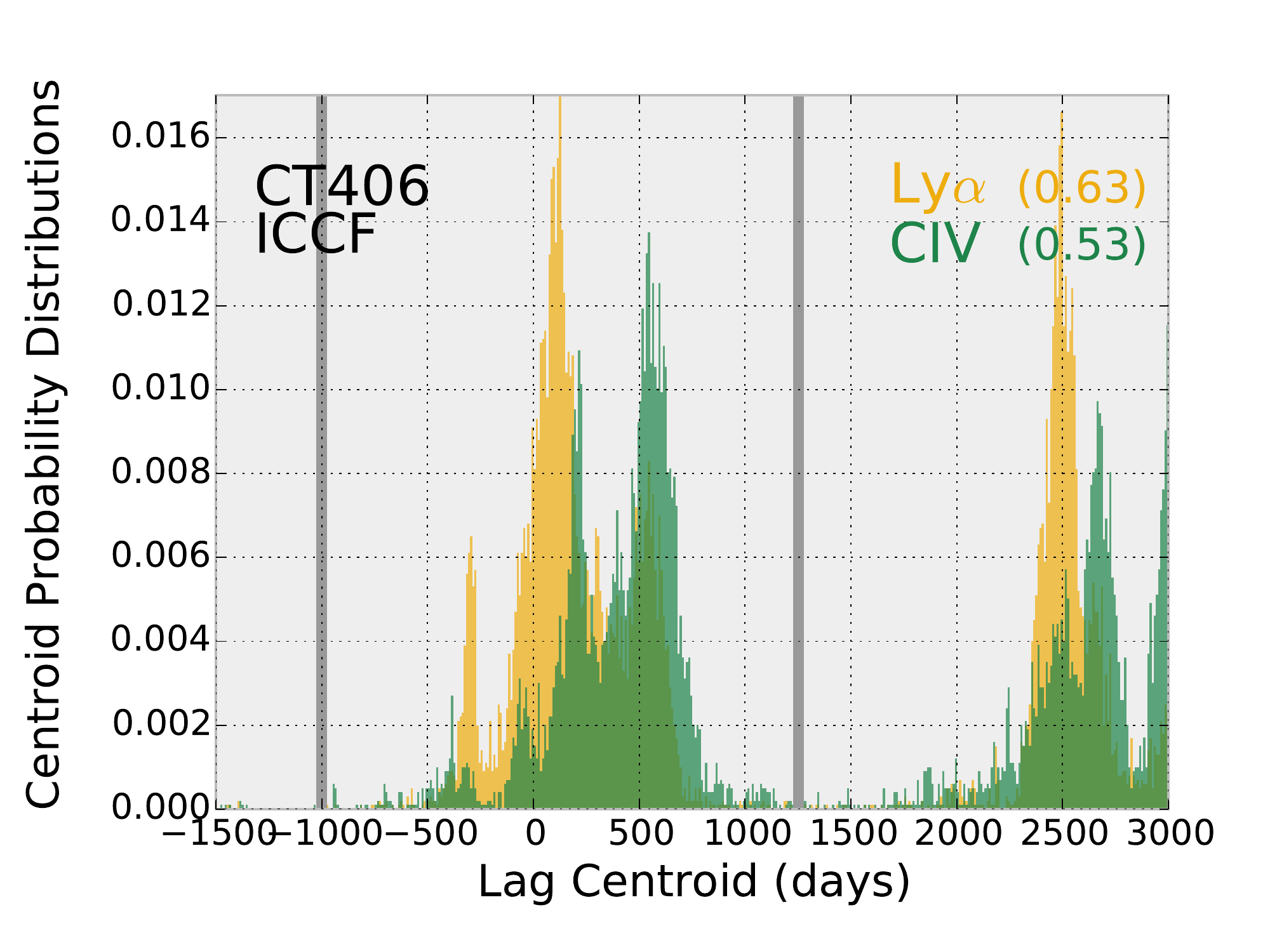}}}
\end{picture}
\end{figure*}

\begin{figure*}
\includegraphics[scale=0.6,trim=0 0 0 0]{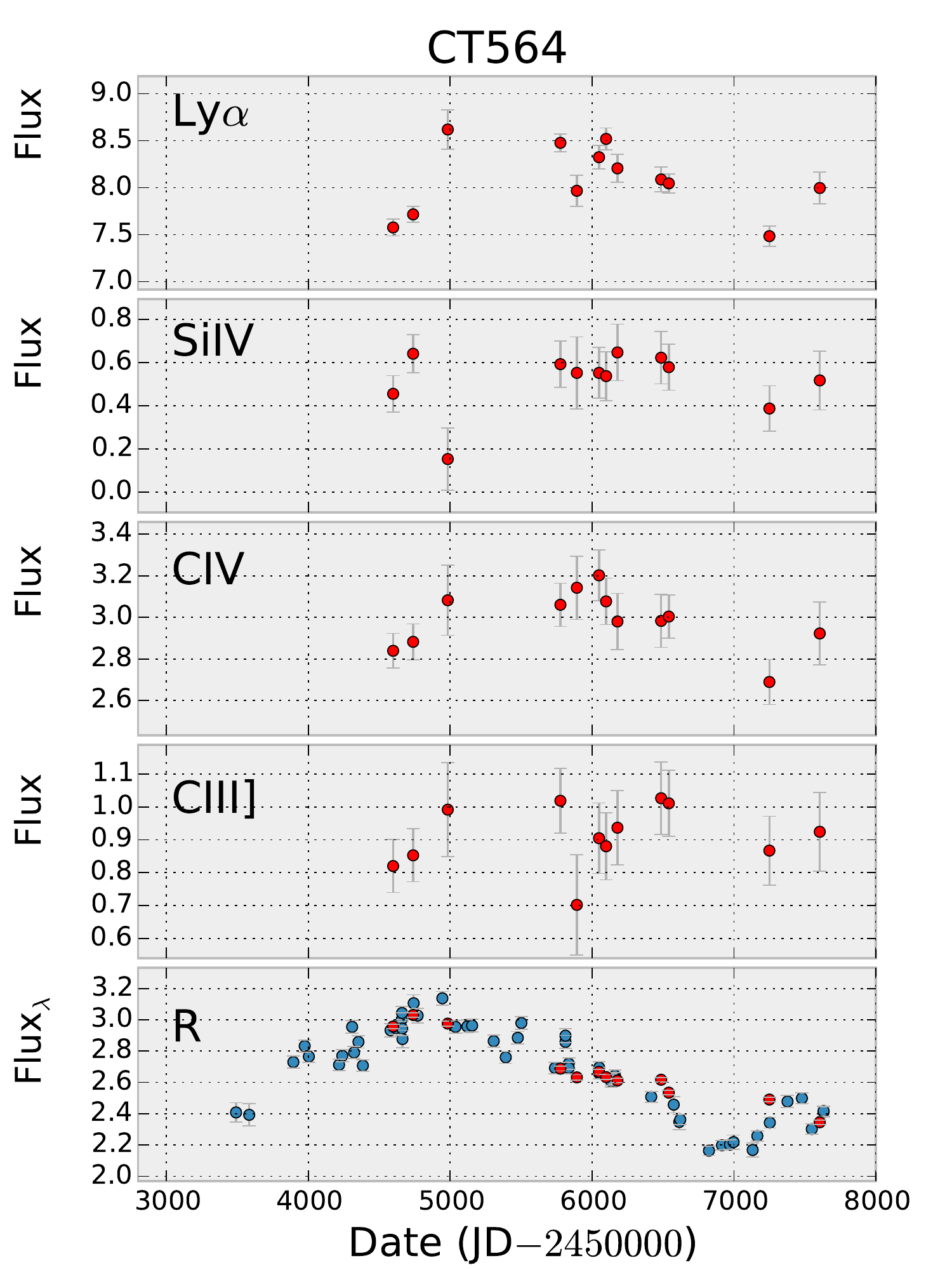}%
\begin{picture}(10,12){
\put(1,6.2){\includegraphics[scale=0.36,trim=0 0 0 0]{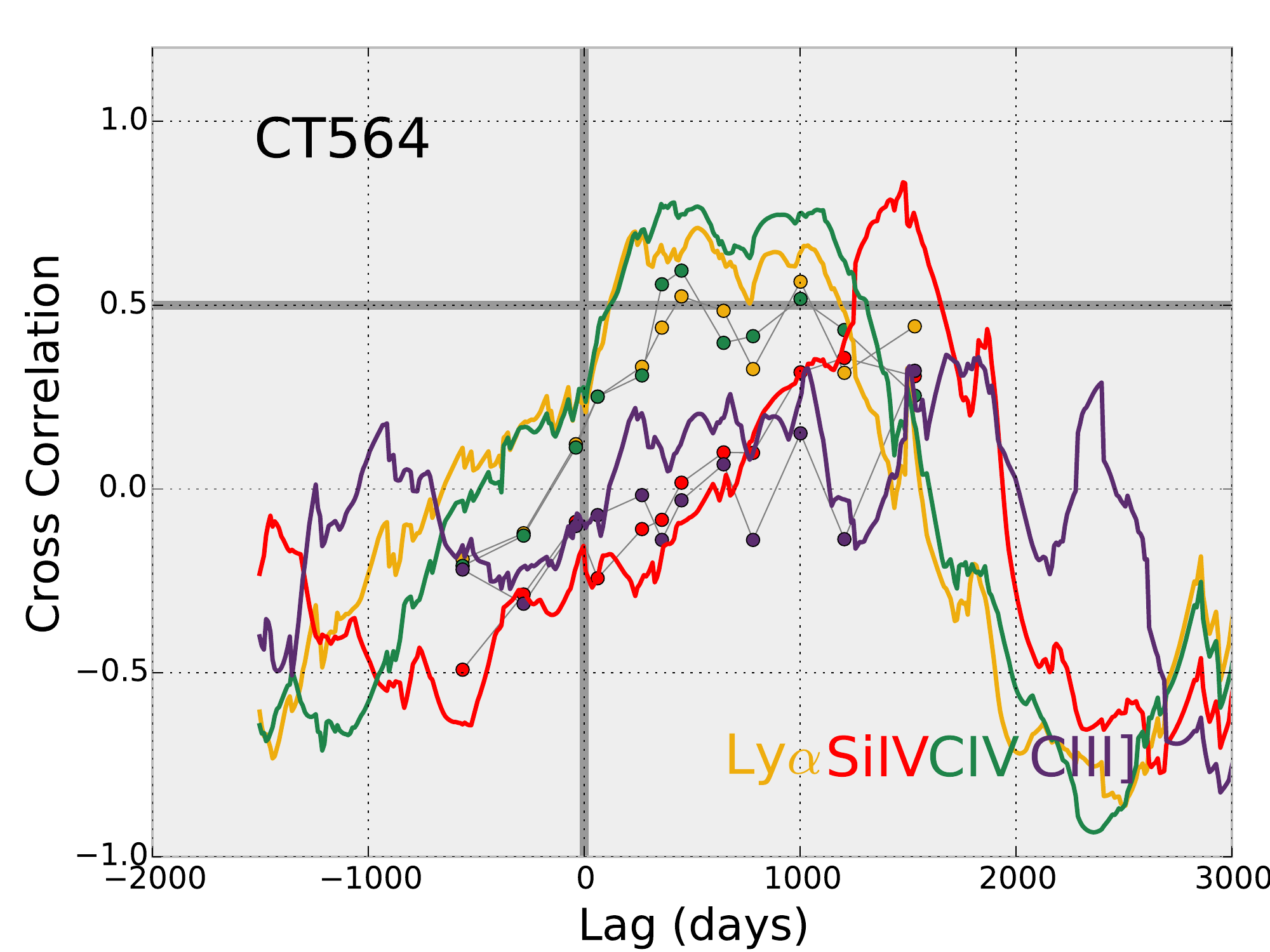}}
\put(0.7,0){\includegraphics[scale=0.40,trim=0 0 0 0]{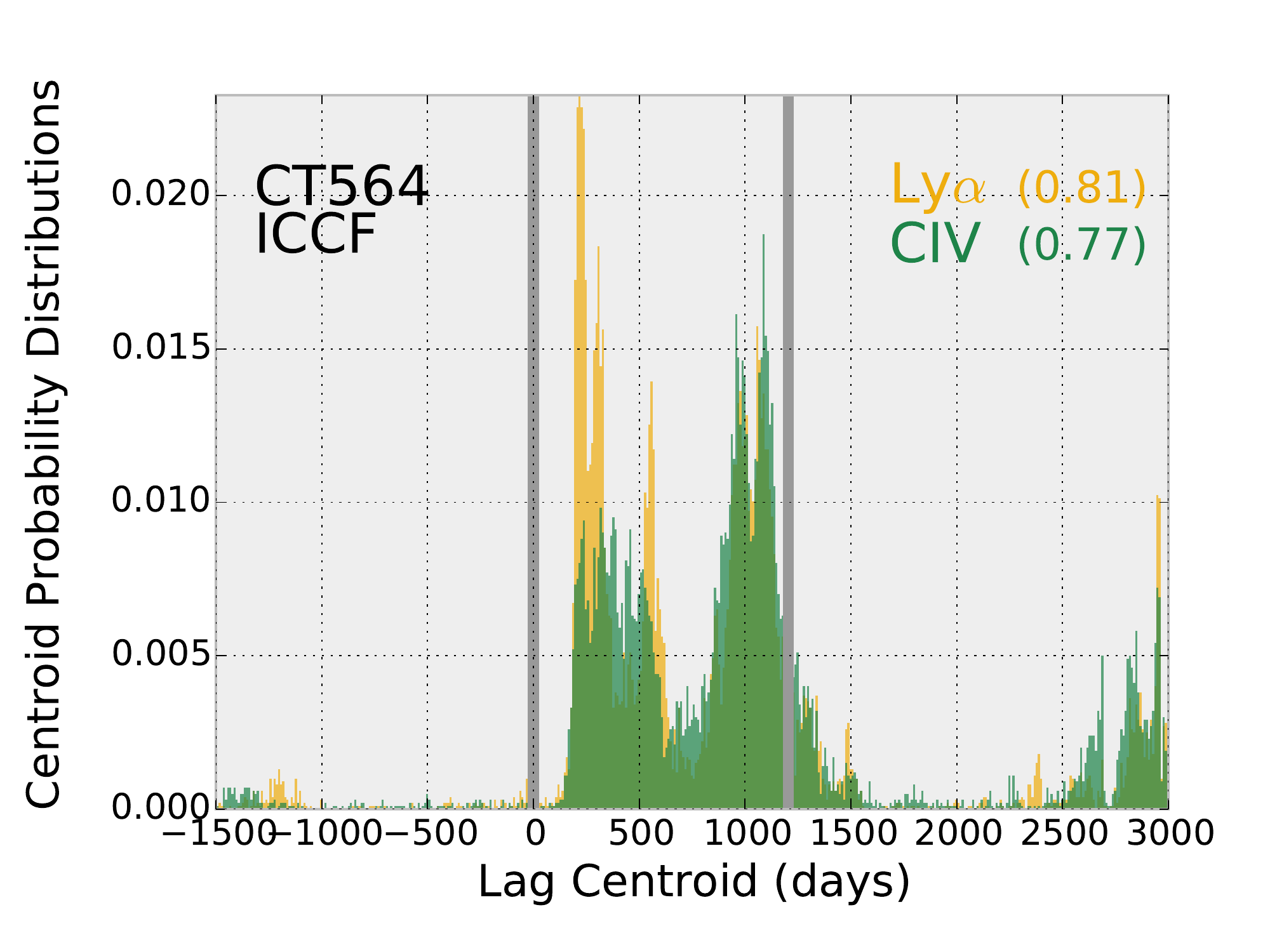}}}
\end{picture}
\includegraphics[scale=0.6,trim=0 0 0 0]{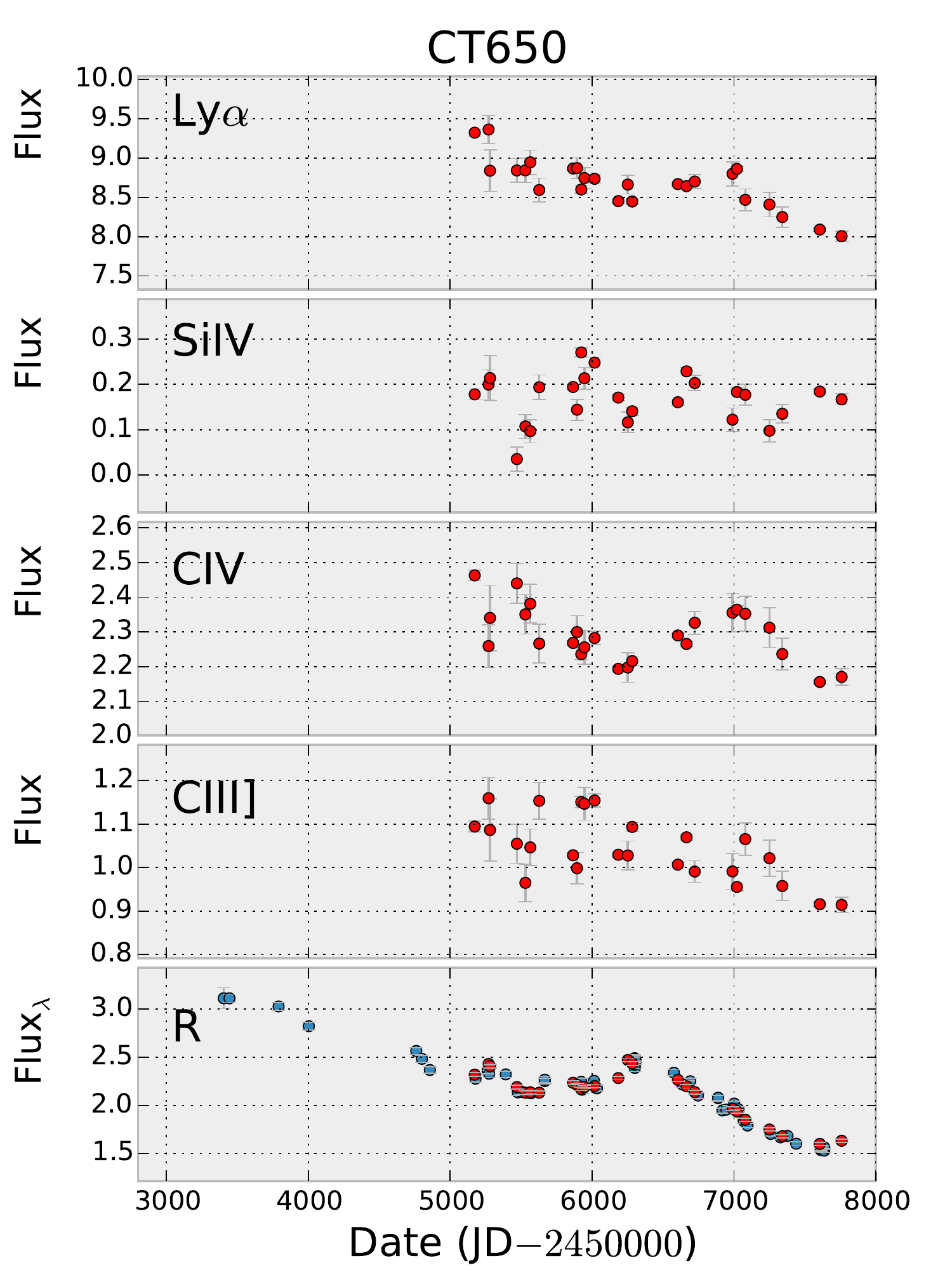}%
\begin{picture}(10,12){
\put(1,6.2){\includegraphics[scale=0.36,trim=0 0 0 0]{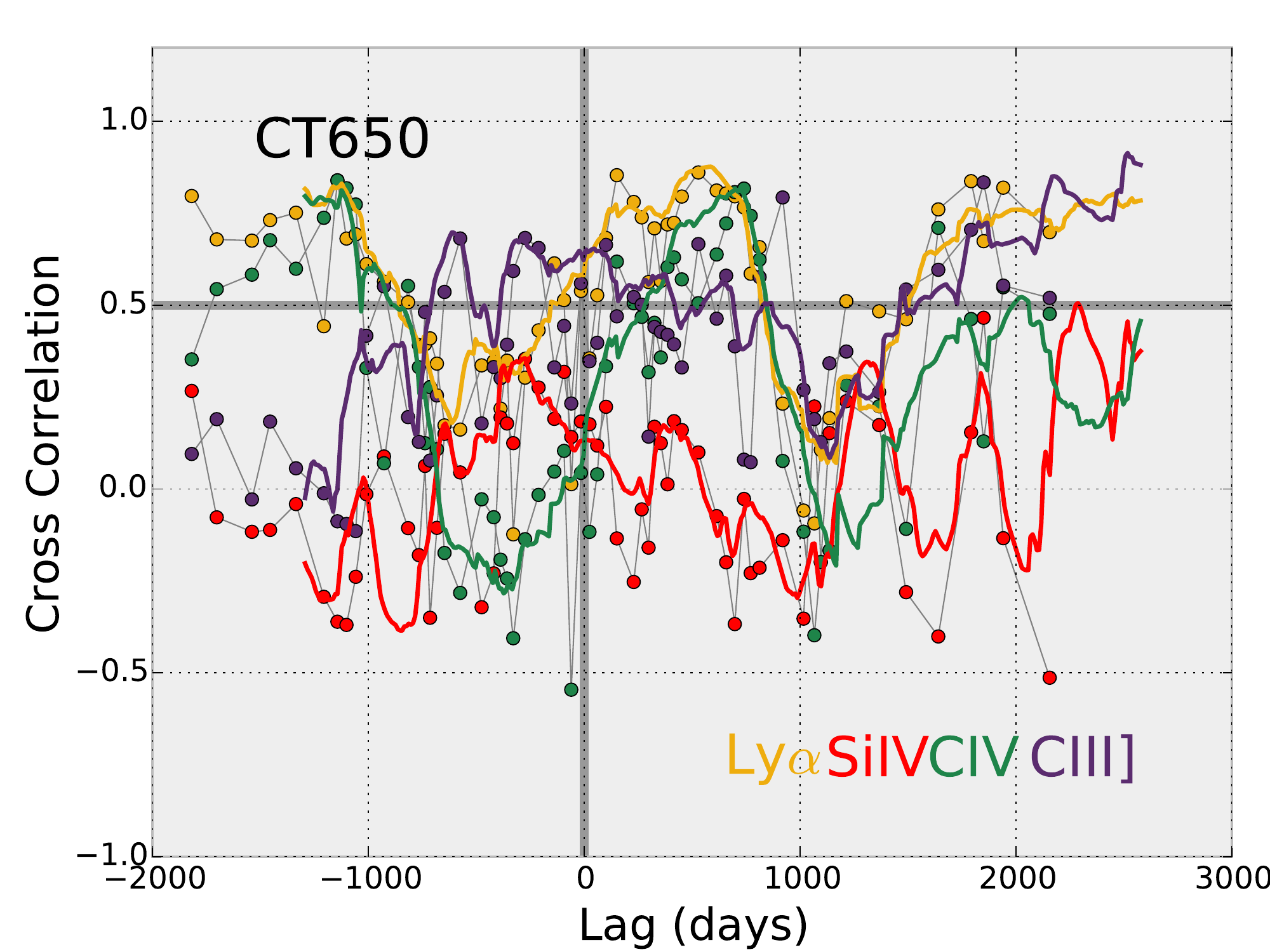}}
\put(0.7,0){\includegraphics[scale=0.40,trim=0 0 0 0]{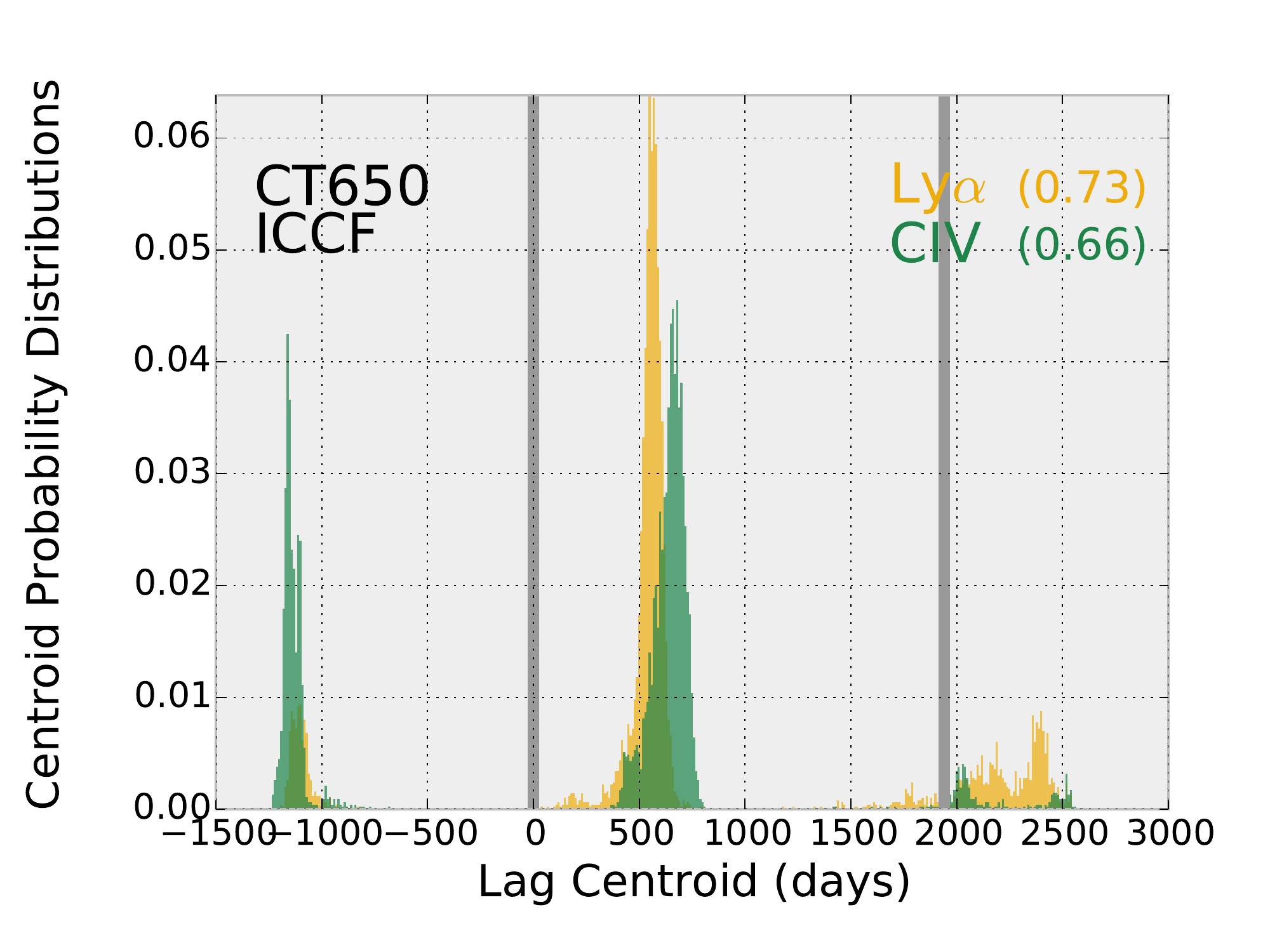}}}
\end{picture}
\end{figure*}

\begin{figure*}
\includegraphics[scale=0.6,trim=0 0 0 0]{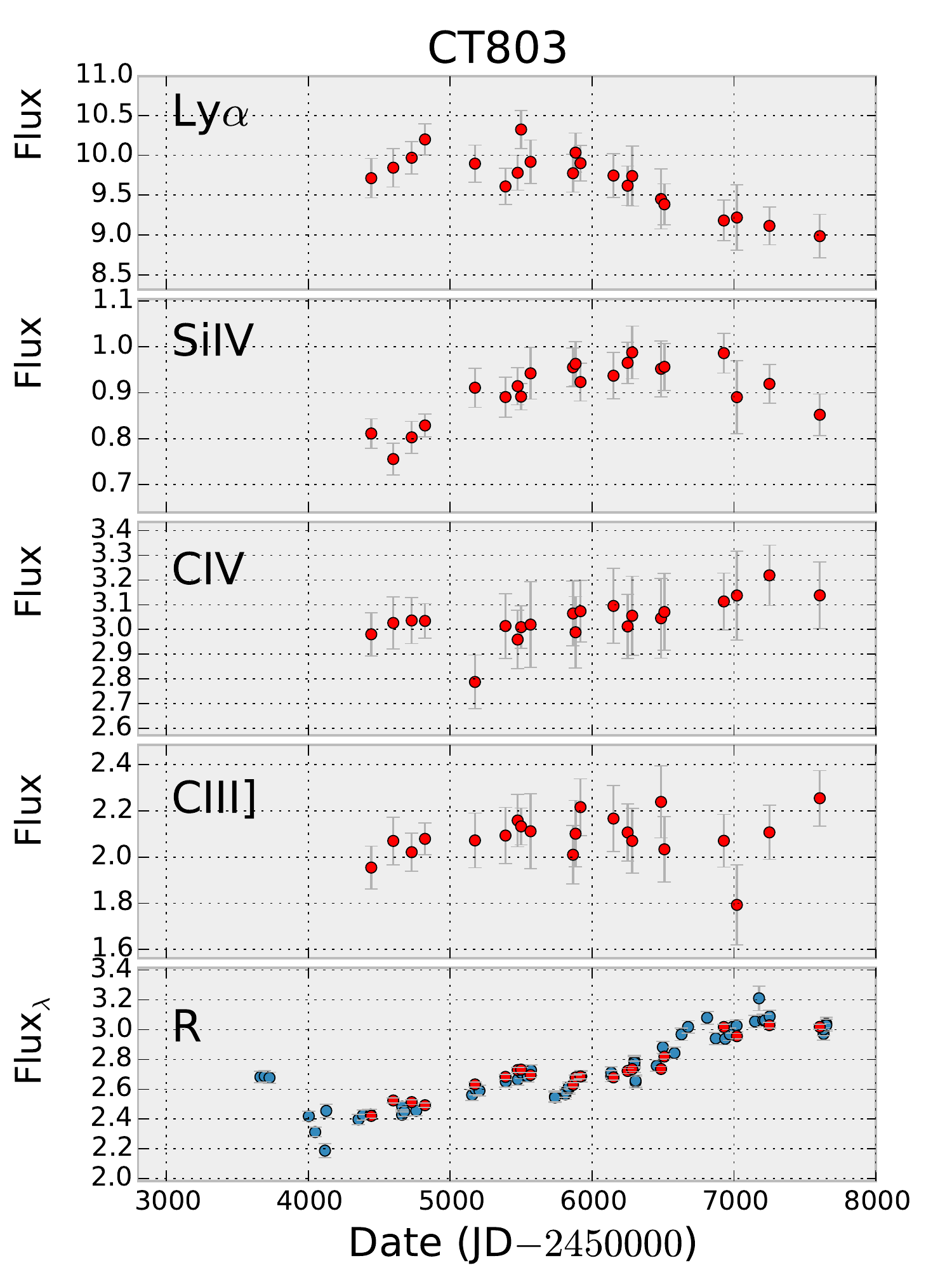}%
\begin{picture}(10,12){
\put(1,3){\includegraphics[scale=0.36,trim=0 0 0 0]{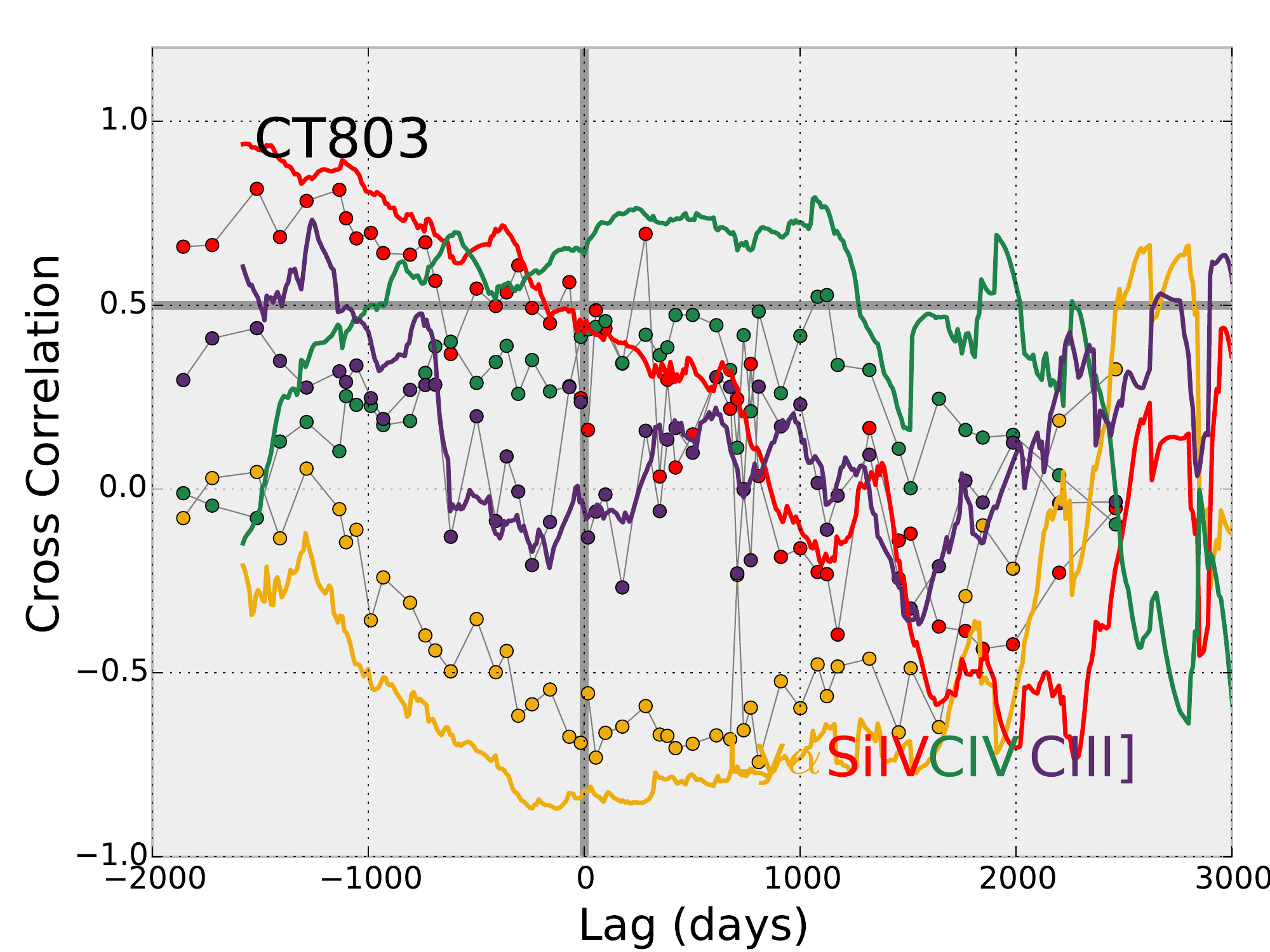}}
\put(0.7,0){}}
\end{picture}
\includegraphics[scale=0.6,trim=0 0 0 0]{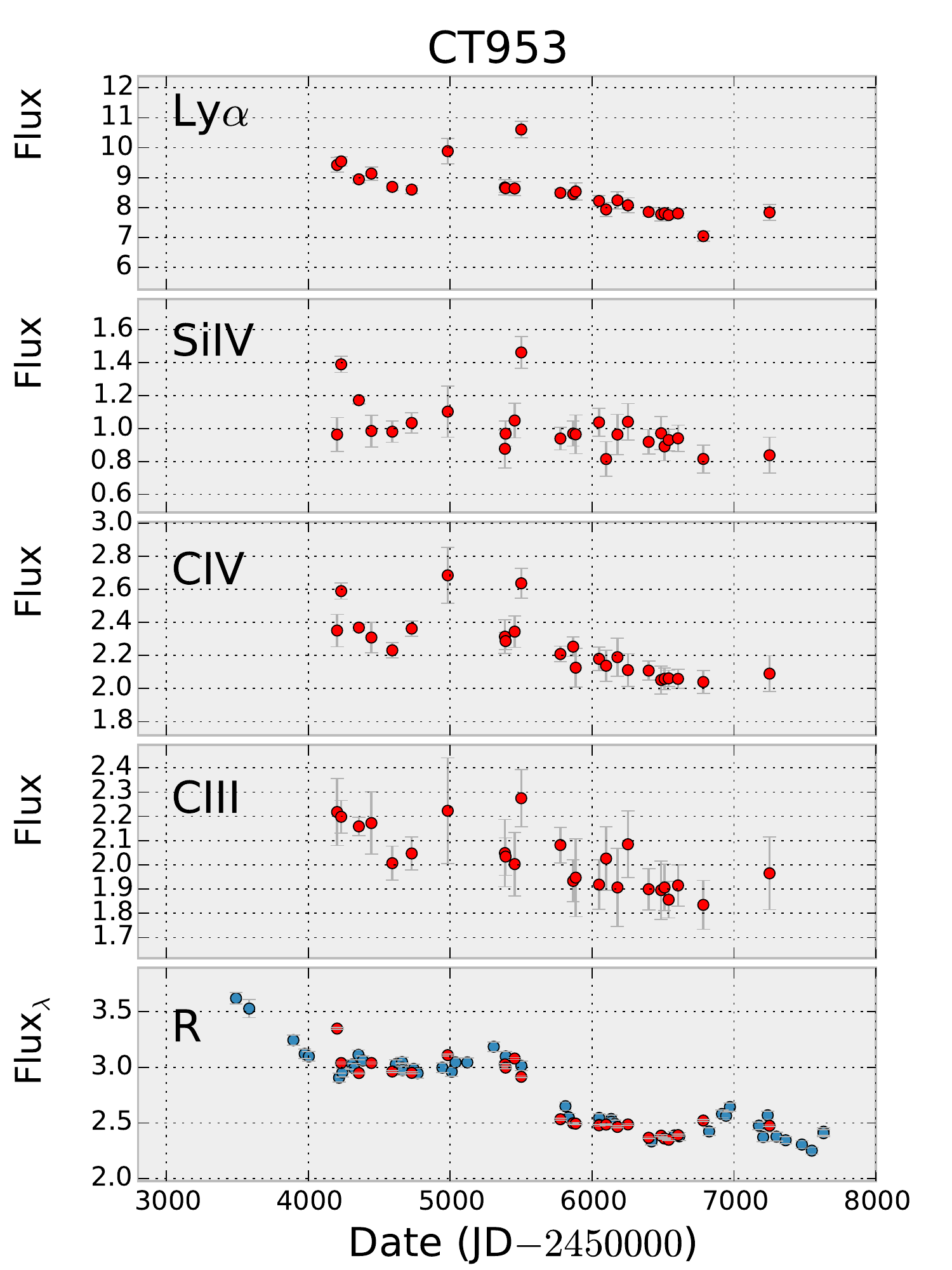}%
\begin{picture}(10,12){
\put(1,6.2){\includegraphics[scale=0.36,trim=0 0 0 0]{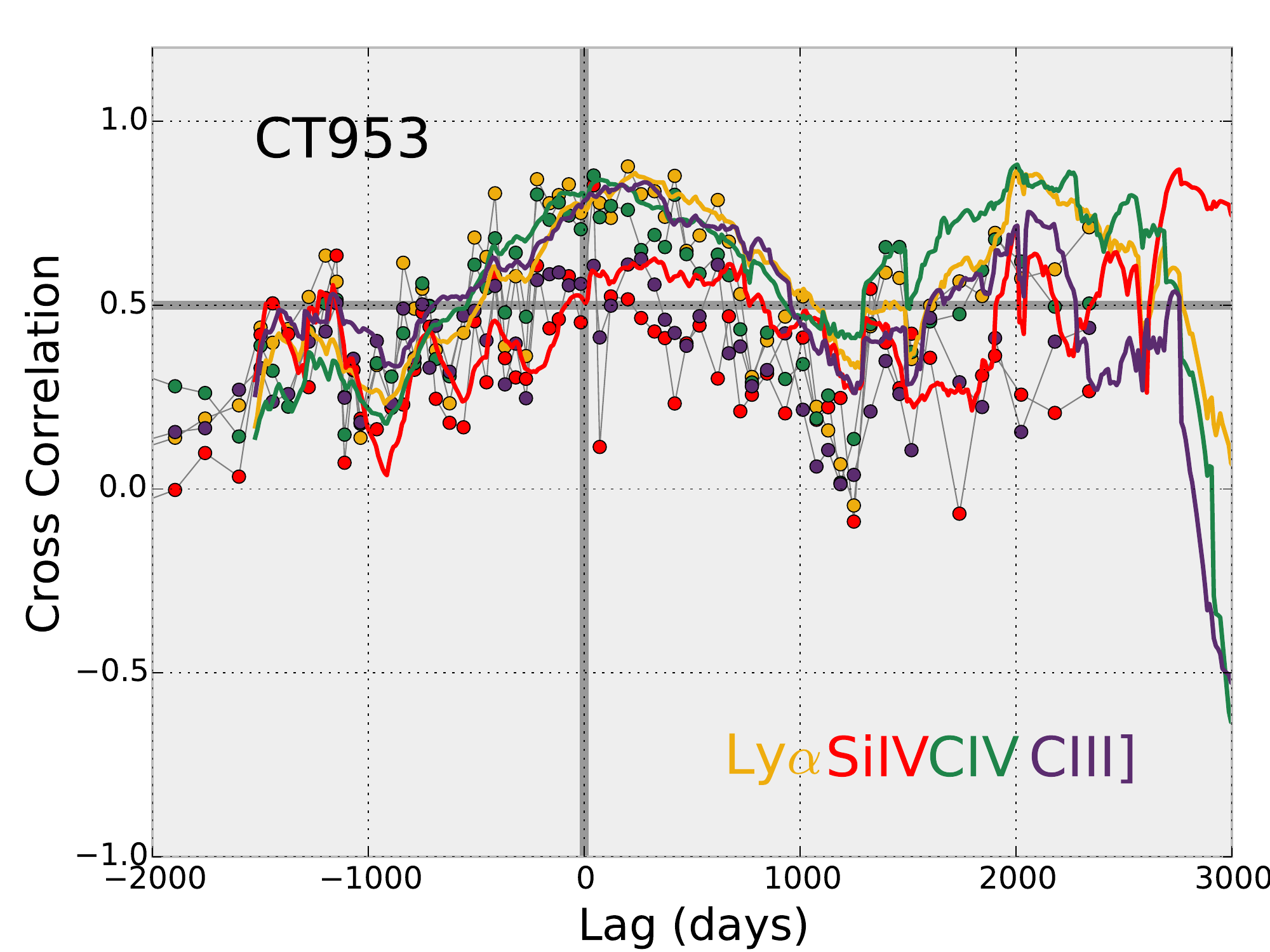}}
\put(0.7,0){\includegraphics[scale=0.40,trim=0 0 0 0]{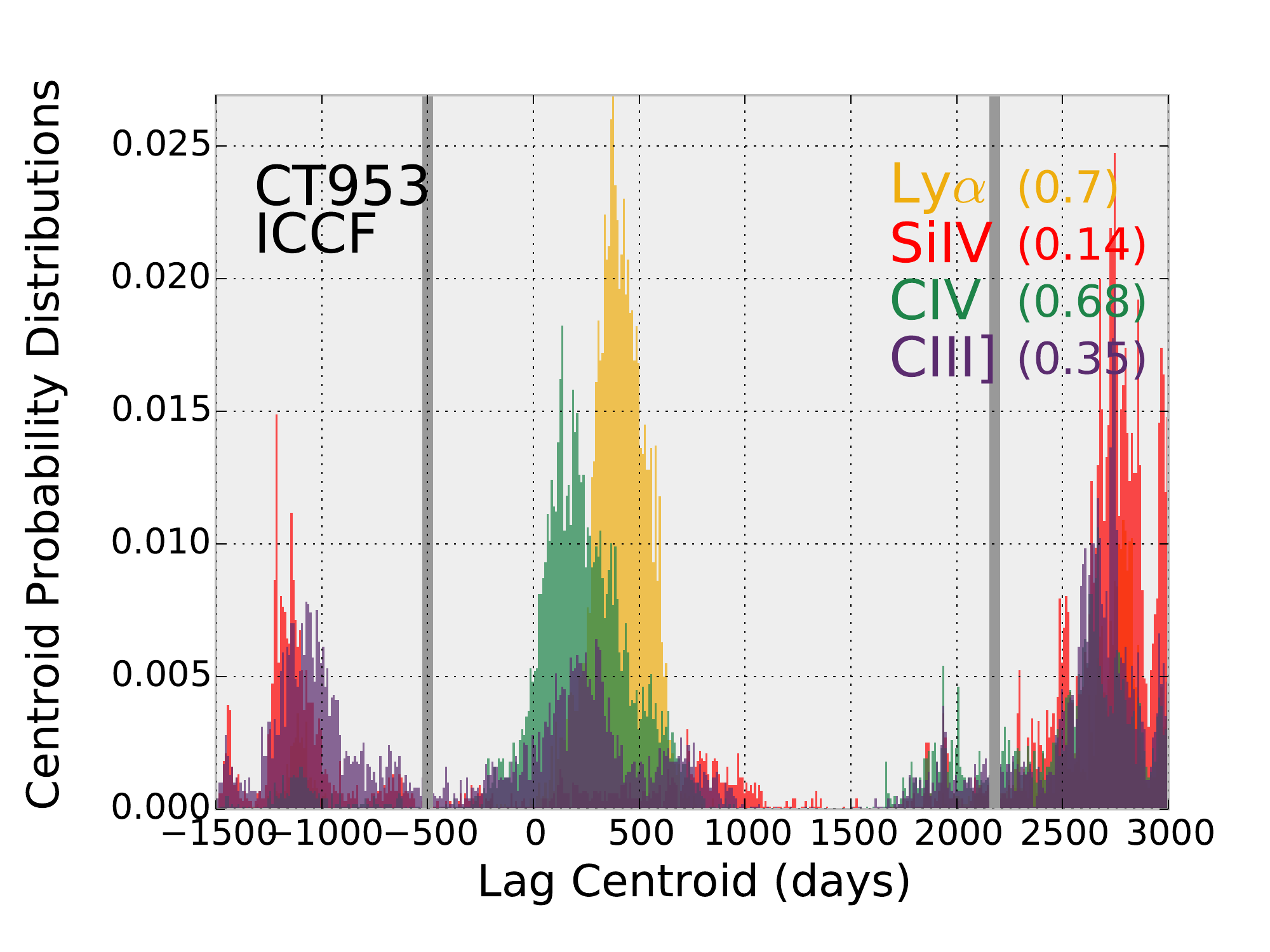}}}
\end{picture}
\end{figure*}

\begin{figure*}
\includegraphics[scale=0.6,trim=0 0 0 0]{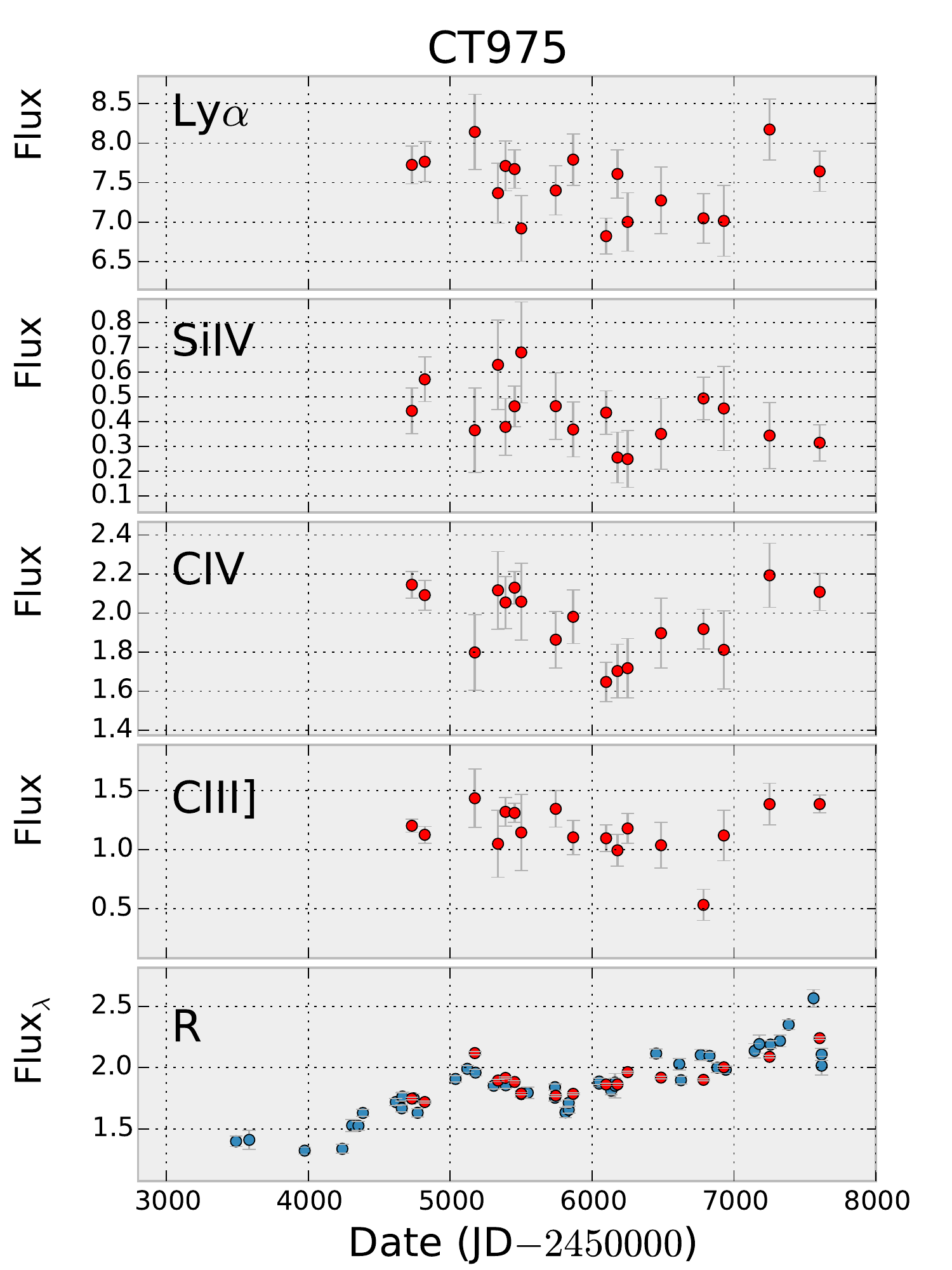}%
\begin{picture}(10,12){
\put(1,6.2){\includegraphics[scale=0.36,trim=0 0 0 0]{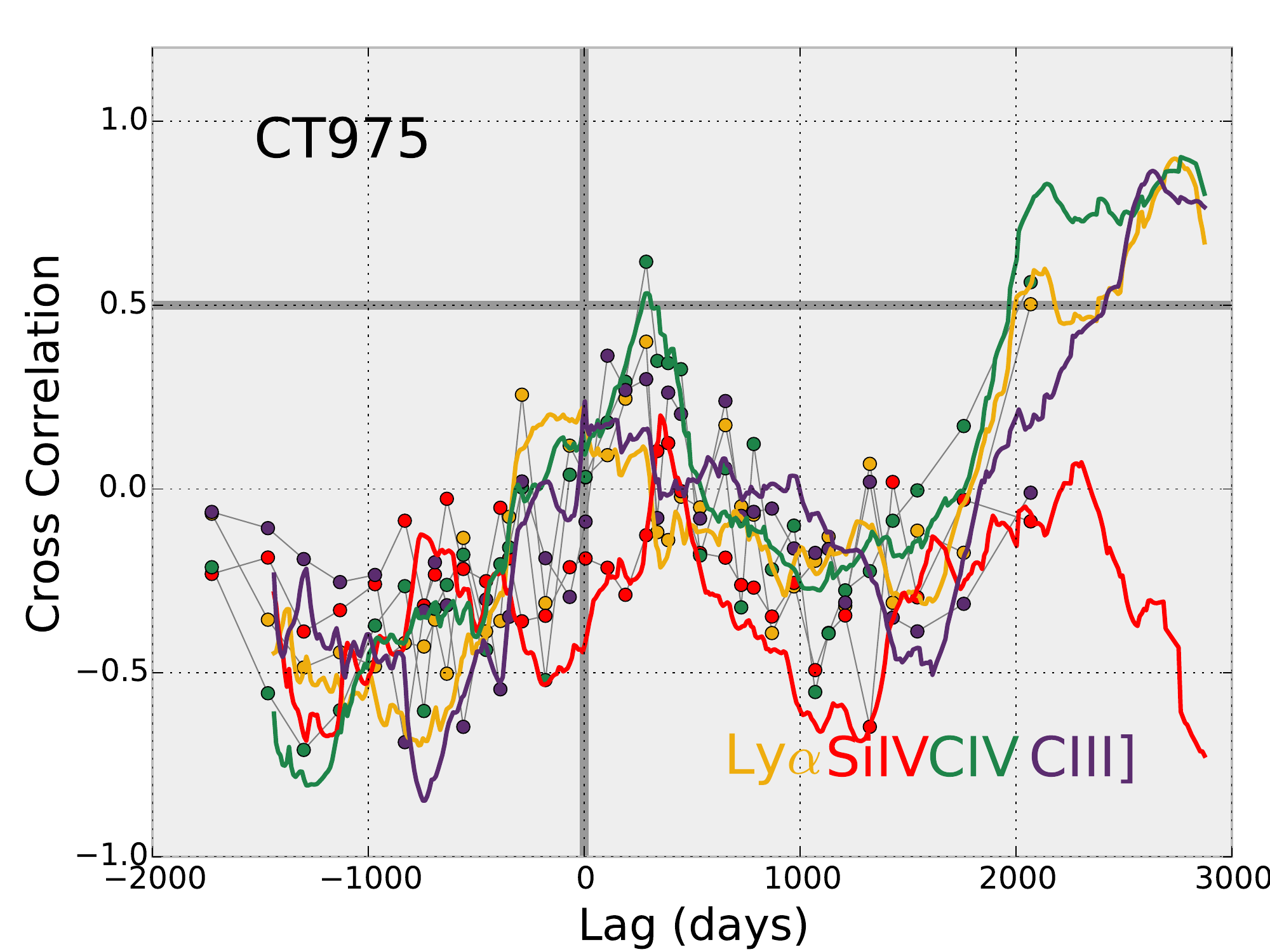}}
\put(0.7,0){\includegraphics[scale=0.40,trim=0 0 0 0]{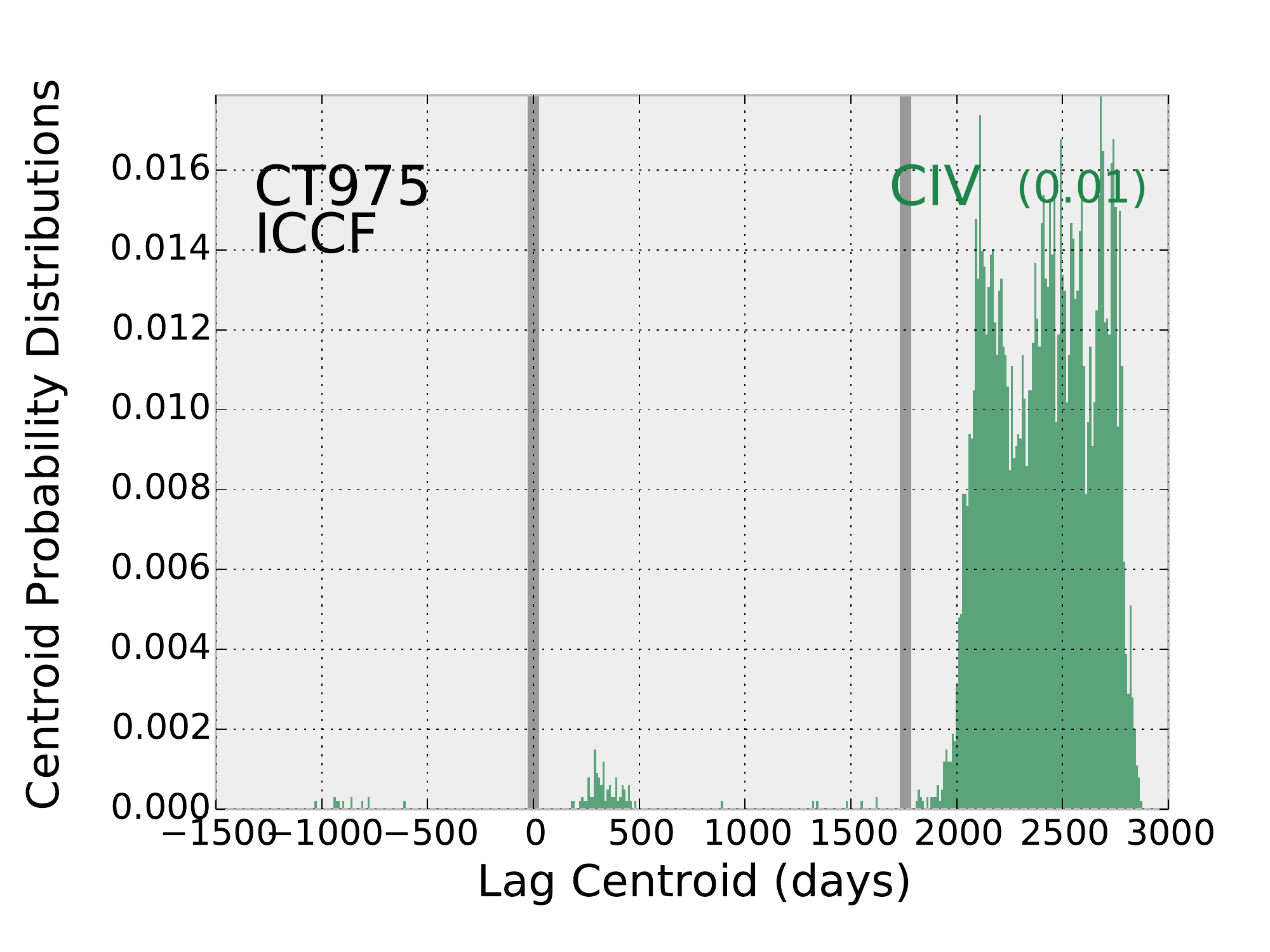}}}
\end{picture}
\includegraphics[scale=0.6,trim=0 0 0 0]{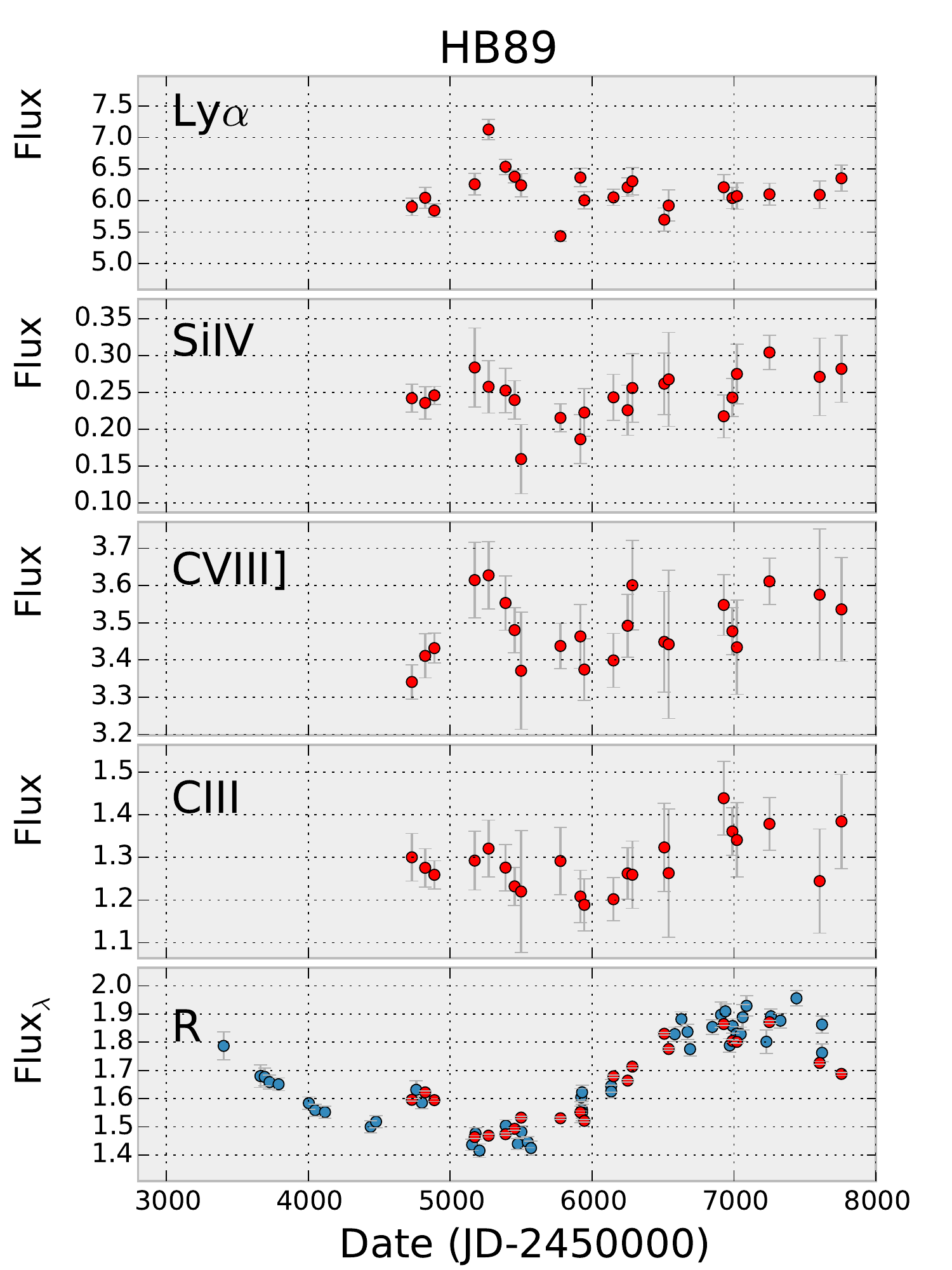}%
\begin{picture}(10,12){
\put(1,6.2){\includegraphics[scale=0.36,trim=0 0 0 0]{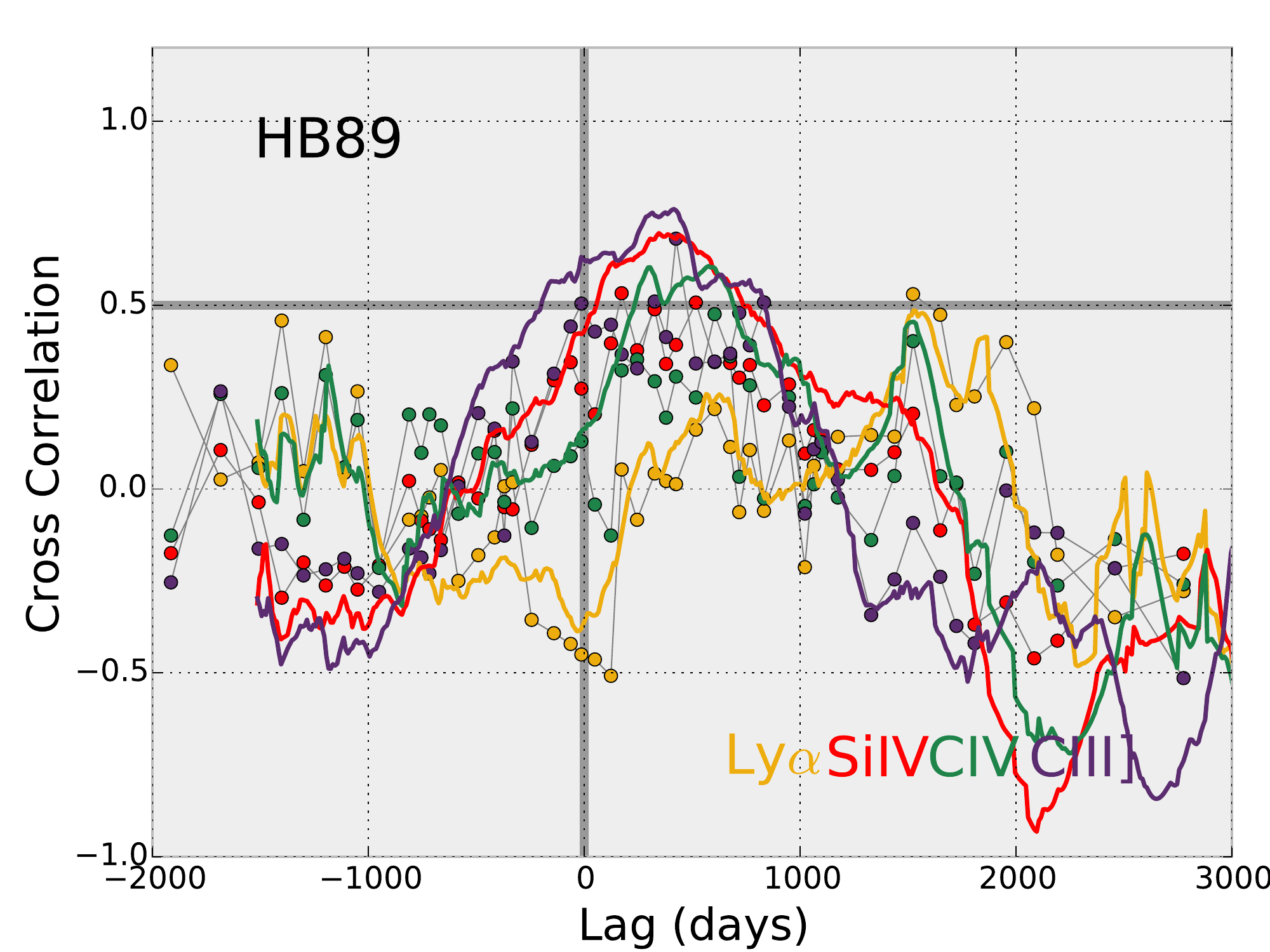}}
\put(0.7,0){\includegraphics[scale=0.40,trim=0 0 0 0]{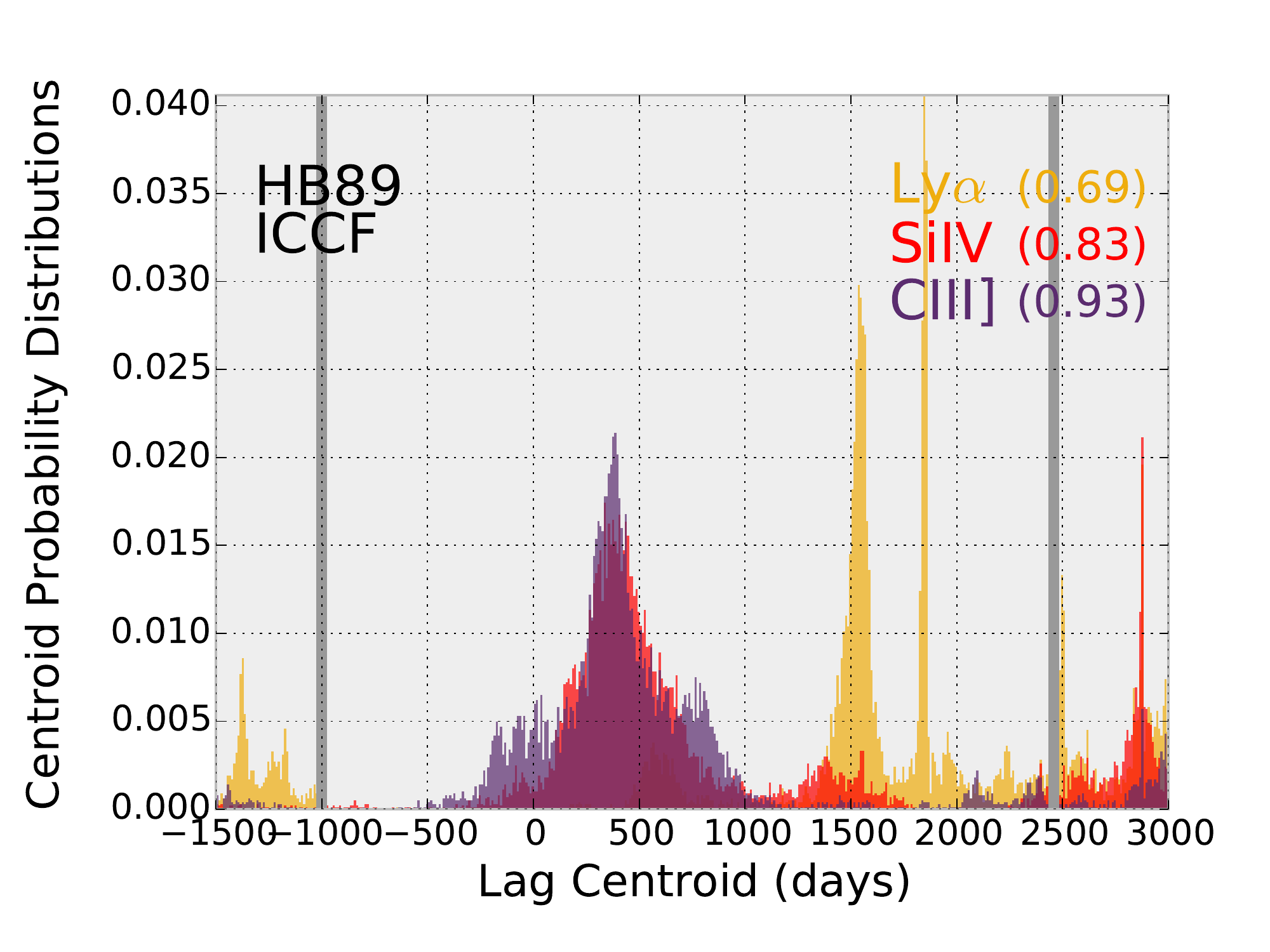}}}
\end{picture}
\end{figure*}

\begin{figure*}
\includegraphics[scale=0.6,trim=0 0 0 0]{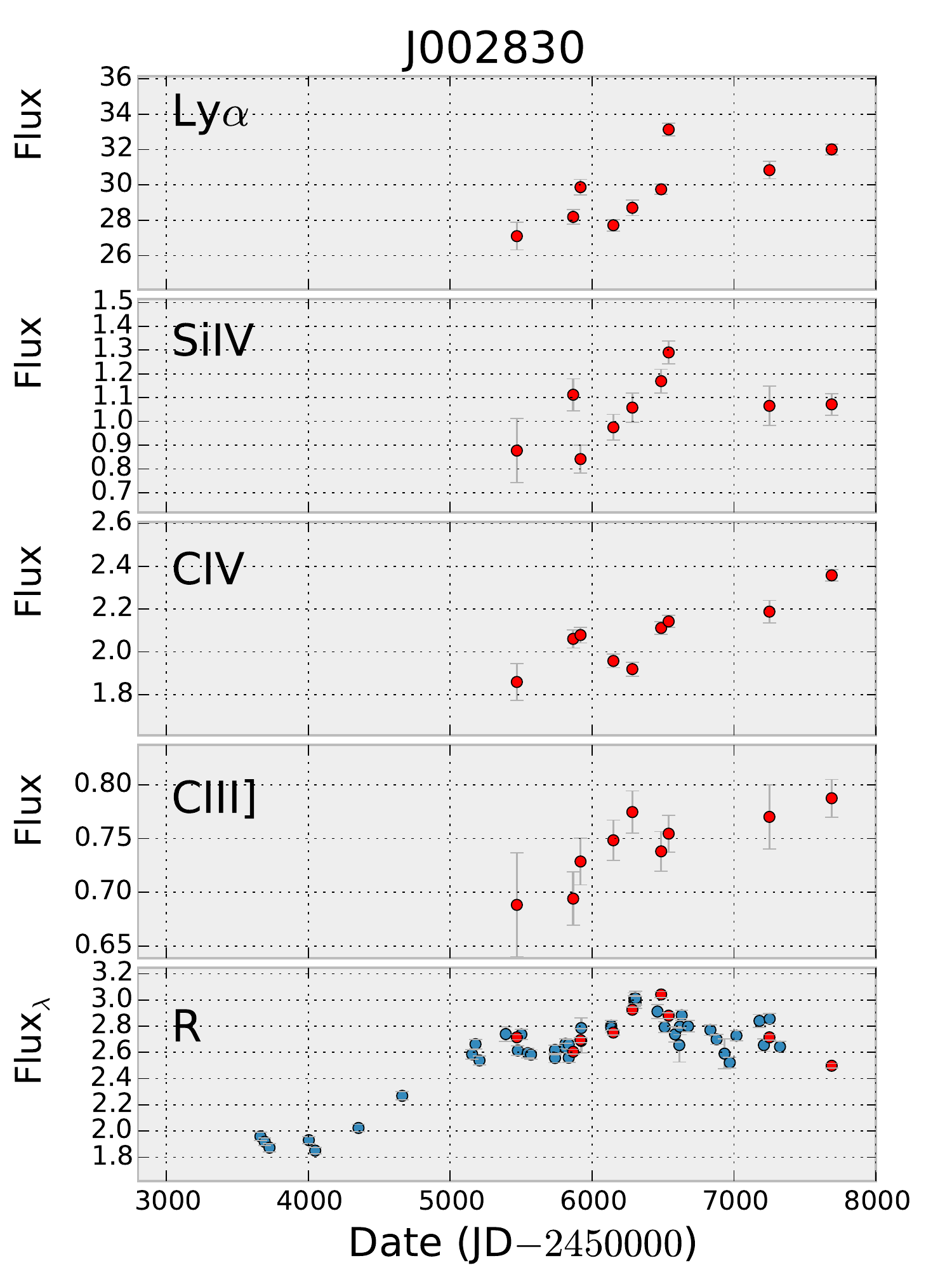}%
\begin{picture}(10,12){
\put(1,3){\includegraphics[scale=0.36,trim=0 0 0 0]{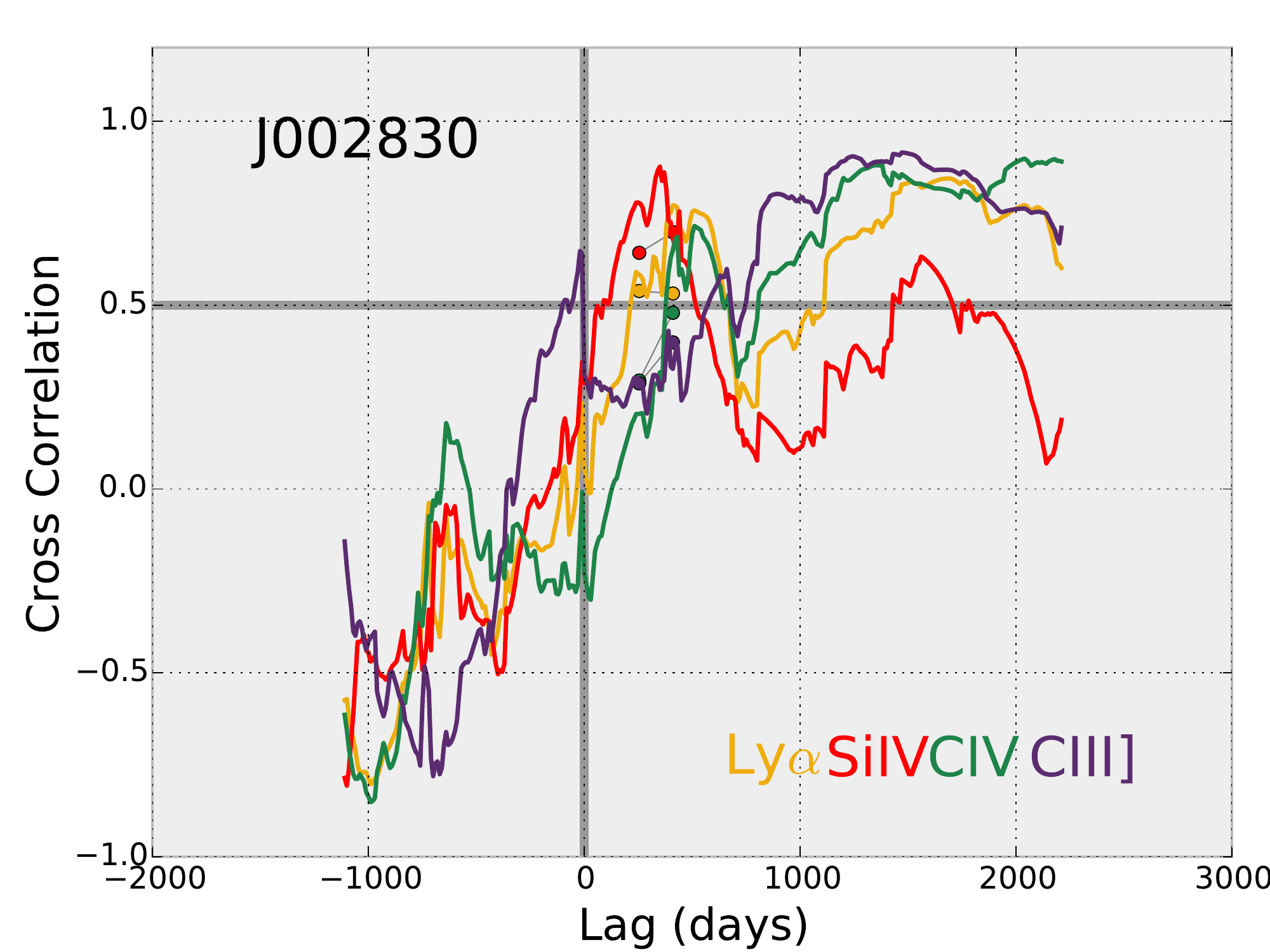}}
\put(0.7,0){}}
\end{picture}
\includegraphics[scale=0.6,trim=0 0 0 0]{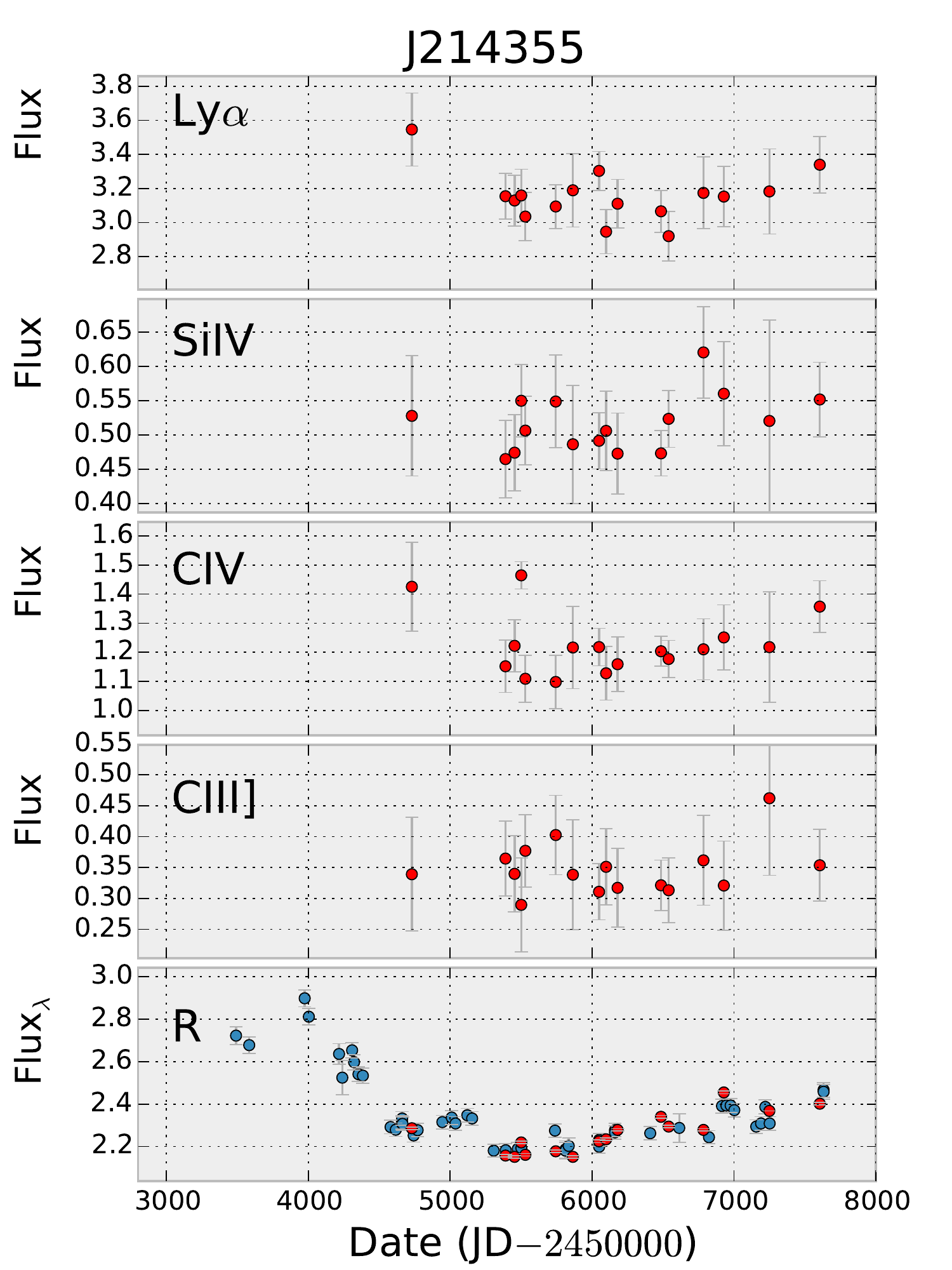}%
\begin{picture}(10,12){
\put(1,6.2){\includegraphics[scale=0.36,trim=0 0 0 0]{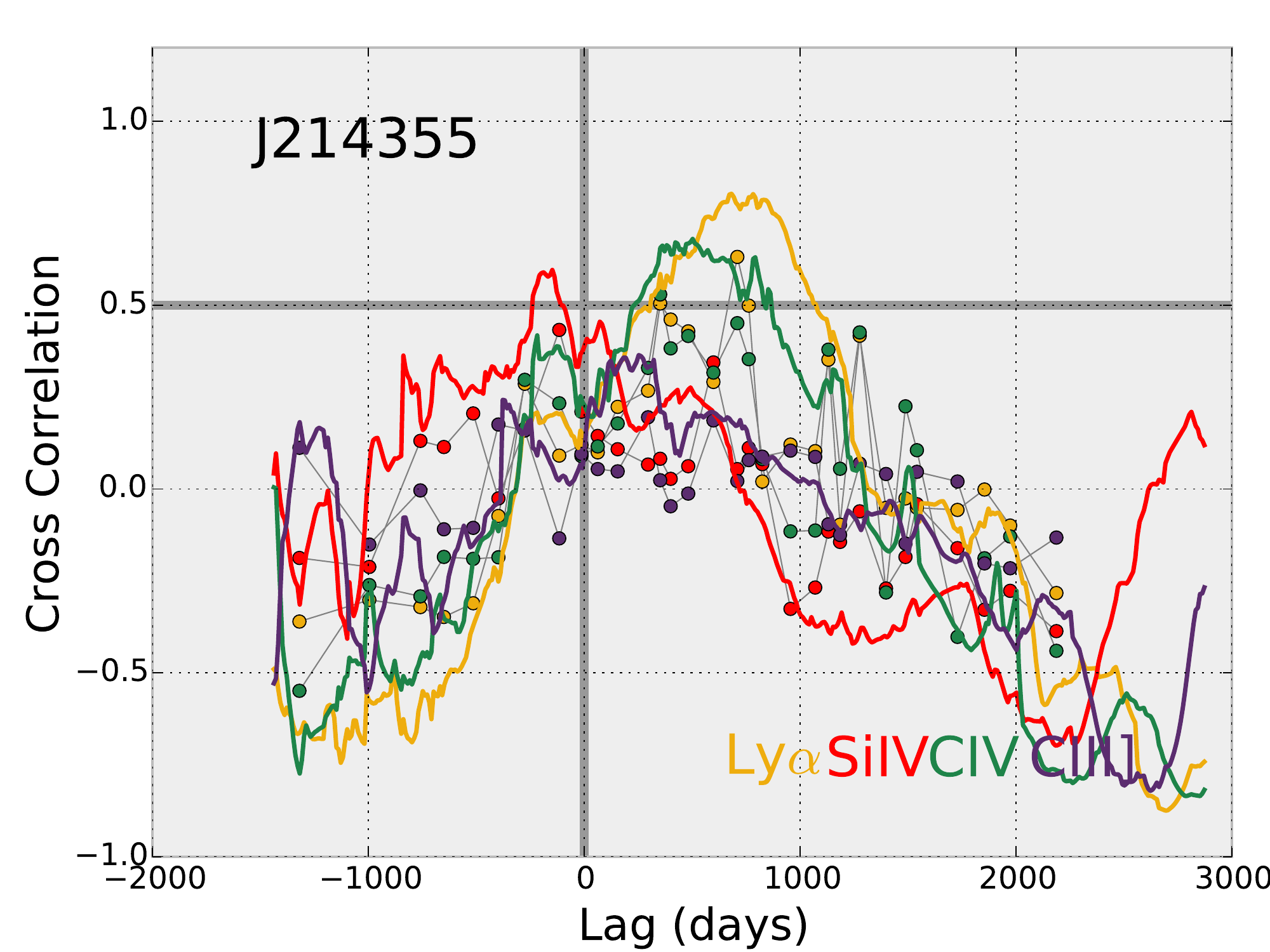}}
\put(0.7,0){\includegraphics[scale=0.40,trim=0 0 0 0]{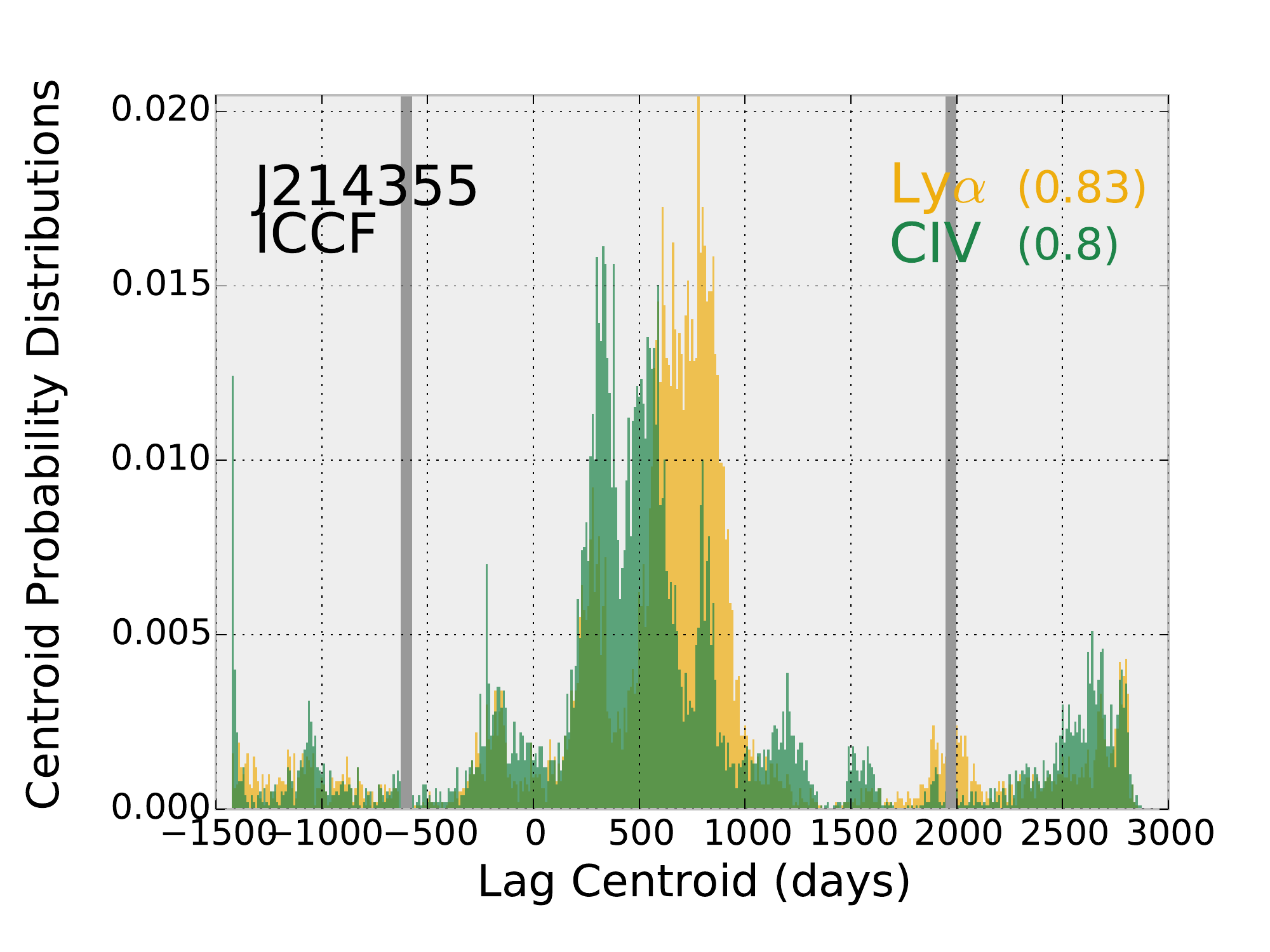}}}
\end{picture}
\end{figure*}

\begin{figure*}
\includegraphics[scale=0.6,trim=0 0 0 0]{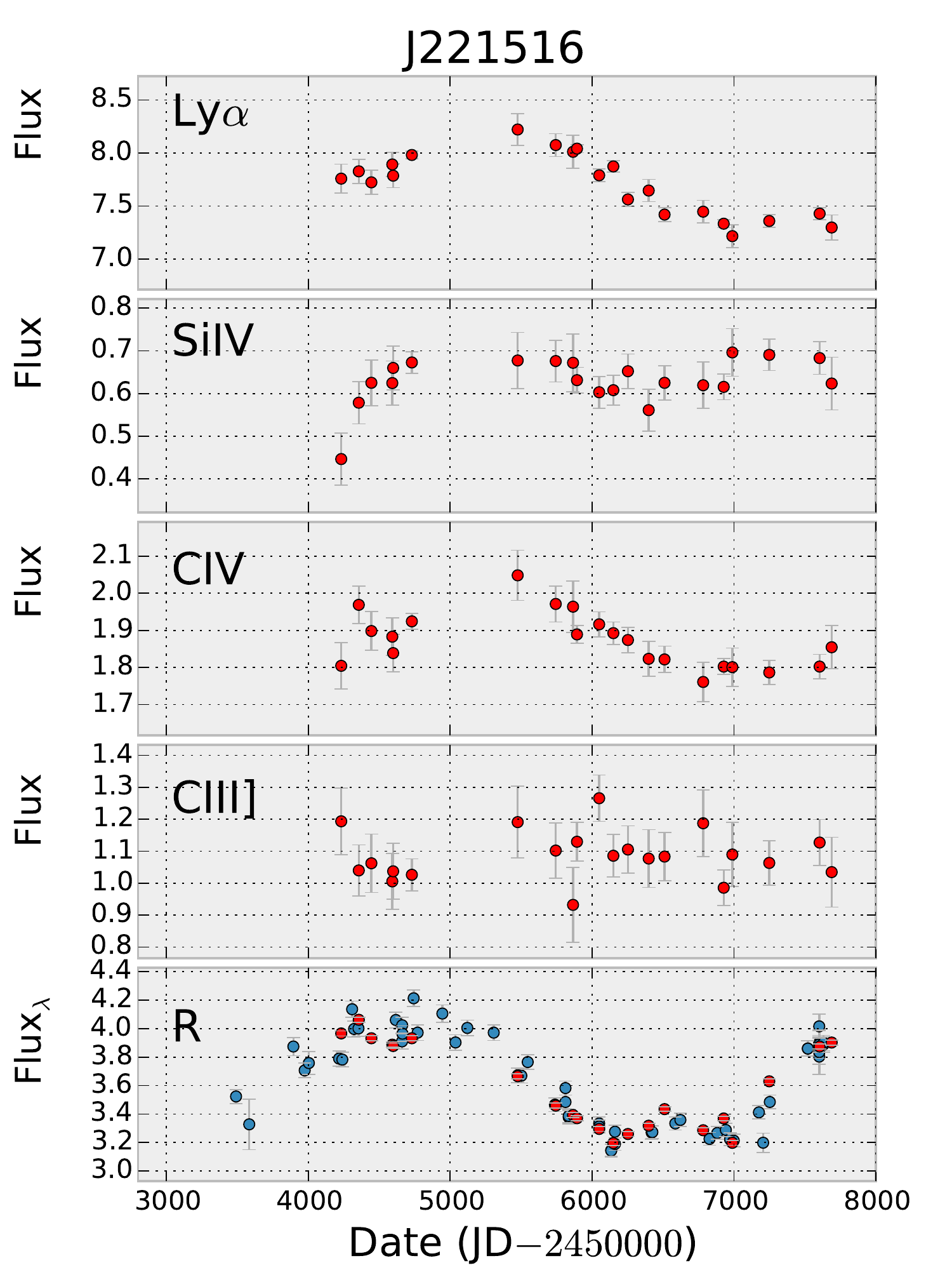}%
\begin{picture}(10,12){
\put(1,6.2){\includegraphics[scale=0.36,trim=0 0 0 0]{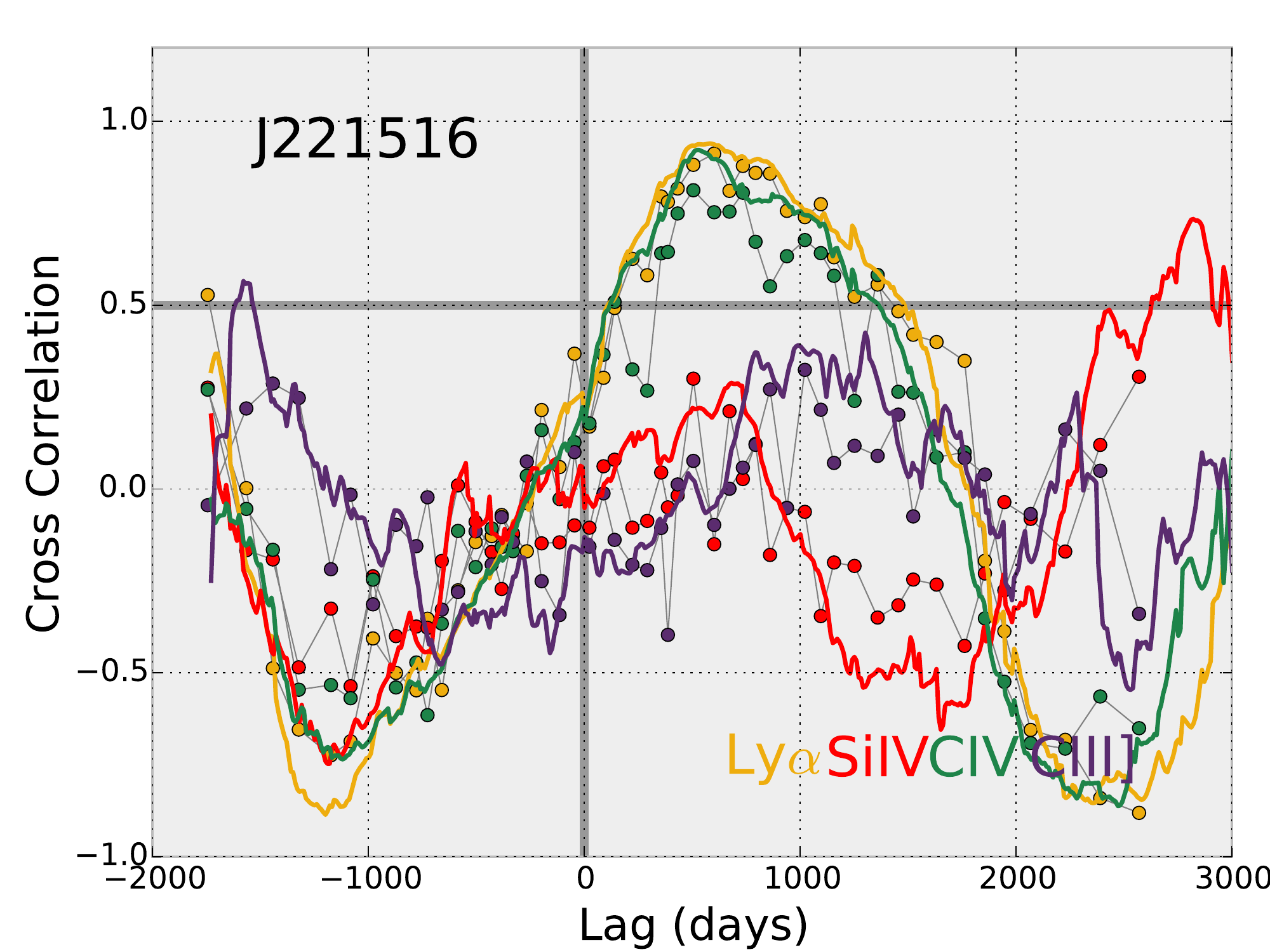}}
\put(0.7,0){\includegraphics[scale=0.40,trim=0 0 0 0]{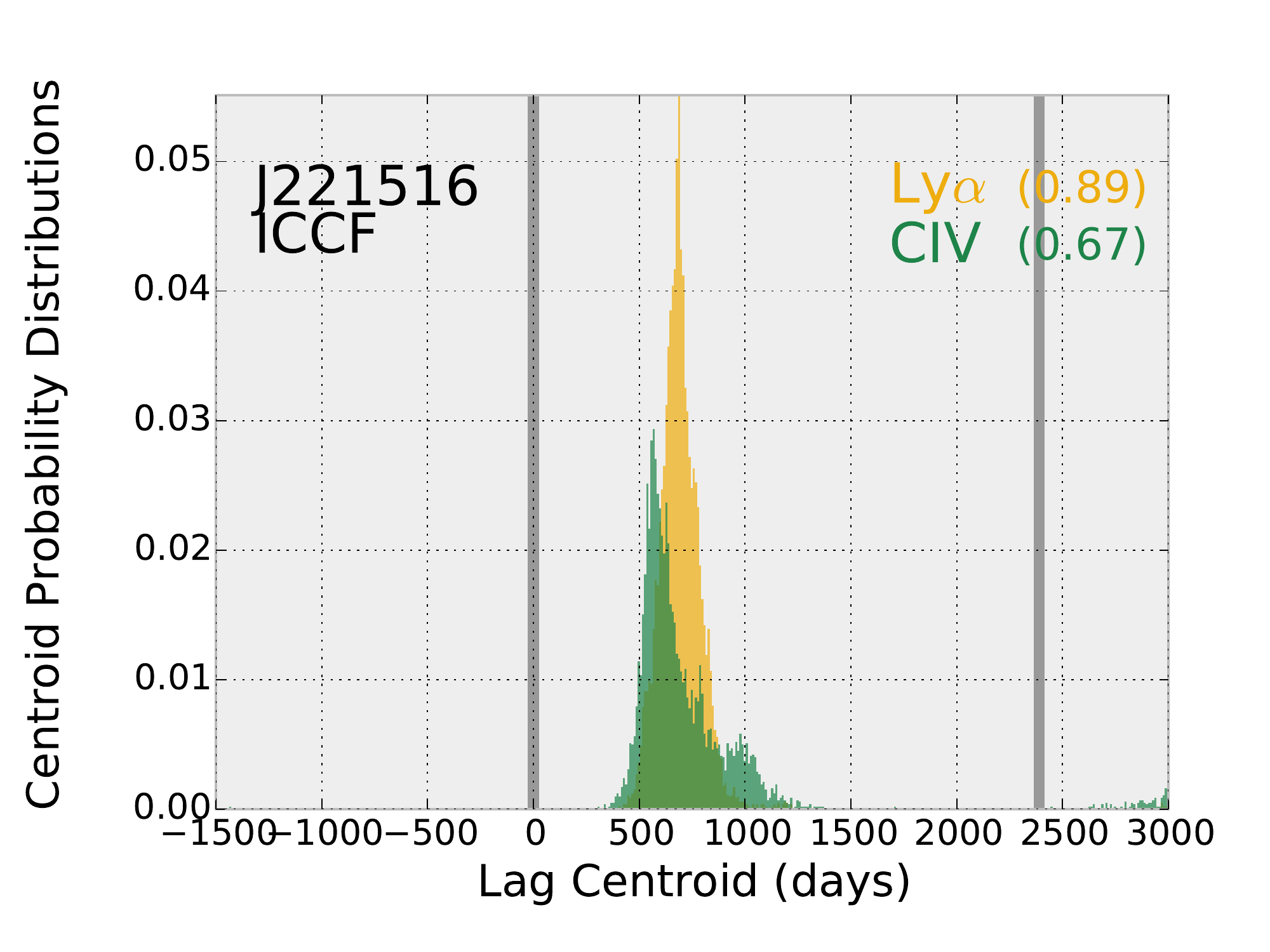}}}
\end{picture}
\includegraphics[scale=0.6,trim=0 0 0 0]{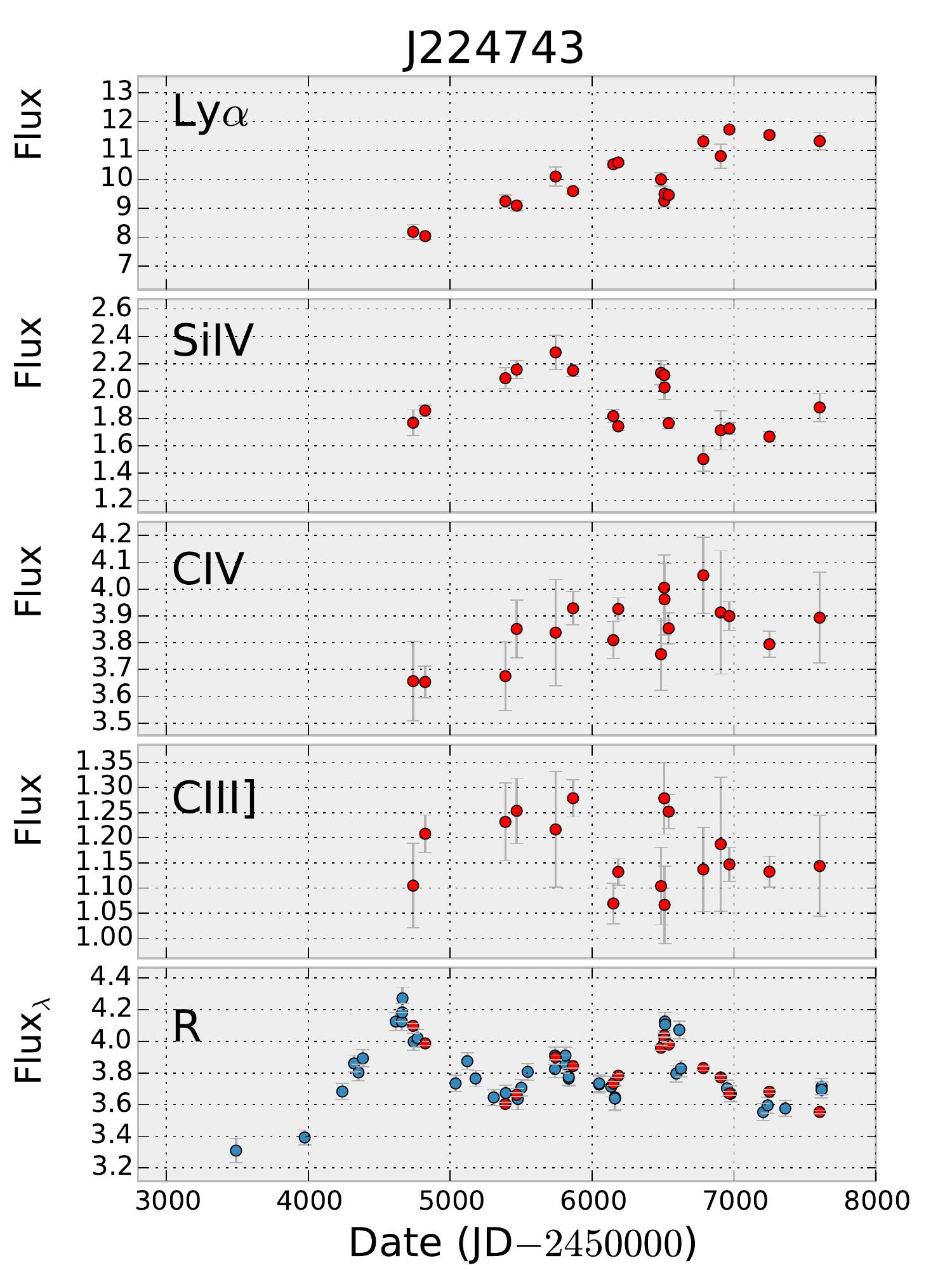}%
\begin{picture}(10,12){
\put(1,3){\includegraphics[scale=0.36,trim=0 0 0 0]{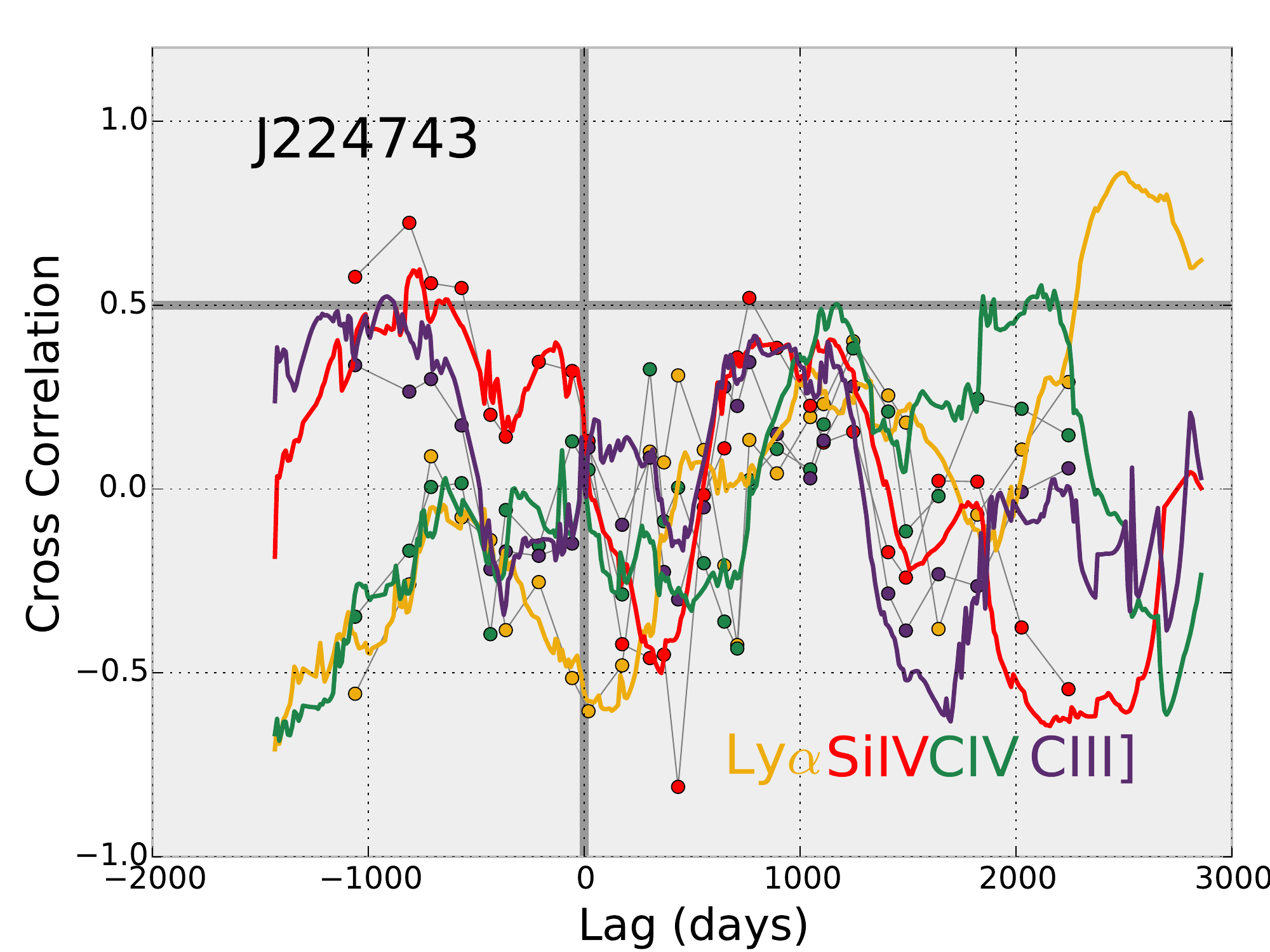}}
\put(0.7,0){}}
\end{picture}
\end{figure*}

Here we report on the $\sim$10-year observer frame light curves for 17
quasars with spectroscopic follow up. The variability properties for
these observations are reported in Table 3. We list the number of
epochs available (with the R-band counting both, the broad-band and
spectroscopic continuum measurements), $R_{max}$, the ratio between
the maximum and minimum flux, $f_{var}$, the normalized variability
amplitude ($\sqrt(\sigma^2\! - \!\delta^2)/\bar{f}$ , where $\delta$
are the photometric uncertainties, $\bar{f}$ is the light curve mean
flux and $\sigma$ its standard deviation (Rodriguez-Pascual et
al.~1997)), $\chi^2$, the reduced chi-square fit to a model with
constant flux, $P_{\chi}$, the probability that the observed $\chi^2$
is due to random errors, and cc indicates whether lag determination
between emission line and continuum light curves was obtained (see
Section 4). Formally, a source is considered variable if $P_{\chi} >
0.95$ and $\bar{f} > 0$. Notice, however, that these statistical
descriptions deal with the amplitude of the variability only, and not
with the structure of the light curves as a function of time. In other
words, while the errors in some light curves are consistent with no
variations, the shape of the light curve might suggest a systematic
flux change with time. This is particularly important for the emission
line light curves, where errors are difficult to quantify
properly. This can be appreciated, for example, in the Ly$\alpha$
light curve of CT1061 which has a $P_{\chi} = 0.4$ and $\bar{f} = 0$
but shows a clear trend of decreasing flux as a function of time.

From Table 3 it can be seen that all R-band light curves show
significant variability and have $P_{\chi} = 1.0$ and $\bar{f} >
0$. Still, there is a range of properties in the variability
structure, with some sources presenting very smooth, slowly varying
fluxes (e.g., CT367) while others go through epochs of a more random,
fast changing fluxes (e.g., CT320).

Most emission line light curves present significant variability.
Adopting $P_{\chi} \geqslant 0.95$ for variable light curves, 15/17
quasars show large Ly$\alpha$ flux fluctuations. CIV follows with
12/17, SiIV with 10/17 and CIII] with 12/17 (plus MgII with 1/1 for
  CT252). In summary, we find that Ly$\alpha$ presents a very high
  probability of showing strong variability. This is in contrast with
  previous results that detected no variations in this line for high
  luminosity sources (Kaspi et al.~2007, Ulrich et al.~1993). CIV is
  also a highly variable line, followed by SiIV and CIII]. This is
    expected as these lines are generally weaker and therefore it
    becomes harder to determine statistically significant
    variations. It should be noticed, however, that our spectroscopic
    sample was selected as those quasars that showed significant
    R-band variability, and therefore it could be biased towards
    highly variable sources. Comparison with previous experiments,
    might therefore, not be very meaningful.

In what follows we divide our sample into two groups: a first group
with `expected' line variability (14/17), i.e., those showing emission
line light curves that agree with the expectations given the continuum
variations, and a second group of those objects with detected line
variability that seems to respond to the continuum changes in
unexpected ways (3/17).

\subsection{Quasars with expected line response}

Most of the monitored quasars show some degree of line variation that
mimics the continuum variability after some elapsed time (see left
panels in Figure 5). This is crucial for the cross correlation
analysis presented in Section 4, as quasars in this group show good
indication that the emission lines are reverberating as a response to
variations in the continuum. Unfortunately, some sources in this group
present rather monotonic flux variations (CT367, CT803, J002830) and
therefore the present light curves are not suitable for
cross-correlation analysis. This will be further characterized in
Section 4.

\subsection{Quasars with unexpected line response}

We find three quasars where the lines response to the continuum
variations are hard to interpret. These are CT320, CT803 and J224743.

In all sources one of emission lines seems to have disengaged from the
observed continuum, while the remaining lines show a pattern of
variability more consistent with the R-band light curve. In the case
of CT320, the anomalous behavior is observed in SiIV, while for CT803
and J224743, it is observed in Ly$\alpha$. As we will see in Section
4, very little correlation is observed in the cross-correlation
analysis of the light curves in these emission lines.

The most likely explanation for the lack of correlation between
continuum and line emission light curves is that in these objects the
{\em observed\/} continuum is not a good counterpart of the ionizing
continuum responsible for the observed line variations. This is not
completely unprecedented, as we will further see in the discussion
section.

\section{Cross Correlation Analysis}

Cross correlation functions (CCFs) for all our targets with
spectroscopic data were determined. All emission line light curves are
used, irrespective of their $P_{\chi}$ values, since as we have seen,
this quantity does not reflect the structure of the variability.

Three methods were employed to examine the degree of correlation
between the continuum and emission line light curves: the interpolated
cross correlation function (ICCF, e.g., Peterson et al.~1998, 2004),
the z-transformed discrete correlation function (ZDCF) as defined by
Alexander (1997), and the JAVELIN method described by Zu et al.~(2011,
2013). We will discuss cross correlation results for the ICCF and ZDCF
methods first, which are presented in the top-right panels of Figure
5, and later comment on the JAVELIN findings.

The ICCF determines the maximum of the CCF between light curves after
interpolating fluxes to a desired cadence. The assumption used is that
the line and continuum fluxes in gaps between two observed points are
properly approximated by a linear interpolation in time between the
two (see e.g., Peterson et al.~2004 and references therein). For our
sample, the ICCF was run for a cadence of 10 days and between $-s/2$
and $s$ days, where $s$ is the time span common to the continuum and
each emission line light curve.

Of the three cross correlation methods described here, ZDCF is the
only one that works solely with the {\em observed\/} values of the
light curves. It is based on the discrete correlation function (DCF)
of Edelson \& Krolik (1988) which uses the available data without
resorting to interpolation, hence not altering the observed light
curves. The DCF method bins the time difference pairs and obtains the
mean correlation coefficient for each bin. A minimum of 11 points is
required in each bin, which determines the length of the correlation,
except for the two bins at the edges where a smaller number of points
is allowed. The ZDCF introduces a `z-transformation' of the DCF
correlation coefficient in order to avoid the inherent skewedness of
the DCF parent distribution function.

\begin{figure*}
\begin{center}
\includegraphics[scale=0.3, trim=30 0 0 0]{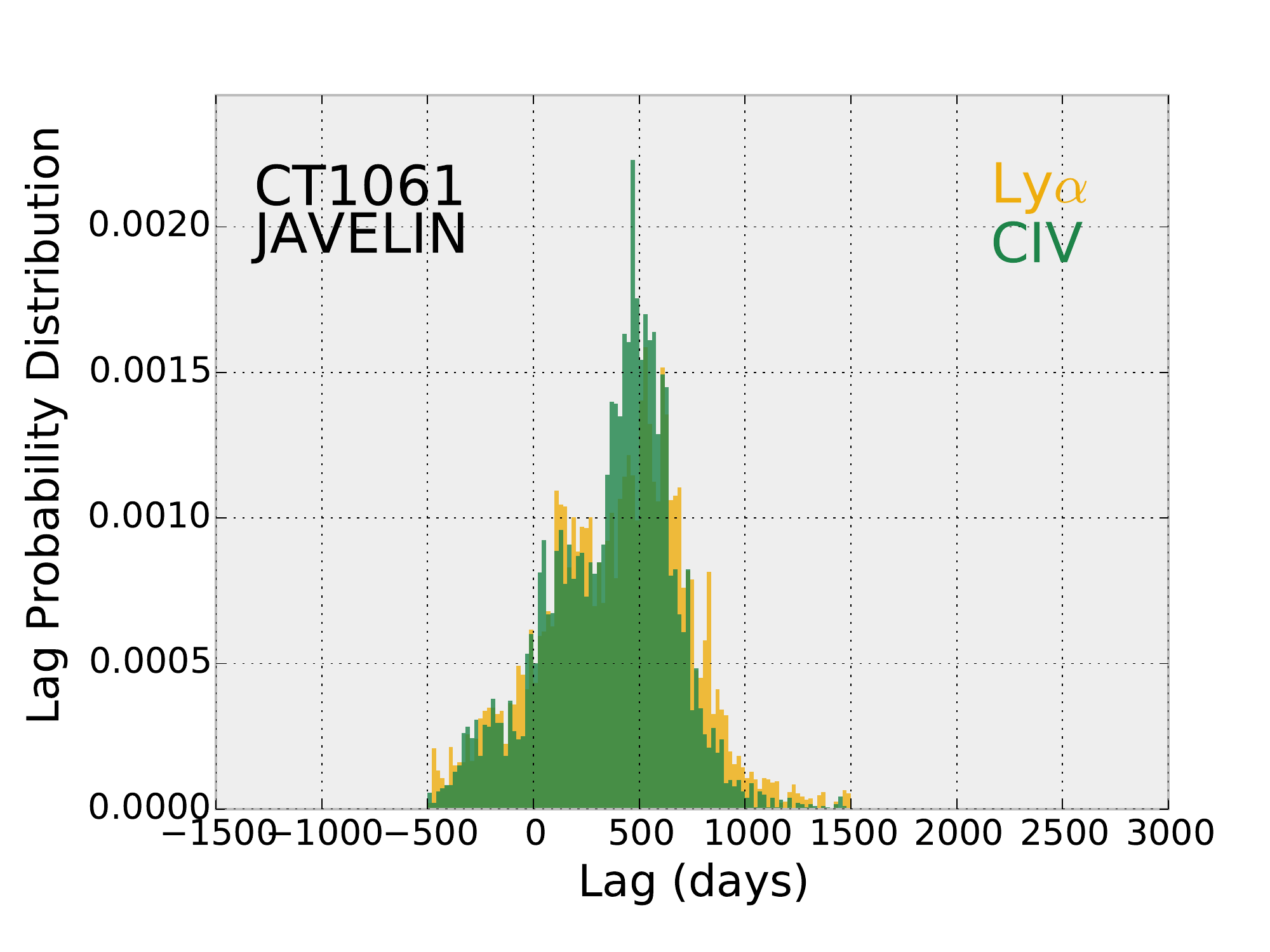}
\includegraphics[scale=0.3, trim=30 0 0 0]{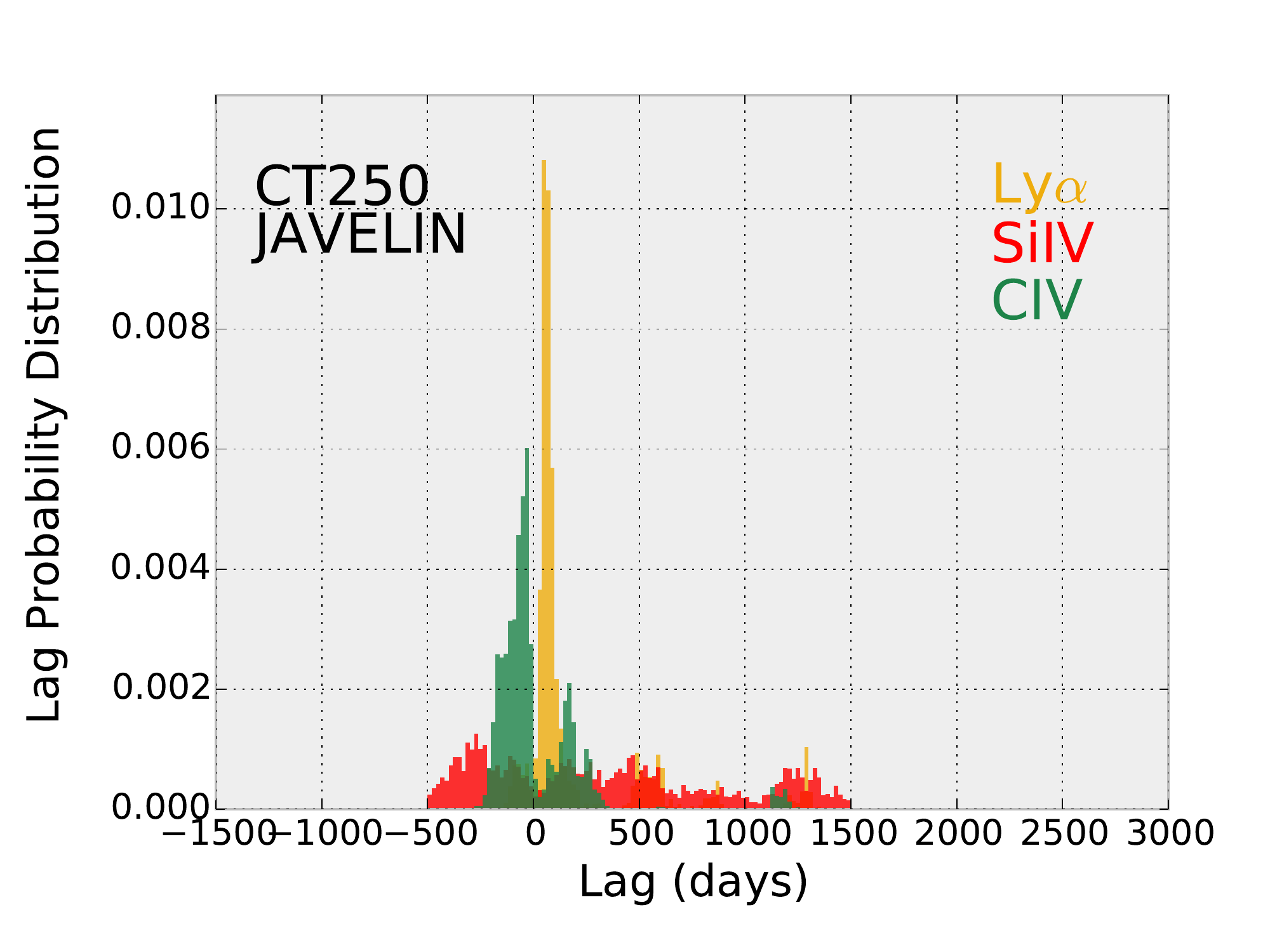}
\includegraphics[scale=0.3, trim=30 0 0 0]{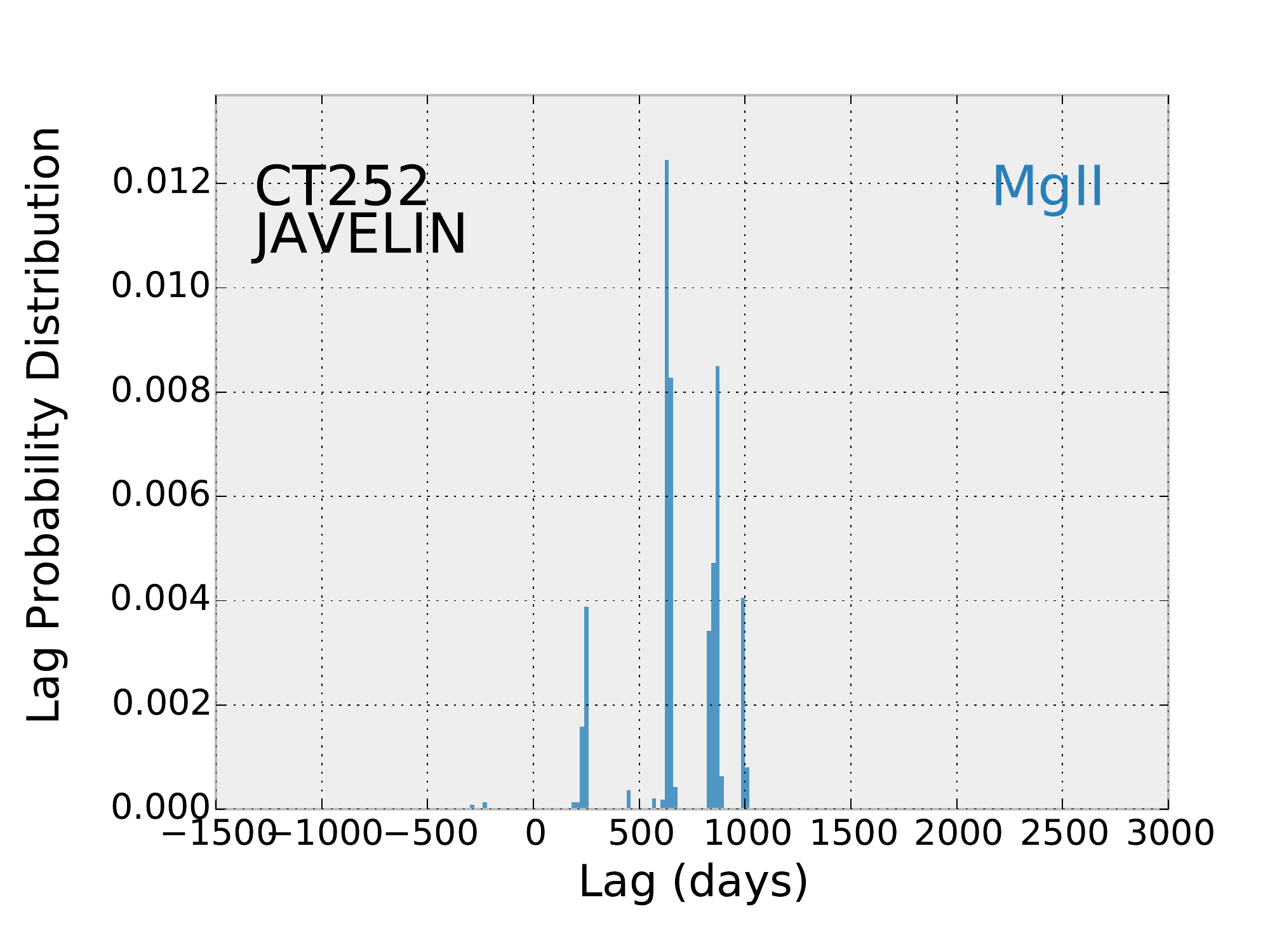}
\includegraphics[scale=0.3, trim=30 0 0 0]{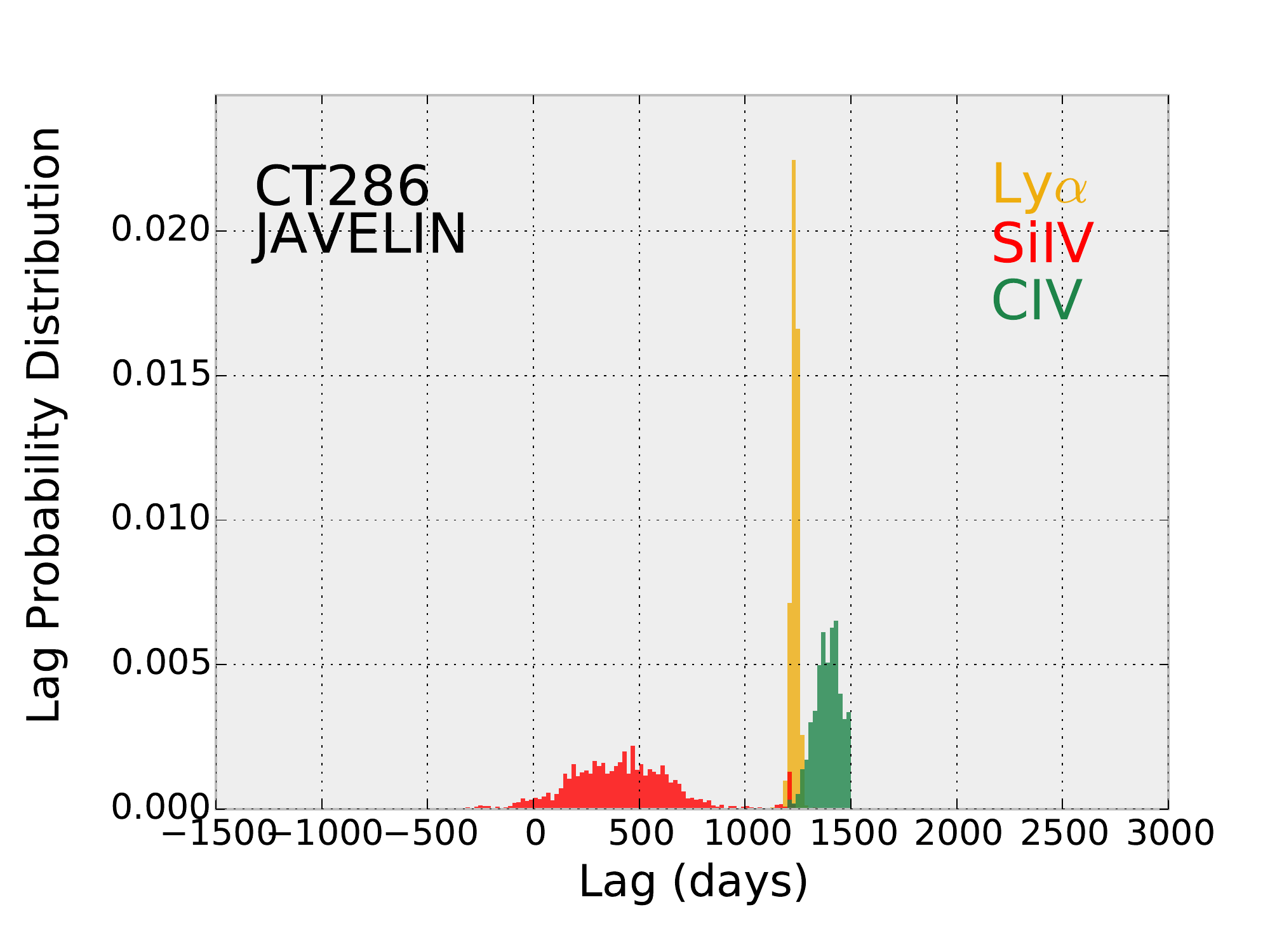}
\includegraphics[scale=0.3, trim=30 0 0 0]{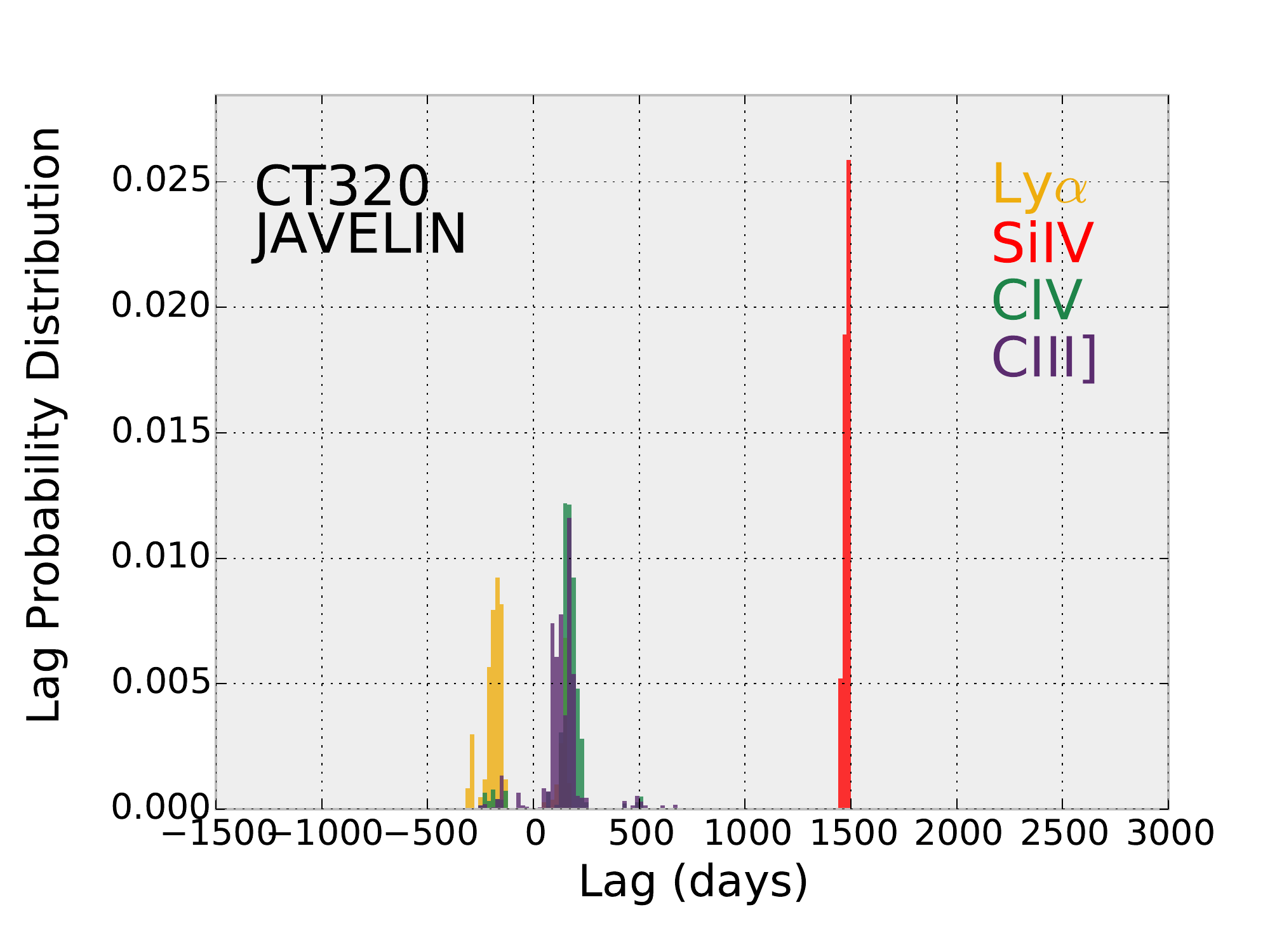}
\includegraphics[scale=0.3, trim=30 0 0 0]{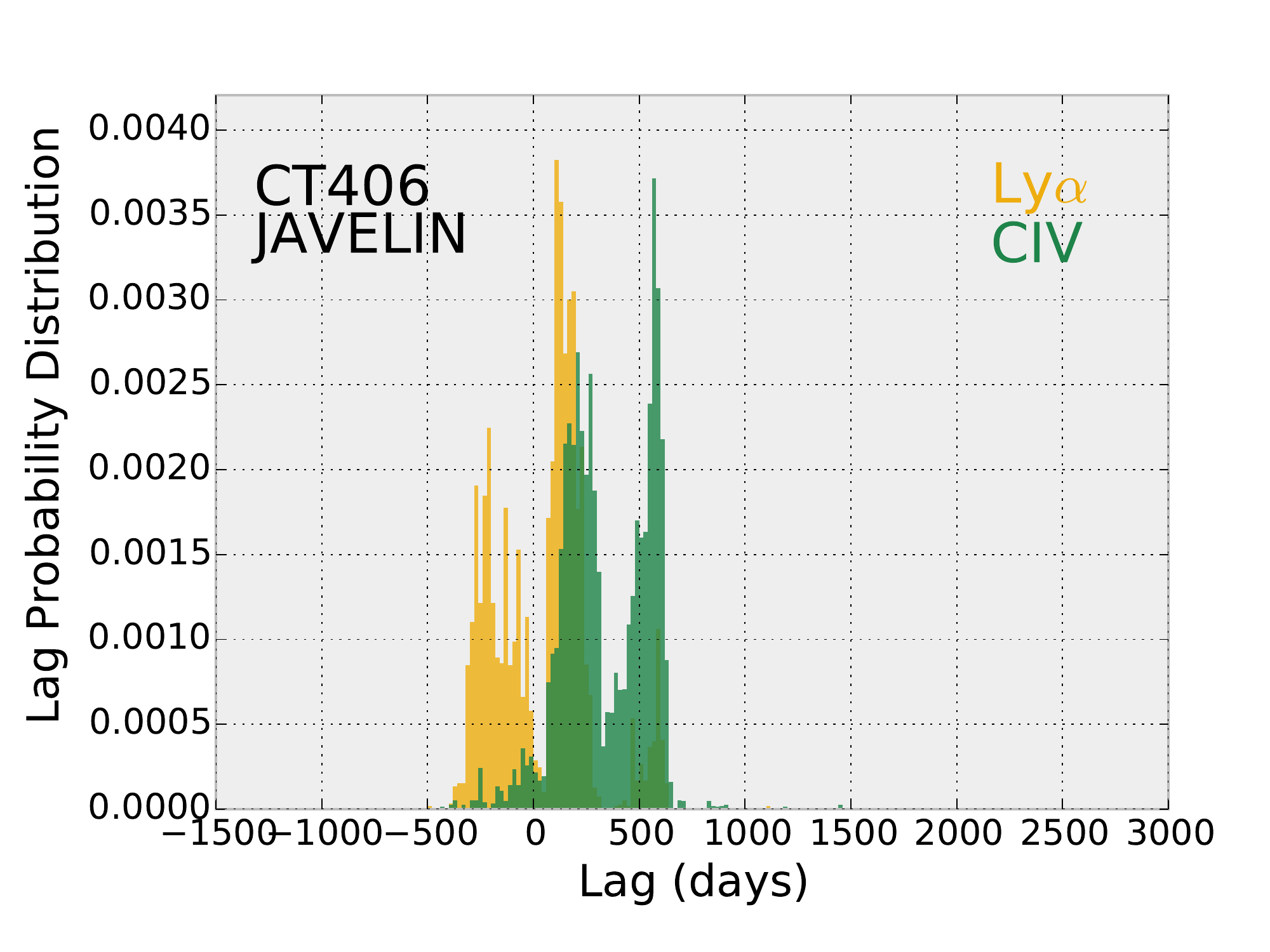}
\includegraphics[scale=0.3, trim=30 0 0 0]{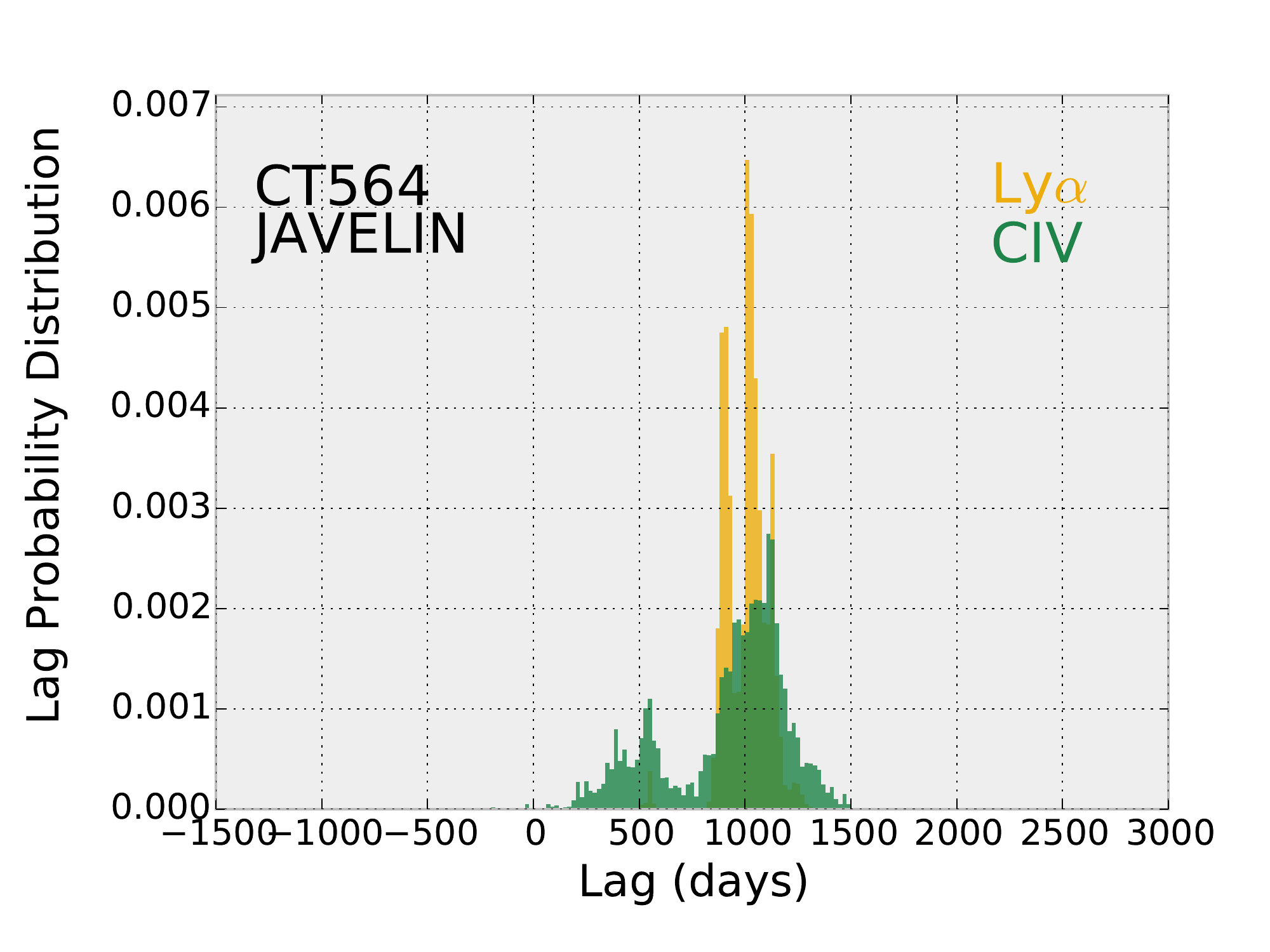}
\includegraphics[scale=0.3, trim=30 0 0 0]{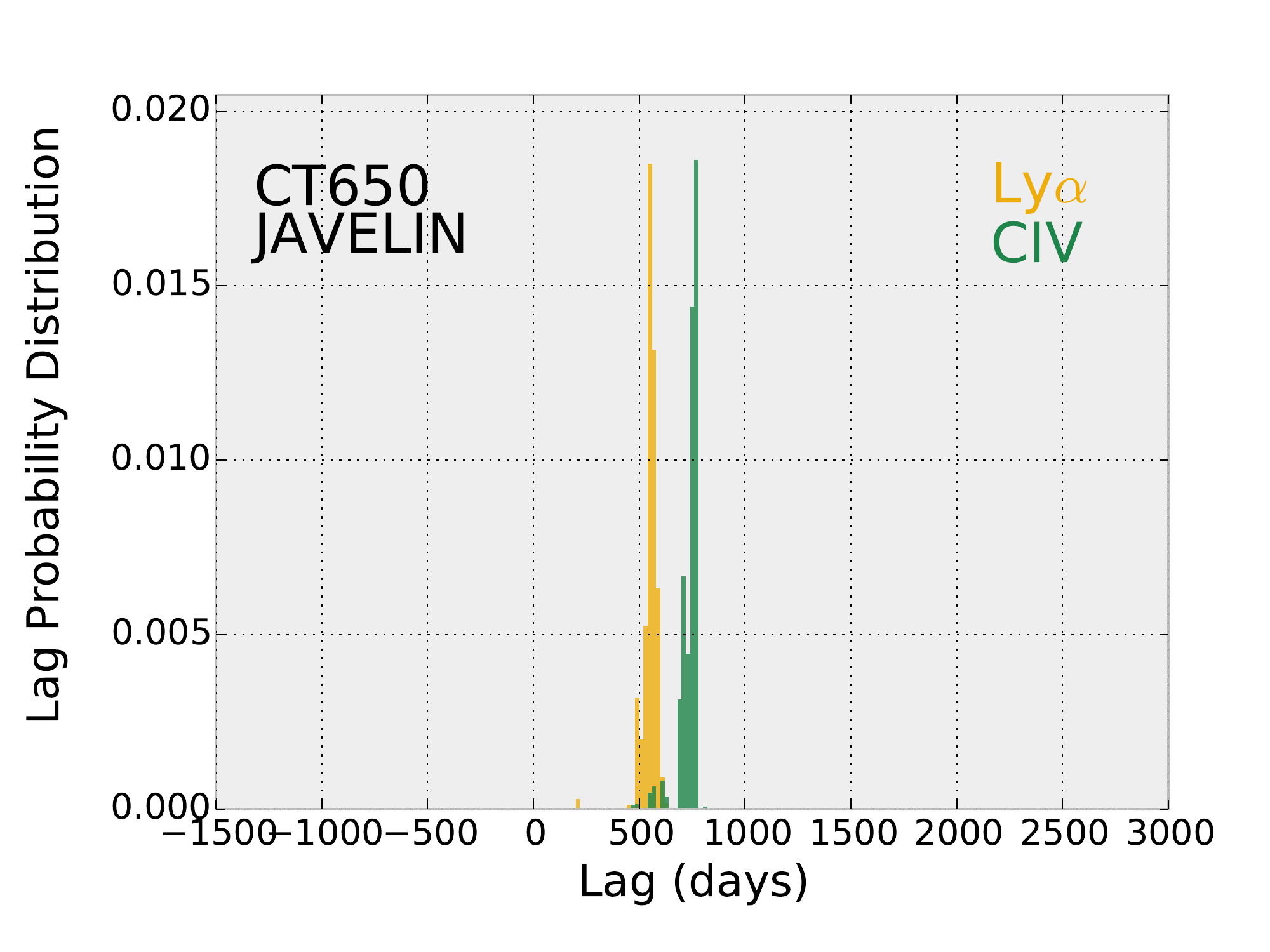}
\includegraphics[scale=0.3, trim=30 0 0 0]{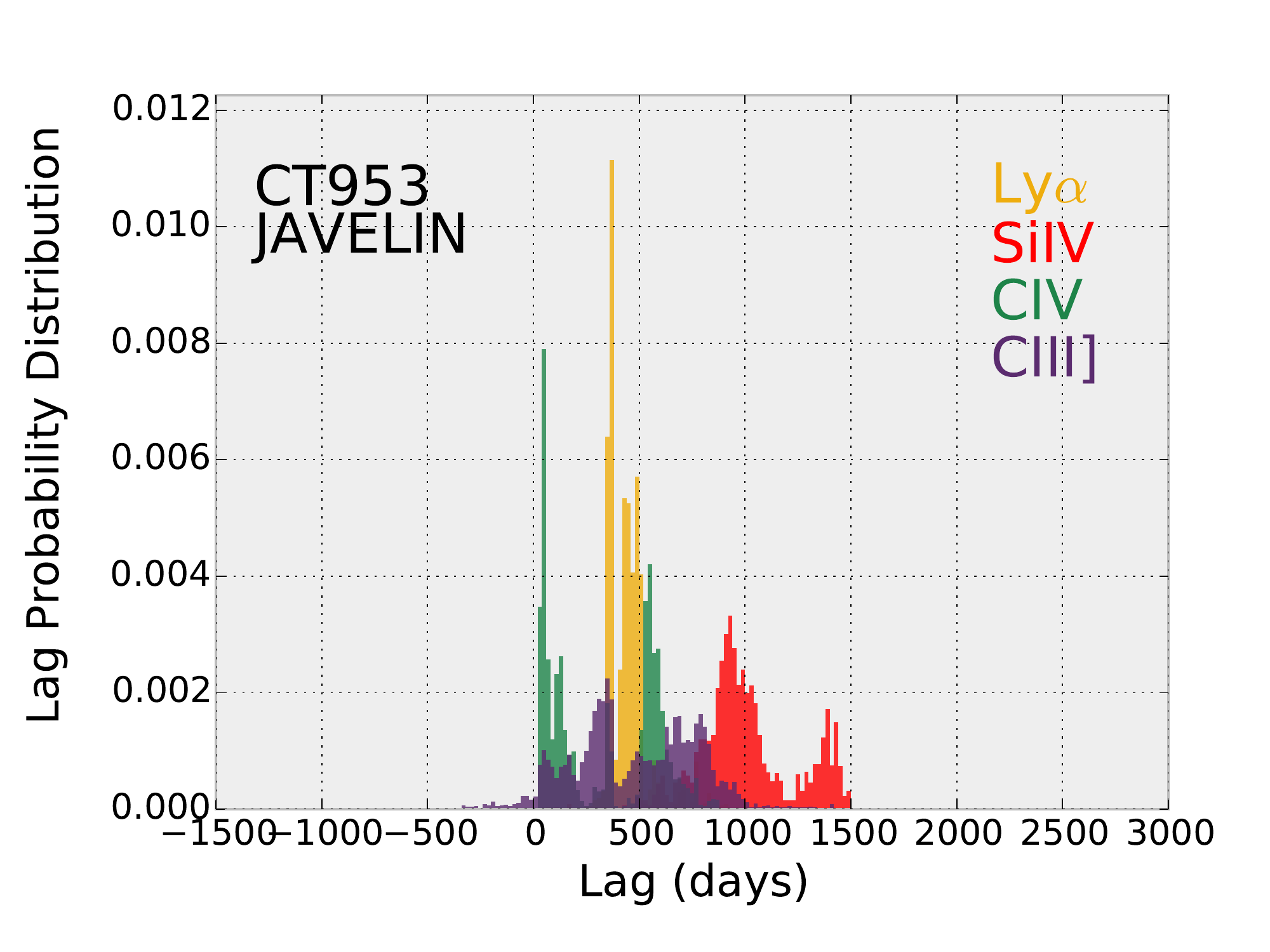}
\includegraphics[scale=0.3, trim=30 0 0 0]{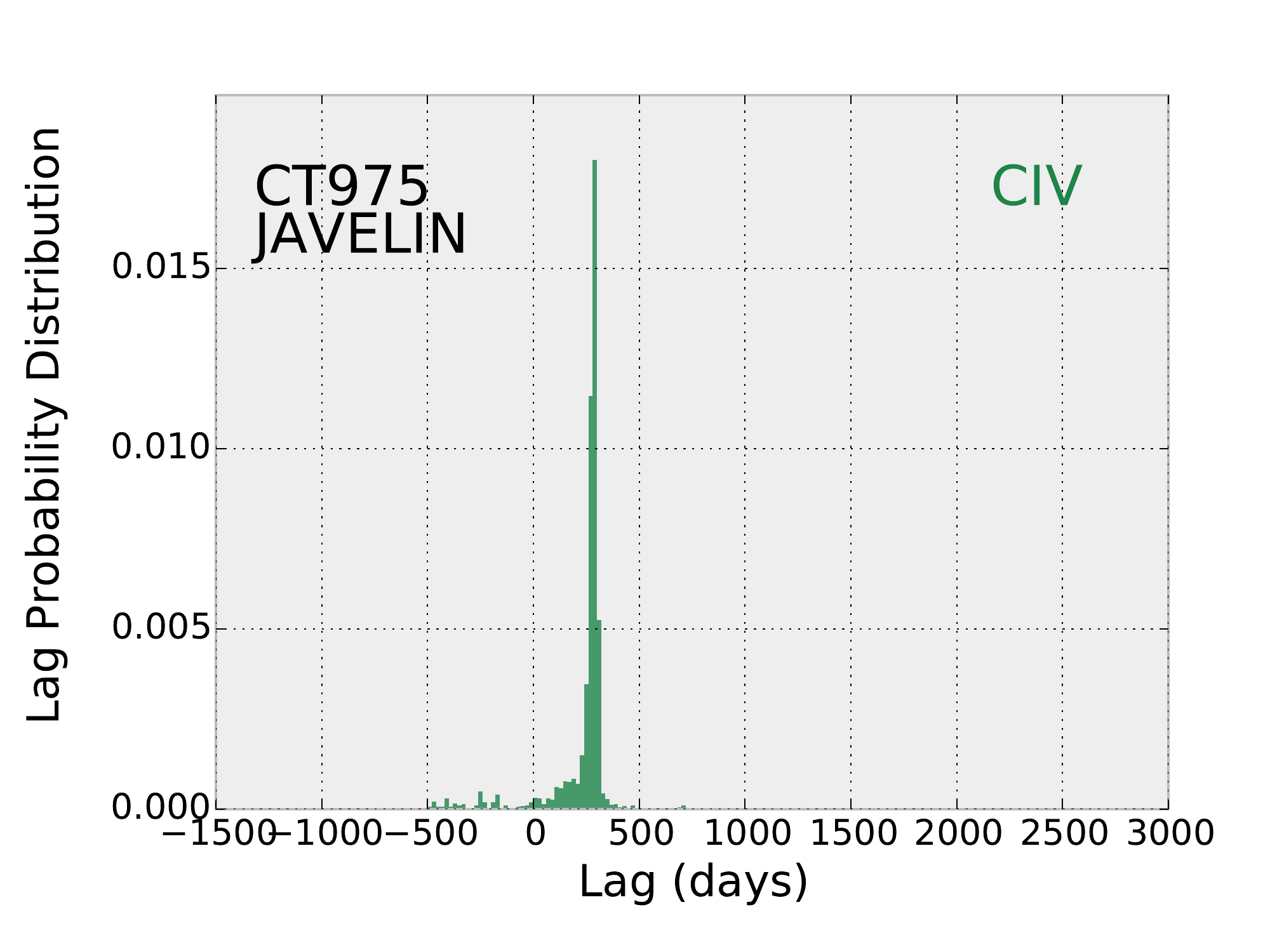}
\includegraphics[scale=0.3, trim=30 0 0 0]{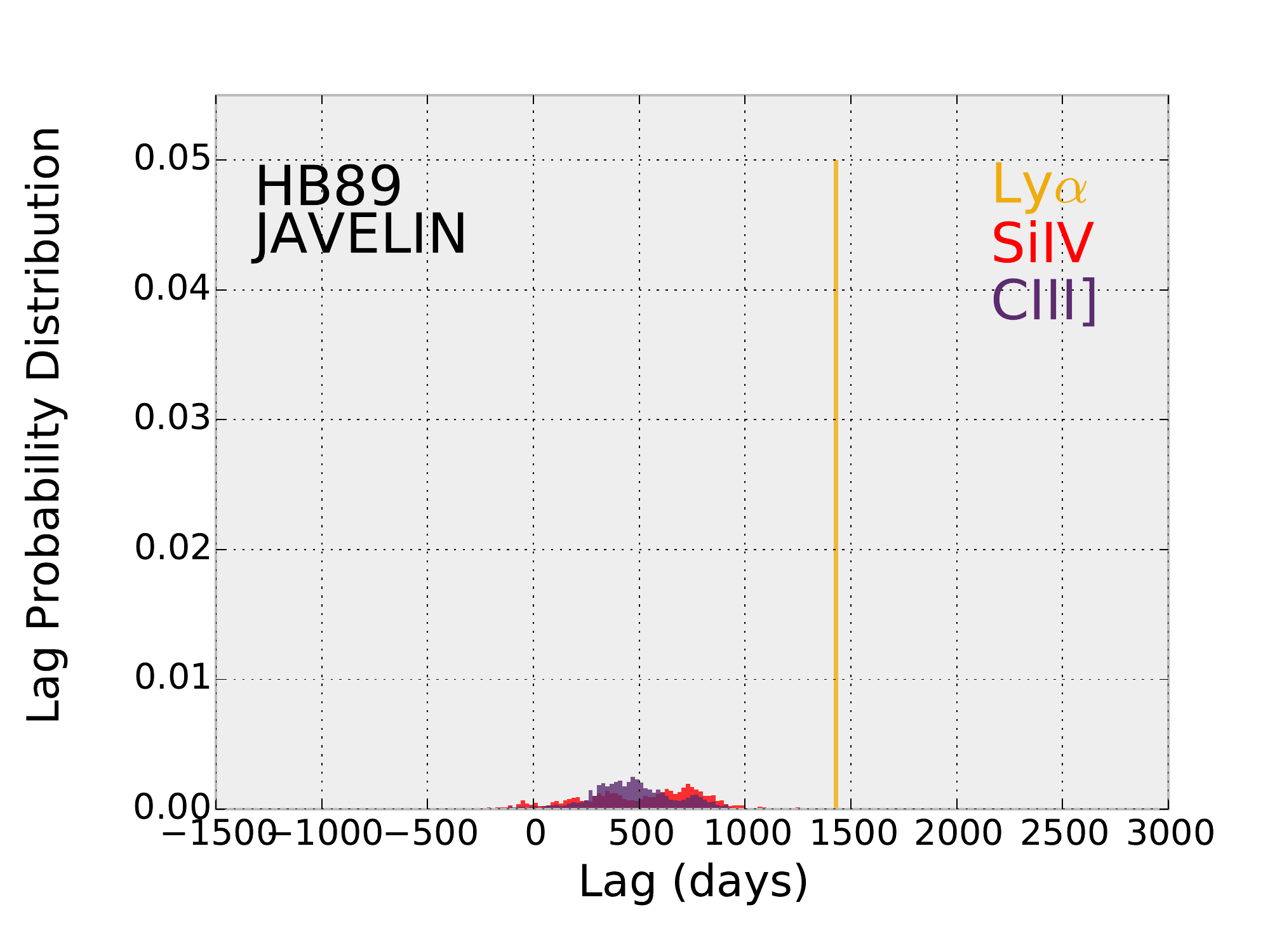}
\includegraphics[scale=0.3, trim=30 0 0 0]{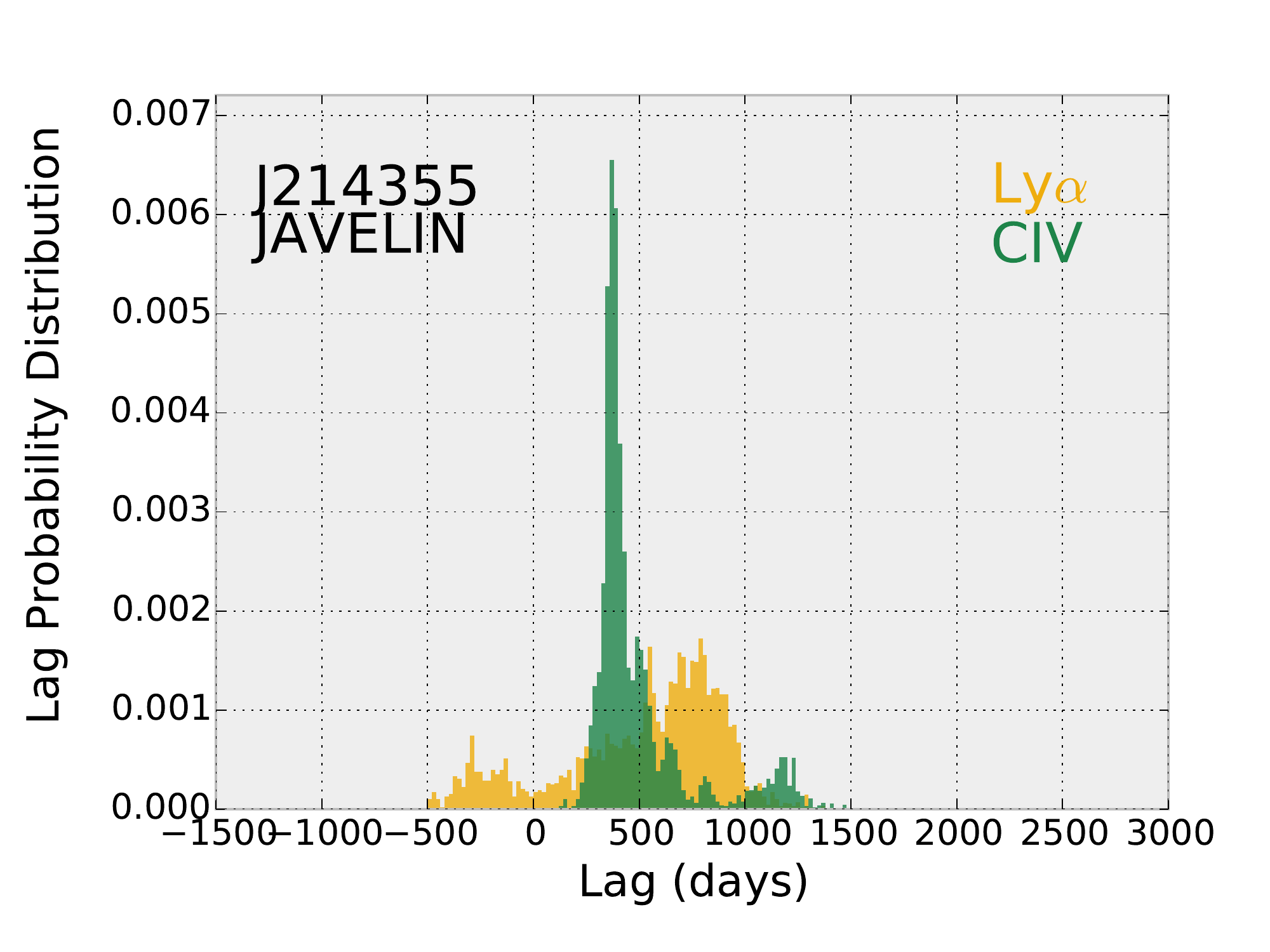}
\includegraphics[scale=0.3, trim=30 0 0 0]{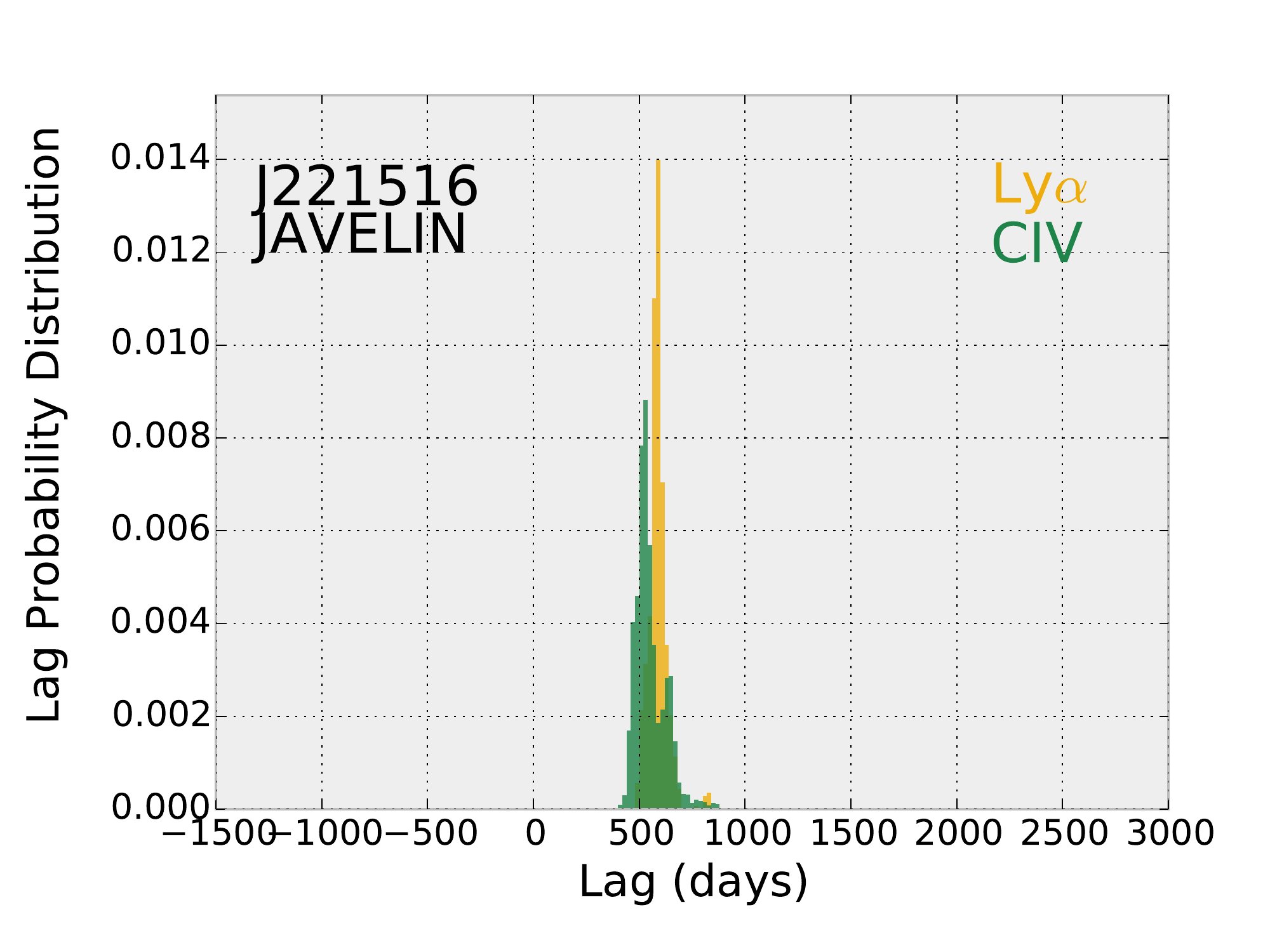}
\caption{Lag Probability Distributions from the JAVELIN
  analysis.}
\end{center}
\end{figure*}

Inspection of the correlations presented in the top-right panels of
Figure 5 shows that in some cases the ICCF and ZDCF follow each other
closely (e.g., Ly$\alpha$ and CIV in CT250), while in other cases
there are significant differences (e.g., SiIV in CT250 and all lines
for J214355), with the ICCF showing higher correlation coefficients
than the ZDCF. This is due to the `extra' information introduced by
the interpolation technique. Also, in some cases the ICCFs are more
extended than the ZDCF (e.g., CT406), because of the requirement of a
minimum number of points per bin by the ZDCF. In other words, those
regions were the ICCFs extend further than the ZDCFs correspond to
regions of sparsely sampled light curves. In summary, ZDCF results
give a more conservative inference of the correlation strength.

Given the above differences between the methods, we will only consider
those emission line light curves that have ICCF and ZDCF cross
correlation coefficient distributions agreeing with each other, and a
well defined peak for positive lags above a cross correlation
coefficient of 0.5 (this is, at least one ZDCF point above 0.5 which
is part of a coherent peak seen in the ZDCF; usually the ICCF will be
found at even higher correlation coefficient levels). We treat
negative peaks as failure to determine physical meaningful lags,
although different explanations, involving a whole different view of
the central region of AGN, could also be considered. Hence, the peaks
to be considered should be found in the upper-right quarter of the
top-right panels in Figure 5. For example, we will consider the
Ly$\alpha$ and CIV light curves for CT1061, but drop the SiIV and
CIII] light curves of this source.

Following these criteria, we drop CT252, CT367, CT803, J002830, and
J224743 from any further analysis. For the remaining sources not all
emission lines will be considered. This is indicated in Table 3.

To estimate the lags, the maximum of the ICCF can be determined in two
ways: finding the peak (or maximum value) of the ICCF (for positive
lags and cross correlations coefficients larger than 0.5), and finding
its centroid (or weighted mean) around the peak above a certain
threshold value. As centroids are more reliable than peaks in flat or
noisy ICCFs, in what follows we adopt the centroids ($\tau_{cent}$) as
the measured lags for those sources showing significant ICCF and ZDCF
peaks, as discussed above. In our case, for the determination of
$\tau_{cent}$, a threshold of 0.85 times the maximum was used.

To determine the lag associated errors, we followed the usual flux
randomization and bootstrapping Monte Carlo technique (also known as
FR/RSS), using a code facilitated by B.~Peterson. Briefly, ICCFs are
computed from light curves constructed after fluxes are randomized
within the observed errors (Flux Randomization -- FR) and 70\%\ of
points are selected from the observed sequences (Random Subset
Selection -- RSS) -- for more details see Peterson et al.~(1998,
2004). 10000 such trials were obtained. As before, the trials had a
cadence of 10 days and run between $-s/2$ and $s$ days. With all
successfully determined centroids, a Cross-Correlation Centroid
Distributions (CCCDs) for objects showing significant ICCF and ZDCF
peaks are presented in the bottom-right panels in Figure 5.

CCCDs in Figure 5 show a wide range of morphologies with sometimes
more than one peak of high statistical probability. This is in
contrast with the CCCDs obtained for many well monitored Seyfert
galaxies, since the presence of many variability `events' helps to
constrain the lags to a singular, well defined peak (e.g., see Clavel
et al.~1991, Wanders et al.~1997, Peterson et al.~2005, Edelson et
al.~2015, Fausnaugh et al.~2016). This degeneracy in the possible lags
seen in the bottom-right panels in Figure 5 cannot be unambiguously
solved unless the light curves could be dramatically extended in
duration. However, it is clear that the observed secondary peaks are
found at the same locations where unfeasible peaks were also observed
in the ICCF-ZDCF distributions shown in the top-right panels in Figure
5 (i.e., regions where the number of ZDCF points is small and hence
the ICCF interpolation not very meaningful; see below) Therefore we
can use the same arguments as before to dismiss them.

For the lag error determinations we truncated the CCCDs at a minimum
and maximum lag and renormalized. The criteria were to leave out
complete peaks that implied negative lags (which are unfeasible),
while negative wings of positive lags were still taken into
account. The upper threshold was given by the largest time bin
computed by the ZDCF algorithm before the final bin (which usually has
less than 11 measurements), i.e., before the light curves become too
sparse for meaningful interpolation and lag determinations. These
thresholds are shown in Figure 5 using vertical gray lines. Lag errors
were finally computed as a $1\sigma$ confidence limit range by
integrating the re-normalized CCCDs from the determined thresholds
until a cumulative value of 0.159 was reached on each end, which
determines $\sigma_{-}$ and $\sigma_{+}$. A final criterion to
consider a lag as reliable is imposed at this stage, with the
requirement that at least 50\%\ of the original CCCD is found within
the defined thresholds (see also Grier et al., 2017). This fraction is
shown in each CCCD presented in Figure 5. This restriction leaves out
the MgII lag for CT252, the CIII] lag for CT953, and the CIV lag for
  CT975. As before this is indicated in the `cc' column in Table
  3. Table 4 presents the final list of lags and their error
  estimates. The MgII lag for CT252, CIII] for CT953, and CIV for
    CT975 are also included in Table 4 but not used for further
    analysis.

We also used JAVELIN to characterize the observed lags (Zu et
al.~2011, 2013). JAVELIN models the light curves as a damped random
walk process (DRWP, also called purely Auto-Regressive, AR(1),
process) as prescribed by Kelly et al.~(2009), i.e., assumes a
particular regime of the power spectral function (with $P_{\nu}
\propto \nu^{\alpha}$ with $\alpha = -2$, breaking to $\alpha = 0$ at
a characteristic frequency) in order to determine a lag and its
significance.

The advantage of JAVELIN over the ICCF method is that the errors
associated to values interpolated between actual observations are
based on the DRWP model, which are usually larger and more realistic
than those obtained from a linear interpolation. However, for
sufficiently well sampled light curves, it has been found that a DRWP
model only applies to about half of AGN of Seyfert and quasar-like
luminosities (Kasliwal et al.~2015). Moreover, the basic assumption of
JAVELIN is that the emission line light curves are the result of the
response to an ionizing continuum which is changing exactly in the
same way as the observed continuum used in the calculations, in
contradiction to some well documented previous results (Goad et
al.~2016), as well as some of the cases presented in this work. In
fact, JAVELIN computes a solution which assumes a transfer function of
the continuum and solves simultaneously for both, the continuum and
line emission interpolated light curves.

We run JAVELIN using the same lag limits used during the ICCF
calculations, implemented 5000 Markov chain Monte Carlo iterations
during the `burn-in' phase, and 10000 iterations during the final
parameter determination step, assumed the default models for
description of the continuum (DRWP, or `Cont\_Model' in JAVELIN
language) and line (`RMap\_Model') light curves, and solved for single
continuum-line pairs at each time. Figure 6 presents the JAVELIN lag
posterior distributions for the same sources presented in Figure 5.

\begin{table*}
\renewcommand{\thetable}{\arabic{table}} \centering
\caption{Cross-correlation analysis}
\begin{tabular}{l|cccc|cccc}
\tablewidth{0pt}
\hline
\hline
       & \multicolumn{4}{c|}{Ly$\alpha$} & \multicolumn{4}{c}{SiIV} \\
       & $\tau_{cent}$ & $\sigma_{-}$ & $\sigma_{+}$ & M$_{\rm BH}$& $\tau_{cent}$ & $\sigma_{-}$ & $\sigma_{+}$ & M$_{\rm BH}$ \\
       & (days) & (days) & (days) & ($10^9$ M$_{\odot}$)& (days) & (days) & (days) & ($10^9$ M$_{\odot}$) \\
\hline
CT1061 & 431 & 239 & 461 & ---  &---&---&---&--- \\
       & (99) & (55) & (106) & --- & --- & --- & --- & ---\\ 
CT250 & 37 & 65 & 155 & ---  & 174 & 652 & 988 & ---  \\
      & (11) & (19) & (45) & --- & (51) & (191) & (290) & ---\\ 
CT286 & 1191 & 78 & 542 & 0.6 $\pm$ 0.2  & 427 & 205 & 395 & 0.7 $\pm$ 0.5  \\
      & (335) & (22) & (153) & --- & (120) & (58) & (111) & ---\\ 
CT320 & -222 & 105 & 365 & ---  & 1818 & 105 & 345 & 3.9 $\pm$ 0.6  \\
      & (-56) & (26) & (92) & --- & (459) & (26) & (87) & ---\\ 
CT406 & 16 & 105 & 505 & ---  &---&---&---&--- \\
      & ( 5) & (29) & (141) & --- & --- & --- & --- & ---\\ 
CT564 & 426 & 193 & 647 & ---  &---&---&---&--- \\
      & (102) & (46) & (155) & --- & --- & --- & --- & ---\\ 
CT650 & 548 & 54 & 56 & 0.16 $\pm$ 0.02  &---&---&---&--- \\
      & (150) & (15) & (15) & --- & --- & --- & --- & ---\\ 
CT953 & 465 & 193 & 87 & 1.8 $\pm$ 0.6  & 779 & 637 & 1153 & ---  \\
      & (127) & (53) & (24) & --- & (213) & (174) & (315) & ---\\ 
CT975 &---&---&---&--- &---&---&---&--- \\
      & --- & --- & --- & --- & --- & --- & --- & ---\\ 
HB89 & 1543 & 171 & 399 & 1.4 $\pm$ 0.3  & 349 & 147 & 693 & ---  \\
     & (438) & (48) & (113) & --- & (99) & (42) & (196) & ---\\ 
J214355 & 724 & 469 & 161 & 1.3 $\pm$ 0.6  &---&---&---&--- \\
     & (187) & (121) & (42) & --- & --- & --- & --- & ---\\ 
J221516 & 637 & 43 & 147 & 0.22 $\pm$ 0.04  &---&---&---&--- \\
     & (185) & (13) & (43) & --- & --- & --- & --- & ---\\ 
\hline
       & \multicolumn{4}{c|}{CIV} & \multicolumn{4}{c}{CIII]} \\
       & $\tau_{cent}$ & $\sigma_{-}$ & $\sigma_{+}$ & M$_{\rm BH}$& $\tau_{cent}$ & $\sigma_{-}$ & $\sigma_{+}$ & M$_{\rm BH}$ \\
       & (days) & (days) & (days) & ($10^9$ M$_{\odot}$)& (days) & (days) & (days) & ($10^9$ M$_{\odot}$) \\
\hline
CT1061 & 397 & 105 & 485 & ---  &---&---&---&--- \\
       & (91) & (24) & (111) & --- & --- & --- & --- & ---\\
CT250 & -24 & 184 & 356 & ---  &---&---&---&--- \\
      & (-7) & (54) & (104) & --- & --- & --- & --- & ---\\
CT286 & 1629 & 327 & 253 & 1.1 $\pm$ 0.2  &---&---&---&--- \\
      & (459) & (92) & (71) & --- & --- & --- & --- & ---\\
CT320 & 217 & 333 & 67 & --- & 162 & 169 & 451 & ---  \\
      & (55) & (84) & (17) & --- & (41) & (43) & (114) & ---\\
CT406 & 411 & 310 & 230 & 0.7 $\pm$ 0.5  &---&---&---&--- \\
      & (115) & (86) & (64) & --- & --- & --- & --- & ---\\
CT564 & 419 & 106 & 674 & ---  &---&---&---&--- \\
      & (100) & (25) & (161) & --- & --- & --- & --- & ---\\
CT650$\star$ & 592 & 38 & 122 & 0.4 $\pm$ 0.1  &---&---&---&--- \\
      & (162) & (10) & (33) & --- & --- & --- & --- & ---\\
CT953$\dag$ & 256 & 204 & 406 & ---  & 417 & 445 & 1355 & ---  \\
      & (70) & (56) & (111) & --- & (114) & (122) & (370) & ---\\
CT975$\dag$ & 289 & 33 & 157 & 0.5 $\pm$ 0.2  &---&---&---&--- \\
     & (77) & ( 9) & (42) & --- & --- & --- & --- & ---\\
HB89 &---&---&---&--- & 343 & 311 & 449 & ---  \\
     & --- & --- & --- & --- & (97) & (88) & (127) & ---\\
J214355 & 493 & 318 & 352 & 0.9 $\pm$ 0.6  &---&---&---&--- \\
     & (128) & (82) & (91) & --- & --- & --- & --- & ---\\
J221516 & 566 & 43 & 337 & 0.3 $\pm$ 0.1  &---&---&---&--- \\
     & (165) & (13) & (98) & --- & --- & --- & --- & ---\\
\hline
      & \multicolumn{4}{c|}{MgII}  \\
      & $\tau_{cent}$ & $\sigma_{-}$ & $\sigma_{+}$ & M$_{\rm BH}$\\
      & (days) & (days) & (days) & ($10^9$ M$_{\odot}$)\\
\hline
CT252$\dag$ & 550 & 170 & 330 & --- \\
            & (190) & (59) & (114) & ---\\
\hline
\hline
\multicolumn{9}{l}{Lags are given in days in the observed frame and, in parenthesis, in the rest frame.}\\
\multicolumn{9}{l}{Black Hole virial masses are given assuming a virial factor of 1.}\\
\multicolumn{9}{l}{M$_{\rm BH}$ values are presented only for objects where $\tau_{cent}/\sigma_{-} > 1$ and $\tau_{cent}/\sigma_{+} > 1$.}\\
\multicolumn{9}{l}{$\star$ Centroid calculations for the CT650 CIV line failed in 60\%\ of the trials.}\\
\multicolumn{9}{l}{$\dag$ The MgII lag for CT252, CIII] for CT953, and CIV for CT975 are not considered reliable}\\
\multicolumn{9}{l}{as they were obtained from less than 50\%\ of the original CCCD distributions.}\\
\end{tabular}
\end{table*}

\begin{figure*}
\begin{center}
\includegraphics[scale=0.42, trim=0 0 0 50]{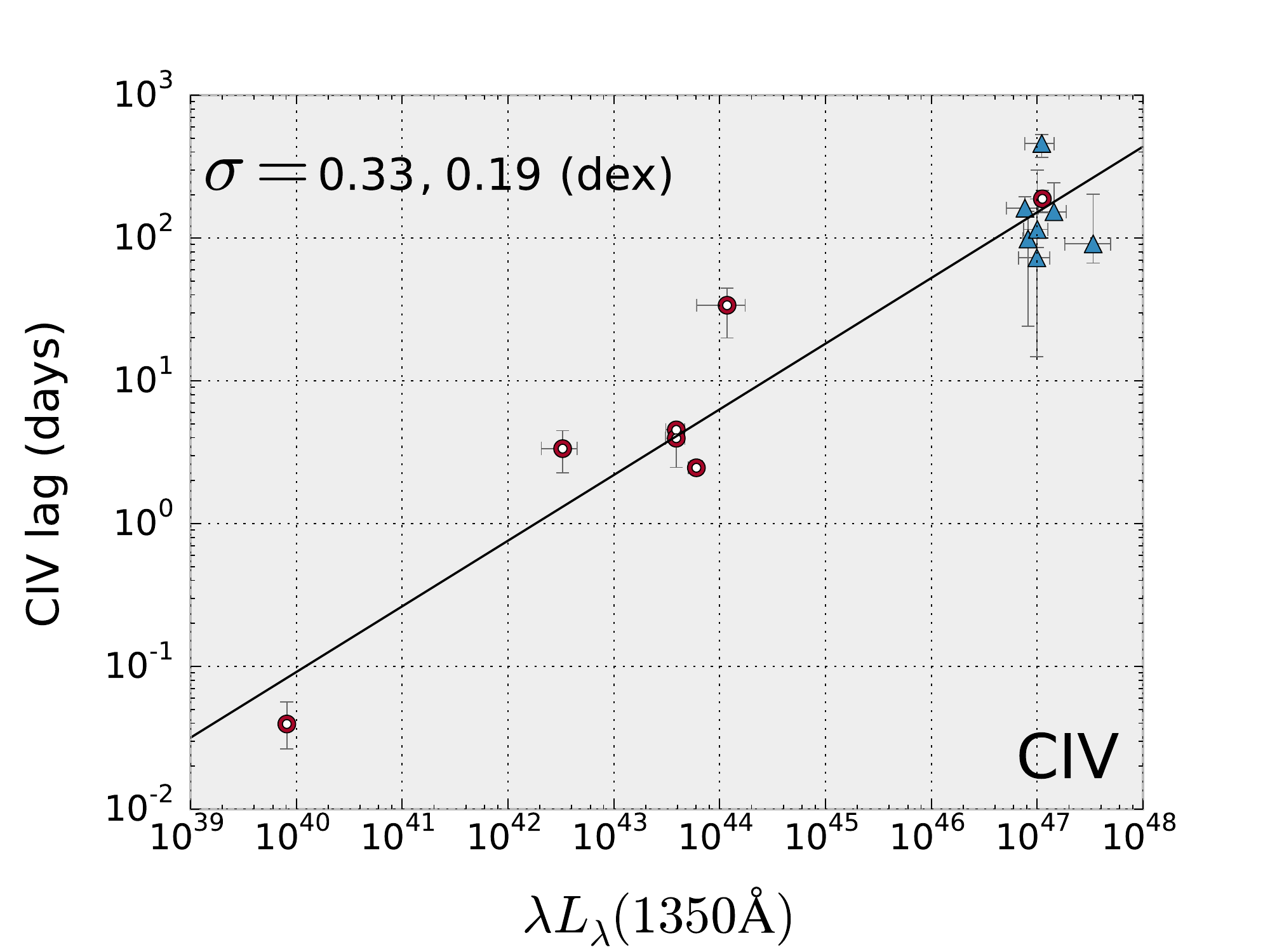}
\includegraphics[scale=0.42, trim=0 0 0 50]{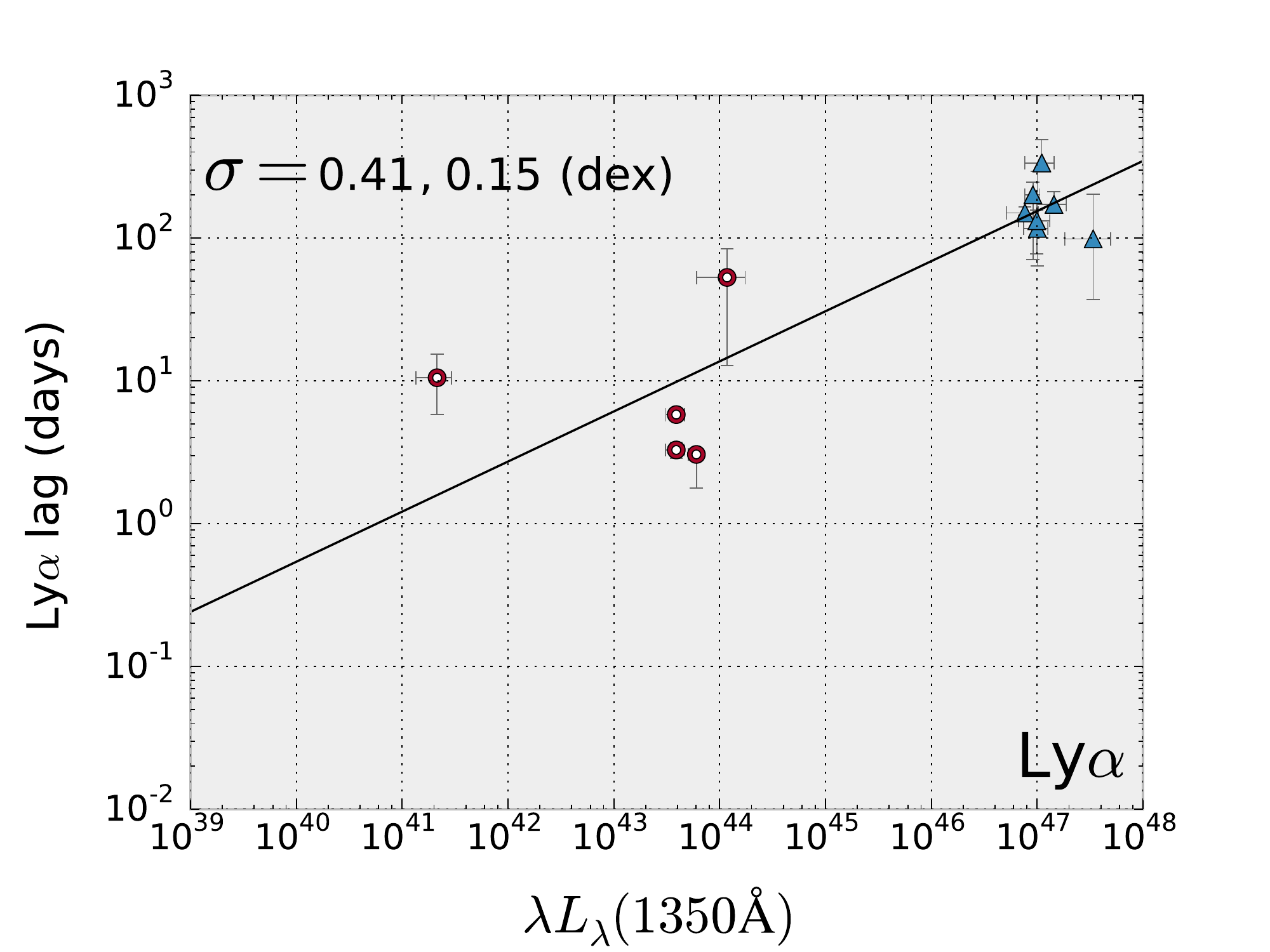}
\includegraphics[scale=0.42, trim=0 0 0 8]{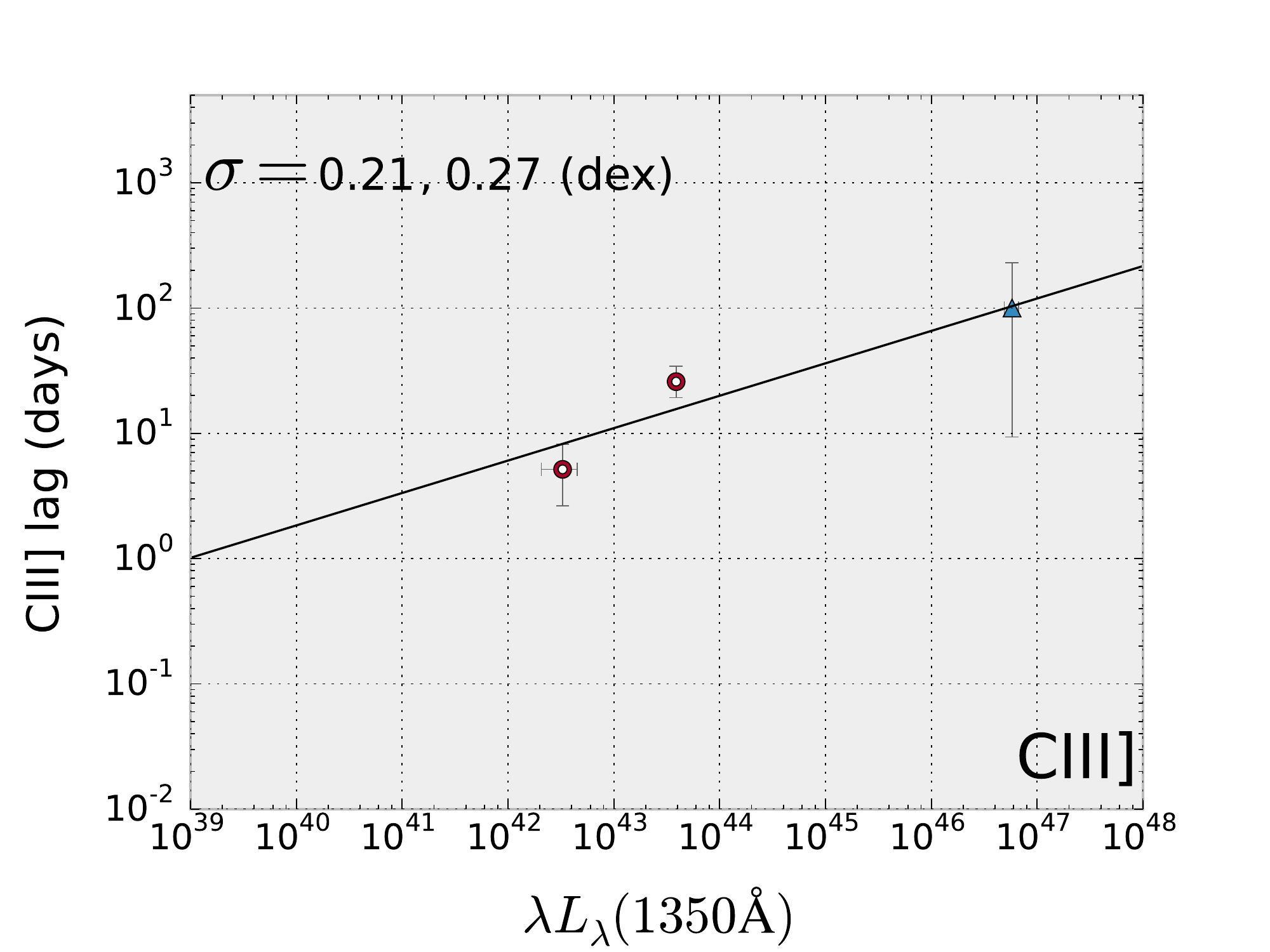}
\includegraphics[scale=0.42, trim=0 0 0 8]{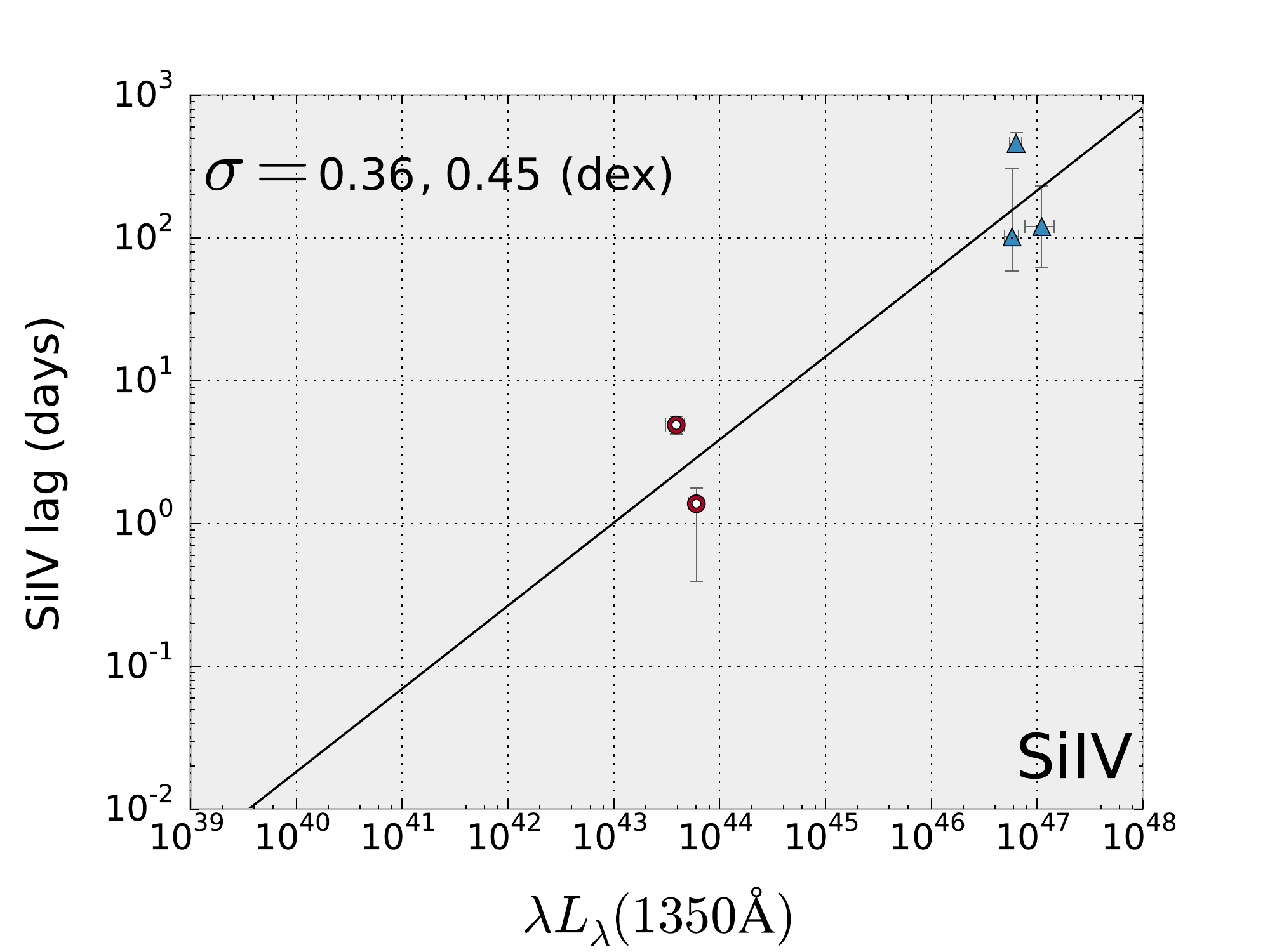}
\caption{Radius--luminosity relations for the Ly$\alpha$, CIV, CIII]
and SiIV emission lines. Our sources are presented with blue solid
triangles, while sources taken from the literature are presented with
red circles. The solid black lines show the linear regression to the
data using a bivariate analysis. Two values of $\sigma$ are shown at
the top left of each panel, where the first value corresponds to the
observed scatter and the second value corresponds to the scatter due
to measurement errors.}
\end{center}
\end{figure*}

Comparison between the histograms presented in Figure 5 and 6 shows
that both methods roughly agree on the best determined lags. However,
JAVELIN tends to present considerably compact probability
distributions for most objects, in some cases with several well
defined, extremely narrow peaks. As already observed by Fausnaugh et
al.~(2016), this is mostly because of the very strong -- and seldom
demonstrated -- assumption that the emission line light curves are a
simple lagged and smoothed version of the continuum emission. In fact,
Fausnaugh et al.~(2016) suggests that the actual dispersion is within
$2\sigma - 3\sigma$ of the JAVELIN quoted errors. In a few cases
(e.g., CT1061), the JAVELIN results are very close to those found by
our ICCF analysis.

JAVELIN results do not show spurious peaks at the edges of the
probability distributions, which is also a consequence of the
assumption of a BLR responding to the continuum light curves.
Finally, it is interesting to notice some cases with contradicting
results between the methods. For example, JAVELIN fails to determine a
MgII lag for CT252, while the ICCF method detects a clear lag around
550 days which however, contains only 5\%\ of the original CCCD
distribution (this is in contrast with the CIV lag for CT975, with a
1\%\ peak observed in the reliable region of the CCCD, which JAVELIN
recovers successfully). At the same time, JAVELIN finds a well behaved
peak at around 1000 days for the SiIV line in CT953, while the ICCF
only finds a very shallow peak.

Since our main aim is to find statistically sound lags for our sample
of luminous quasars, for the analysis and discussion in the next
sections we will use the more conservative ICCF and CCCD-based lags
and error estimates as representative line lags for our sources.

\section{Radius--Luminosity relations}

The radius--luminosity relation represents a milestone for the
determination of the masses of black holes hosted by AGN. Here we will
update the radius--luminosity relation for the CIV emission line first
presented by Peterson et al.~(2005, 2006) and later extended to higher
luminosities by Kaspi et al.~(2007) and Trevese et.~(2014). To our
knowledge, no determination of the radius--luminosity relations for
the Ly$\alpha$ has previously been attempted, as no results for
objects above $\lambda L_{\lambda} (1350 {\rm \AA}) = 10^{45}$ ergs
s$^{-1}$ had been reported until now.

In what follows we only construct radius--luminosity relations for
objects where the lags are inconsistent with zero at a $1\sigma$
level, this is, when $\tau_{cent}/\sigma_{-} > 1$. UV $\lambda
L_{\lambda} (1350 {\rm \AA})$ luminosities were obtained from the mean
spectra of each quasar and are reported in Table 1. The errors
represent the rms variation observed in the R-band continuum light
curves. Scatter in the radius--luminosity correlations is given in
each plot, where the first value corresponds to the observed scatter
and the second value corresponds to the scatter due to measurement
errors.

A linear regression was determined for each radius--luminosity
relation using the bivariate method (BCES) of Akritas \& Bershady
(1996), which takes into account errors in both, the lags and the
luminosities. As our lag error determinations are not symmetrical, we
use the mean of both confidence limits as a first guess for the lag
error of each data point and iterate so that the final error bar
considered ($\sigma_{-}$ or $\sigma_{+}$) is determined by whether the
points are found above or below the best fit solution. Convergence was
always found after a few iterations.

\subsection{The CIV Radius--Luminosity relation}

For luminosities below $\lambda L_{\lambda} (1350 {\rm \AA}) =
10^{45}$ ergs s$^{-1}$, CIV lag measurements are compiled in Peterson
et al.~(2005, 2006) and Metzroth et al.~(2006), and a very recent
determination for NGC5548 is found in De Rosa et al.~(2015). Two
sources above this luminosity limit have been published: S5\,0836+71
by Kaspi et al.~(2007), and PG1247+267 by Trevese et al.~(2014). Using
the ICCF method it is not possible, however, to determine the lag for
PG1247+267, as Trevese et al.~(2014) also pointed out, and therefore
we will not include this source in our CIV radius--luminosity
determination. Our work adds to the list seven new high luminosity
sources, namely, CT1061, CT286, CT564, CT650, CT953, J214355 and
J221516.

In Figure 7 we present the results from the linear regression to the
CIV radius--luminosity relation together with the lag and luminosity
measurements. Following Kaspi et al.~(2007), we write the CIV
radius--luminosity in the following way:

\begin{equation}
\frac{R_{\rm CIV}}{10\ {\rm lt-days}} = (0.22\pm0.10) \left [ \frac{\lambda L_{\lambda} (1345{\rm \AA})}{10^{43}\ {\rm erg\ s}^{-1}} \right ]^{(0.46\pm0.08)}
\end{equation}

The updated CIV radius--luminosity relation is very close to that
reported by Kaspi et al.~(2007), albeit with slightly larger
uncertainties. This is because we have included the mean from two lag
measurements of NGC4151 taken from Metzroth et al.~(2006), which
helped to bridge the luminosity gap between the more luminous Seyfert
galaxies and the dwarf Seyfert NGC4395 ($\lambda L_{\lambda} (1350
{\rm \AA}) \sim 10^{40}$ ergs s$^{-1}$), but added more dispersion to
the relation. Also, the two measurements for the NGC4395 lag have been
averaged, reducing the weight of the measurements at the very low
luminosity end.

We remind the reader that a well determined radius--luminosity
relation for CIV does not solve the issues of using CIV as a well
calibrated mass estimator for AGN, as the main problems with this line
is the difficulty to determine a velocity that would reflect a
virialized component of the velocity field of the line (see further
discussion in Section 6.3 and e.g., Baskin \& Laor 2005; Denney et
al.~2012, 2016; Mej\'ia-Restrepo et al.~2016, and references therein).

\subsection{The Ly$\alpha$ Radius--Luminosity relation}

From our sample, CT1061, CT286, CT564, CT650, CT953, J214355 and
J221516 show significant lags and well determined uncertainties and
will be considered for the determination of the Ly$\alpha$
radius--luminosity relation.

In order to have homogenous measurements, we re-determined the
Ly$\alpha$ lags for the Seyfert galaxies monitored by the
International AGN
Watch\footnote{http://www.astronomy.ohio-state.edu/$\sim$agnwatch/}
using the ICCF FR/RSS code that we employed with our sources.  This is
motivated by the improvements introduced to the FR/RSS code following
Peterson et al.~(2004), since all AGN Watch results were obtained
previous to that date. We found significant lag constraints for
NGC3783 ($3.5^{+1.6}_{-2.0}$ days), NGC7469 ($2.0^{+0.2}_{-1.3}$
days), Fairall 9 ($9.4^{+5.1}_{-4.9}$ days), and 3C390.3
($61^{+33}_{-42}$ days) which were originally published by Reichert et
al.~(1994), Wanders et al.~(1997), Rodriguez-Pascual et al.~(1997) and
O'Brien et al.~(1998), respectively. Finally, we added the recent
results for NGC5548 from De Rosa et al.~(2015) which gives a lag of
$5.9^{+0.3}_{-0.3}$ days during the `non-anomalous' period of the
campaign.

The analytical expressions found for the radius--luminosity relation
using the linear regression is as follows:

\begin{equation}
\frac{R_{\rm Ly\alpha}}{10\ {\rm lt-days}} = (0.61\pm0.80) \left [ \frac{\lambda L_{\lambda} (1345{\rm \AA})}{10^{43}\ {\rm erg\ s}^{-1}} \right ]^{(0.35\pm0.19)}
\end{equation}

Unfortunately, the Ly$\alpha$ radius--luminosity relation is not well
constrained at the mid and low luminosity end, as the International
AGN Watch sources show a very large dispersion and no very
low-luminosity AGN has been monitored for this line.

\subsection{The CIII] and SiIV Radius--Luminosity relations}

A CIII] radius--luminosity relation was determined using lag
  measurements for NGC4151 (Metzroth et al.~2006) and a new analysis
  of the NGC5548 light curves originally presented by Clavel et
  al.~(1991), which gave a lag of $26.3^{+8.3}_{-7.0}$ days. This work
  adds HB89 as the only quasar from our sample that meets the
  requirement $\tau_{cent}/\sigma_{-} > 1$.

To determine a SiIV radius-luminosity relation, we re-analyzed the
historical data for NGC7496 and 3C390.3 (Wanders et al.~1997 and
O'Brien et al.~1998, respectively), but could only determine a
significant lag of $1.4^{+0.4}_{-1.1}$ days for NGC7496. We add the
recent determination for NGC5548 from De Rosa et al.~(2015). CT286,
CT320 and HB89 lag measurements from our sample are included. These
relationships are presented in Figure 7. The resulting analytical
expressions are as follows:

\begin{equation}
\frac{R_{\rm CIII\\]}}{10\ {\rm lt-days}} = (1.10 \pm 0.77) \left [ \frac{\lambda L_{\lambda} (1350{\rm \AA})}{10^{43}\ {\rm erg\ s}^{-1}} \right ]^{(0.26 \pm 0.16)}
\end{equation}

\begin{equation}
\frac{R_{\rm SiIV}}{10\ {\rm lt-days}} = (0.10 \pm 0.10) \left [ \frac{\lambda L_{\lambda} (1350{\rm \AA})}{10^{43}\ {\rm erg\ s}^{-1}} \right ]^{(0.58 \pm 0.16)}
\end{equation}

As with Ly$\alpha$, the zero point of the SiIV radius--luminosity
relation is very ambiguous because of the large dispersion observed in
the Seyfert regime and the lack of any measurement for very low
luminosity sources.

\section{Discussion}

\subsection{Sources with unexpected line variability}

We have seen that 3/17 of our sources, i.e., $18^{+14}_{-9}$\%
assuming Poisson statistics (Gehrels 1986), show unexpected line
variability, where the Ly$\alpha$ or SiIV emission line light curves
do not seem to follow that of the observed UV continuum.

One possibility is that in these objects these emission lines never
responds to the observed continuum. Another possibility, is that this
is a transitional behavior due to changes in the BLR properties or the
ionizing source.

Disengagement of the line response has not been generally reported
among the Seyfert galaxies that have been subject to RM campaigns,
although a non-linear response of CIV to the continuum variations was
recognized in early observations of NGC5548 (see Maoz 1994). In fact,
a clearer `anomalous' behavior was seen in the very recent monitoring
of NGC5548, where a departure of the line emission light curves is
observed during $\sim 1/3$ of the $\sim$ 170 day long campaign and
where high ionization emission lines, in particular CIV, show a
stronger disagreement with the continuum light curve than low
ionization emission lines (Goad et al.~2016).

It seems that this behavior is more common in our sample of high
luminosity quasars, even though the number of variability `events'
observed in the quasar light curves is usually smaller than those
typically observed during seasonal monitoring campaigns of Seyfert
galaxies. However, this result is based on limited data, while the
different monitoring cadence, lengths of the campaigns and wide ranges
in BH masses and accretion rates (which might ultimately drive the
variability of the sources), complicate the comparison beyond the
scope of this paper.

It is interesting to notice, however, that while in NGC5548 this
behavior is strongest in the CIV line (with an ionization potential of
47.9 eV), in CT320, CT803 and J224743 is SiIV and Ly$\alpha$ that
behave in an anomalous way. This might suggest that the cause is a
different ionization continuum, as predicted for accretion disks
around BHs of different masses but similar accretions rates (e.g.,
Davis \& Laor, 2011), or seen in the relation between the
$\alpha_{ox}$ index (indicative of the fractional output in the X-ray
and optical bands) with AGN luminosity (e.g., Vignali et al., 2003).

\subsection{Location of the line emitting regions}

\begin{figure}
\begin{center}
\includegraphics[scale=0.42, trim=0 0 0 50]{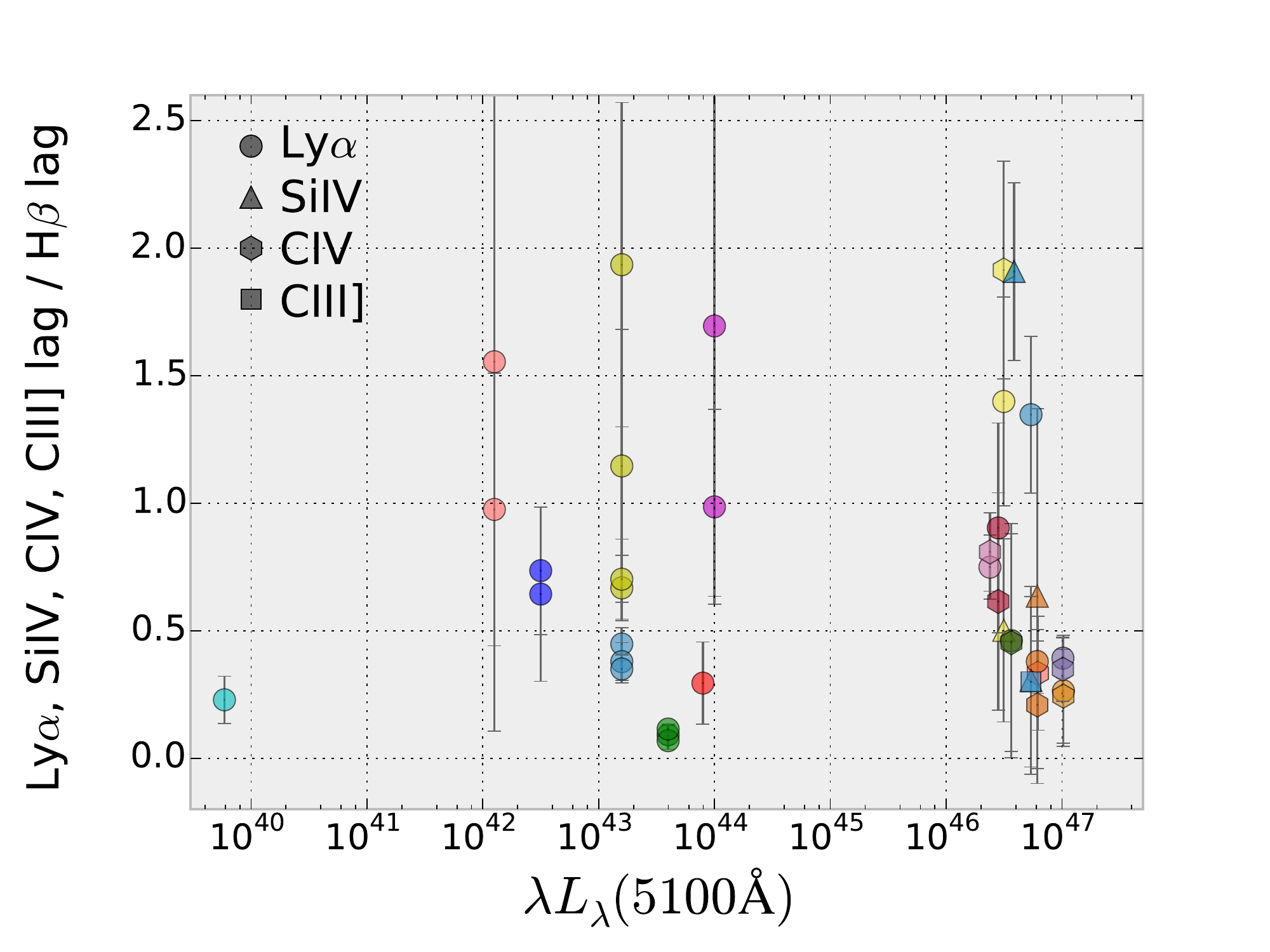}\\
\caption{Ratio of Ly$\alpha$, SiIV, CIV and CIII] to H$\beta$ lags.
The Ly$\alpha$, SiIV, CIV and CIII] lags are those presented in Table
  4 while the H$\beta$ lags were predicted using the $\lambda
  L_{\lambda} (5100$\AA) vs H$\beta$ radius--luminosity correlation
  found in Bentz et al.~(2013) and the inferred 5100\AA\ luminosities
  (see text). Different shapes correspond to different emission lines
  (see legend), while different colors correspond to different
  sources. Seyfert galaxies are: NGC3783, NGC4151, NGC5548, NGC7469
  and 3C390.3, as stated before. We have also added data from the 1989
  campaign of NGC5548 and published by Clavel et al.~(1991), as it
  includes the CIII] emission line (after re-calculating the lags
    using the ICCF FR/RSS code that we employed with our sources). The
    objects reported in this work are clustered at the high luminosity
    end of the diagram. Errors in the Ly$\alpha$, SiIV, CIV and CIII]
      lines were assumed as the average of the lower and upper
      $1\sigma$ confidence limits in the lag measurements, while the
      scatter in the H$\beta$ radius--luminosity correlation for the
      Clean2 sample from Bentz et al.~(2013) was taken as the H$\beta$
      lag errors.}
\end{center}
\end{figure}

\begin{figure}
\begin{center}
\includegraphics[scale=0.42, trim=0 0 0 0]{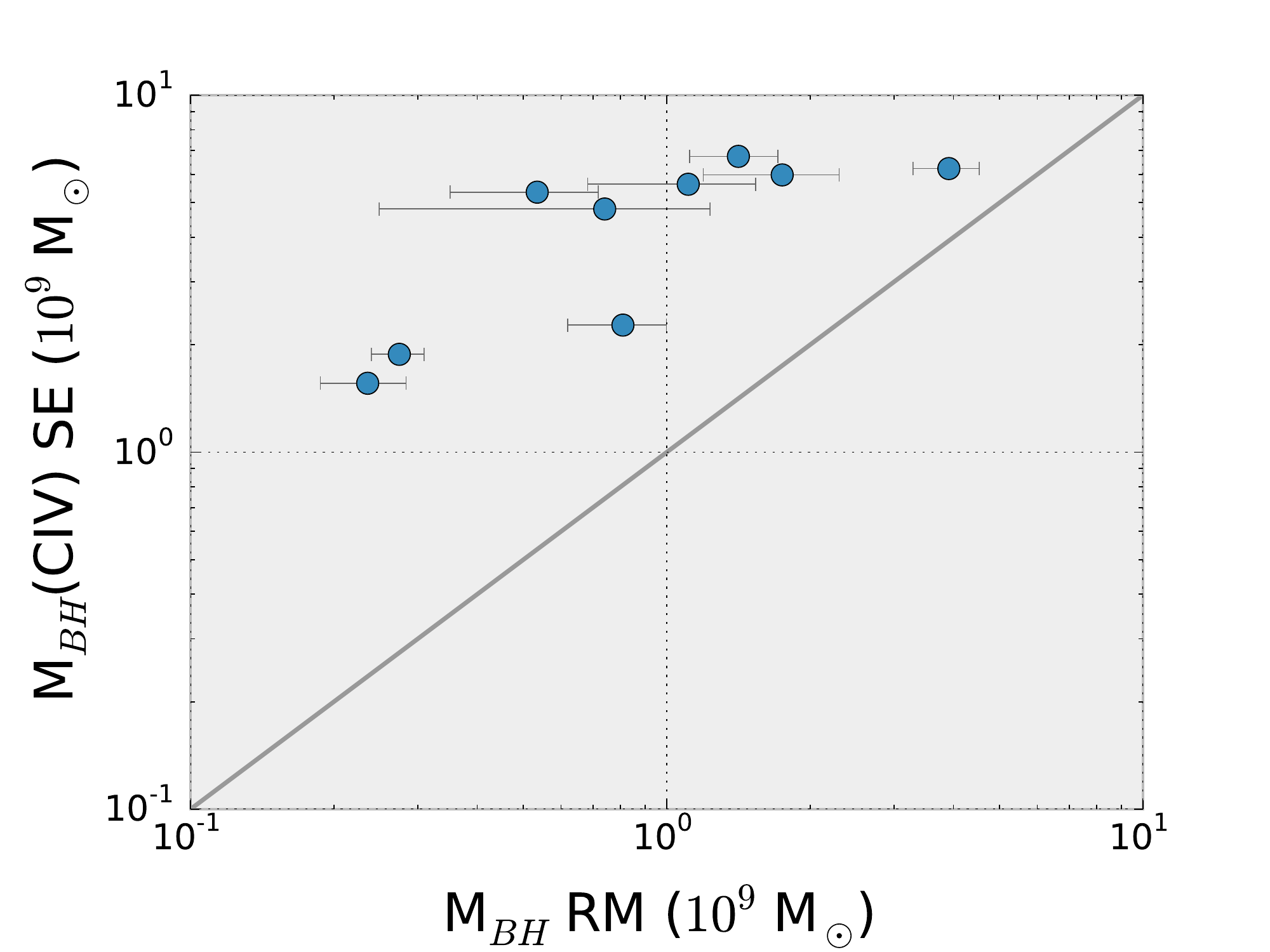}\\
\caption{Comparison between Single--Epoch (SE) and Reverberation
  Mapping (RM) based BH masses, as presented in Tables 1 and 4. The
  thick gray line corresponds to the 1:1 relation.}
\end{center}
\end{figure}

One of the most significant and early results from RM in nearby
Seyferts was the discovery that the BLR has a non-negligible radius
and that different lines form at different distances from the central
BH. From our cross correlation analysis we can revisit this result and
try to extended it to a wider luminosity range. 

Figure 8 presents the ratio of Ly$\alpha$, SiIV, CIII] and CIV lags to
  H$\beta$ lags for Seyfert galaxies monitored by the International
  AGN Watch and quasars from this work deemed reliable in Table 4.
  H$\beta$ lags were determined using the $\lambda L_{\lambda}
  (5100$\AA) vs H$\beta$ radius--luminosity correlation for the Clean2
  sample in Bentz et al.~(2013). Ly$\alpha$, SiIV, CIII] and CIV lags
    and associated confidence limits were taken from Table 4. Errors
    for these lines were assumed as the average of the lower and upper
    $1\sigma$ limits. Errors in the H$\beta$ lags were taken as the
    scatter reported for the radius--luminosity correlation determined
    by Bentz et al.~(2013), i.e., $\sigma^2 = 0.018$. 5100\AA\ fluxes
    for our objects were obtained in the same way as in Section 2.1.2.

In general, we find a range of ratios that span up to a factor
6. However, most line lags are consistent with their emitting regions
being interior to the predicted location of H$\beta$. There is also no
clear stratification among the four lines for which we have determined
lags in this work, suggesting that they are all produced at similar
distances from the central black hole. There is no evidence for a
clear dependency of the ratios with luminosity. This can be quantified
comparing the lag cumulative distributions of Seyfert and quasars for
{\em all\/} line ratios shown in Figure 8 using a KS test, which
yields a p-value of 0.23, suggesting that the two distributions are
very similar. This indicates that sources spanning 5 orders of
magnitude in luminosity present a homologous BLR structure.

Notice the change in line ratios between the modern results for
NGC5548 (in blue) taken from De Rosa et al.~(2015), when compared with
the historic values (in yellow) published by Clavel et
al.~(1991). This seems to be evidence of a restructuration of the BLR
in this source, with the line emitting regions moving further in,
despite a very small change in the UV and optical luminosities between
these epochs. This also includes the $H\beta$ location, as shown in
Pei et al.~(2017), with the lag about five times shorter than expected
based on past measurements and the H$\beta$ radius--luminosity
correlation from Bentz et al.~(2013).

Another interesting result is that all radius--luminosity relation
slopes in Equations (1)--(5) are consistent at the $2\sigma$ level
with the na\"ively expected value of 0.5, as predicted by the
assumption of a photoionized BLR where the mean ionization parameter
and mean density at the peak emissivity for a certain line remain
constant (see Bentz et al.~2013 for further discussion). Notice,
however, that some of the slopes, such as that of the CIII] line, have
  rather large errors.

\subsection{Black hole masses from RM and SE methods}

In Table 4 we give BH masses for sources with lags determined at a
$1\sigma$ level in the uncertainties at both sides of the probability
CCCD distribution, this is, $\tau_{cent}/\sigma_{-} > 1$ and
$\tau_{cent}/\sigma_{+} > 1$. Masses were obtained as M$_{\rm BH}^{\rm
  RM} = f \times \tau_{cent} \times c \times {\rm FWHM}^2$, where $f$
is the virial factor, $\tau_{cent}$ is the ICCF lag centroid reported
in Table 4, and the FWHMs were measured from the mean spectra of each
quasar and found in Table 2. Because of the noisy nature of the
emission lines in the rms spectra of our objects (see Figure 3), we
only measured the FWHMs from the total mean flux spectra. We have
adopted a virial factor $f = 1$, also assumed for the SE
determinations. A good agreement is seen for RM masses obtained using
different lines, with all results consistent within a $2\sigma$ level.

Inspection of Tables 1 and 4 shows that RM masses (M$_{\rm BH}^{\rm
  RM}$) are systematically smaller than those obtained using the CIV
single--epoch (SE) method (M$_{\rm BH}^{\rm SE}$), which were
determined using M$_{\rm BH}^{\rm SE} = 10^{6.353} \times
(L_{1450})^{0.599} \times {\rm FWHM}^2$ as presented in
Mej\'ia-Restrepo et al.~(2016). Again, the FWHM values are those
presented in Table 1. The results can be seen in Figure 9, where the
mean of the RM masses is plotted when lags from more than one line are
available for a single object. Our results suggest that, on average,
M$_{\rm BH}^{\rm SE}$ are overestimated when compared with M$_{\rm
  BH}^{\rm RM}$. The discrepancy is probably driven by the non-virial
behavior of the CIV line width. The SE cross-calibration of CIV
assumes that the region responsible for the emission of this line
obeys $R_{\rm H\beta} / R_{\rm CIV}$ =
(FWHM(H$\beta$)/FWHM(CIV))$^{-2}$ (since SE masses obtained from
H$\beta$ and CIV must satisfy M$_{\rm BH}^{\rm SE}$(H$\beta$) =
M$_{\rm BH}^{\rm SE}$(CIV)). As we just determined for our quasars,
$R_{\rm CIV} / R_{\rm H\beta} \sim 0.5$ and we would expect that
(FWHM(H$\beta$)/FWHM(CIV))$^2 \sim 0.5$ or FWHM(CIV) $\sim
1.4\ \times$ FWHM(H$\beta$) for a virialized system. However,
typically it is found that CIV is narrower than this, suggesting a
dominant non-virialized component in many sources (see Trakhtenbrot
\& Netzer, 2012).

\section{Summary}

For the first time we have presented RM results for a substantial
number of luminous quasars found at $z \ga 2$. From our results we can
summarize the following:

\begin{itemize}

\item Out of 17 quasars with spectroscopic follow up 14 show that all
  their emission line light curves seem to reverberate in response to
  the variations observed in the continuum light curves, while 3/17
  show peculiar behavior in one emission line. This might suggests
  that the observed R-band continuum in these three sources does not
  follow the changes that the ionizing continuum experiences.

\item Reliable lag measurements are determined for 11 quasars for the
  Ly$\alpha$ emission line, 5 quasars for the SiIV emission line, 11
  for the CIV emission line, and 2 quasars for the CIII] emission line
    (Table 4). However, only $\sim 1/2$ of the determined lags have
    $\tau_{cent}/\sigma > 1$ and are therefore sufficiently
    constrained to allow for the determination of BH masses.

\item Radius--luminosity relations for Ly$\alpha$, SiIV, CIV, and
  CIII] are presented using our data and previous lag
    determinations. Slopes are well constrained for all
      correlations and are found to be less than $2\sigma$ away from
      the predicted value of 0.5, although the Ly$\alpha$ and CIII]
    relations present large fractional errors.

\item We find that the regions responsible for the emission of
  Ly$\alpha$, SiIV, CIV, CIII] are commonly interior to that producing
    H$\beta$. At the same time, there is no clear stratification among
    them. This is found to be the case across 5 orders of magnitude in
    continuum luminosity.

\item For those quasars with lags determined at a $1\sigma$ level (for
  both, $\sigma_{-}$ and $\sigma_{+}$), we determined BH masses using
  the observed emission line FWHMs. The RM masses are systematically
  smaller than those determined from single--epoch CIV calibrations.
  This is consistent with a significant non-virialized component to
  the CIV line profile.

\end{itemize}

\acknowledgements 

This project would had never been possible without the support of the
Chilean National TAC (CNTAC) which during more than 10 years allocated
hundreds of nights of telescope time to conduct our reverberation
campaign. We also thank Brad Peterson and Mario Hamuy for facilitating
code used in the analysis of the data. Finally, we also thank the
anonymous referee for the suggestion to use MC to determine line
light curve errors. PL acknowledges support by Fondecyt along all
these years, and in particular to Project \#1161184.

\appendix

Here we present the emission line and R-band continuum light curves
for the quasar CT1061. The remaining light curves can be downloaded
from the electronic journal database. In all tables dates are
expressed in JD$-2450000$ days, and fluxes in units of $10^{-14}$ ergs
s$^{-1}$ cm$^{-2}$ for the line measurements and $10^{-16}$ ergs
s$^{-1}$ cm$^{-2}$ \AA$^{-1}$ for the R-band light curves.

\begin{table*}
\renewcommand{\thetable}{\arabic{table}} \centering
\caption{Emission Line and R-band continuum light curves for CT1061}
\begin{tabular}{ccc|ccc|ccc|ccc|ccc}
\hline \hline
\multicolumn{3}{c|}{Ly$\alpha$} & \multicolumn{3}{c|}{SiIV} & \multicolumn{3}{c|}{CIV} & \multicolumn{3}{c|}{CIII]} & \multicolumn{3}{c}{R} \\
JD & flux & error & JD & flux & error & JD & flux & error & JD & flux & error & JD & flux & error \\
\hline
5270.8 & 27.19 & 0.44 & 5270.8 & 2.06 & 0.26 & 5270.8 & 8.12 & 0.12 & 5270.8 & 4.15 & 0.35 & 3403.9 & 5.50 & 0.08\\
5280.7 & 27.39 & 0.53 & 5280.7 & 1.90 & 0.32 & 5280.7 & 8.19 & 0.16 & 5280.7 & 3.96 & 0.31 & 3444.8 & 5.48 & 0.08\\
5658.6 & 27.04 & 0.46 & 5658.6 & 2.06 & 0.24 & 5658.6 & 8.01 & 0.13 & 5658.6 & 3.99 & 0.24 & 3490.7 & 5.55 & 0.08\\
5917.8 & 27.28 & 0.53 & 5917.8 & 2.04 & 0.26 & 5917.8 & 8.03 & 0.17 & 5917.8 & 4.20 & 0.29 & 3781.7 & 5.70 & 0.08\\
5945.8 & 27.26 & 0.50 & 5945.8 & 2.12 & 0.22 & 5945.8 & 8.12 & 0.14 & 5945.8 & 4.00 & 0.28 & 3846.7 & 5.52 & 0.08\\
6015.7 & 26.72 & 0.54 & 6015.7 & 1.92 & 0.26 & 6015.7 & 7.86 & 0.19 & 6015.7 & 3.73 & 0.30 & 3894.5 & 5.58 & 0.08\\
6017.7 & 27.42 & 0.50 & 6017.7 & 2.10 & 0.29 & 6017.7 & 8.10 & 0.17 & 6017.7 & 3.73 & 0.44 & 4140.9 & 5.48 & 0.08\\
6048.6 & 26.95 & 0.45 & 6048.6 & 1.95 & 0.29 & 6048.6 & 7.96 & 0.13 & 6048.6 & 4.36 & 0.33 & 4154.8 & 5.53 & 0.08\\
6298.8 & 26.83 & 0.37 & 6298.8 & 2.00 & 0.22 & 6298.8 & 8.04 & 0.12 & 6298.8 & 4.00 & 0.26 & 4168.9 & 5.58 & 0.08\\
6723.8 & 26.65 & 0.36 & 6723.8 & 1.90 & 0.20 & 6723.8 & 7.84 & 0.08 & 6723.8 & 4.46 & 0.31 & 4258.6 & 5.63 & 0.08\\
6783.6 & 26.68 & 0.53 & 6783.6 & 1.92 & 0.24 & 6783.6 & 8.00 & 0.21 & 6783.6 & 4.12 & 0.30 & 4272.5 & 5.69 & 0.08\\
7020.8 & 26.20 & 0.40 & 7020.8 & 1.90 & 0.24 & 7020.8 & 7.46 & 0.25 & 7020.8 & 3.77 & 0.34 & 4505.8 & 6.21 & 0.09\\
7759.0 & 26.30 & 0.47 & 7759.0 & 1.96 & 0.27 & 7759.0 & 7.75 & 0.11 & 7759.0 & 3.75 & 0.24 & 4579.7 & 6.36 & 0.09\\
--- & --- & --- & --- & --- & --- & --- & --- & --- & --- & --- & --- & 4580.5 & 6.22 & 0.09\\
--- & --- & --- & --- & --- & --- & --- & --- & --- & --- & --- & --- & 4856.7 & 6.58 & 0.09\\
--- & --- & --- & --- & --- & --- & --- & --- & --- & --- & --- & --- & 4889.8 & 6.71 & 0.09\\
--- & --- & --- & --- & --- & --- & --- & --- & --- & --- & --- & --- & 4910.8 & 6.63 & 0.09\\
--- & --- & --- & --- & --- & --- & --- & --- & --- & --- & --- & --- & 4944.7 & 6.76 & 0.09\\
--- & --- & --- & --- & --- & --- & --- & --- & --- & --- & --- & --- & 5010.5 & 6.78 & 0.09\\
--- & --- & --- & --- & --- & --- & --- & --- & --- & --- & --- & --- & 5235.8 & 6.48 & 0.09\\
--- & --- & --- & --- & --- & --- & --- & --- & --- & --- & --- & --- & 5264.7 & 6.50 & 0.09\\
--- & --- & --- & --- & --- & --- & --- & --- & --- & --- & --- & --- & 5270.8 & 6.70 & 0.04\\
--- & --- & --- & --- & --- & --- & --- & --- & --- & --- & --- & --- & 5274.7 & 6.60 & 0.09\\
--- & --- & --- & --- & --- & --- & --- & --- & --- & --- & --- & --- & 5280.7 & 6.65 & 0.04\\
--- & --- & --- & --- & --- & --- & --- & --- & --- & --- & --- & --- & 5306.7 & 6.55 & 0.09\\
--- & --- & --- & --- & --- & --- & --- & --- & --- & --- & --- & --- & 5628.8 & 6.16 & 0.08\\
--- & --- & --- & --- & --- & --- & --- & --- & --- & --- & --- & --- & 5628.8 & 6.20 & 0.09\\
--- & --- & --- & --- & --- & --- & --- & --- & --- & --- & --- & --- & 5653.8 & 6.21 & 0.09\\
--- & --- & --- & --- & --- & --- & --- & --- & --- & --- & --- & --- & 5653.8 & 6.25 & 0.09\\
--- & --- & --- & --- & --- & --- & --- & --- & --- & --- & --- & --- & 5658.6 & 6.21 & 0.04\\
--- & --- & --- & --- & --- & --- & --- & --- & --- & --- & --- & --- & 5666.7 & 6.32 & 0.09\\
--- & --- & --- & --- & --- & --- & --- & --- & --- & --- & --- & --- & 5666.7 & 6.25 & 0.09\\
--- & --- & --- & --- & --- & --- & --- & --- & --- & --- & --- & --- & 5738.5 & 6.12 & 0.08\\
--- & --- & --- & --- & --- & --- & --- & --- & --- & --- & --- & --- & 5738.5 & 6.13 & 0.08\\
--- & --- & --- & --- & --- & --- & --- & --- & --- & --- & --- & --- & 5917.8 & 6.01 & 0.04\\
--- & --- & --- & --- & --- & --- & --- & --- & --- & --- & --- & --- & 5929.7 & 6.01 & 0.08\\
--- & --- & --- & --- & --- & --- & --- & --- & --- & --- & --- & --- & 5929.7 & 5.96 & 0.08\\
--- & --- & --- & --- & --- & --- & --- & --- & --- & --- & --- & --- & 5945.8 & 5.95 & 0.04\\
--- & --- & --- & --- & --- & --- & --- & --- & --- & --- & --- & --- & 6013.6 & 5.79 & 0.08\\
--- & --- & --- & --- & --- & --- & --- & --- & --- & --- & --- & --- & 6013.6 & 5.76 & 0.08\\
--- & --- & --- & --- & --- & --- & --- & --- & --- & --- & --- & --- & 6015.7 & 5.54 & 0.04\\
--- & --- & --- & --- & --- & --- & --- & --- & --- & --- & --- & --- & 6017.7 & 5.86 & 0.04\\
--- & --- & --- & --- & --- & --- & --- & --- & --- & --- & --- & --- & 6033.6 & 5.79 & 0.08\\
--- & --- & --- & --- & --- & --- & --- & --- & --- & --- & --- & --- & 6033.6 & 5.75 & 0.08\\
--- & --- & --- & --- & --- & --- & --- & --- & --- & --- & --- & --- & 6048.6 & 5.85 & 0.04\\
--- & --- & --- & --- & --- & --- & --- & --- & --- & --- & --- & --- & 6049.5 & 5.72 & 0.08\\
--- & --- & --- & --- & --- & --- & --- & --- & --- & --- & --- & --- & 6049.5 & 5.70 & 0.08\\
--- & --- & --- & --- & --- & --- & --- & --- & --- & --- & --- & --- & 6298.7 & 5.53 & 0.08\\
--- & --- & --- & --- & --- & --- & --- & --- & --- & --- & --- & --- & 6298.7 & 5.44 & 0.08\\
--- & --- & --- & --- & --- & --- & --- & --- & --- & --- & --- & --- & 6298.8 & 5.54 & 0.04\\
--- & --- & --- & --- & --- & --- & --- & --- & --- & --- & --- & --- & 6303.7 & 5.54 & 0.08\\
--- & --- & --- & --- & --- & --- & --- & --- & --- & --- & --- & --- & 6303.8 & 5.63 & 0.11\\
--- & --- & --- & --- & --- & --- & --- & --- & --- & --- & --- & --- & 6306.7 & 5.60 & 0.08\\
--- & --- & --- & --- & --- & --- & --- & --- & --- & --- & --- & --- & 6306.7 & 5.50 & 0.08\\
--- & --- & --- & --- & --- & --- & --- & --- & --- & --- & --- & --- & 6400.6 & 5.33 & 0.07\\
--- & --- & --- & --- & --- & --- & --- & --- & --- & --- & --- & --- & 6723.8 & 5.19 & 0.04\\
--- & --- & --- & --- & --- & --- & --- & --- & --- & --- & --- & --- & 6783.6 & 5.14 & 0.04\\
--- & --- & --- & --- & --- & --- & --- & --- & --- & --- & --- & --- & 7020.8 & 5.26 & 0.04\\
--- & --- & --- & --- & --- & --- & --- & --- & --- & --- & --- & --- & 7109.6 & 5.28 & 0.07\\
--- & --- & --- & --- & --- & --- & --- & --- & --- & --- & --- & --- & 7123.6 & 5.24 & 0.07\\
--- & --- & --- & --- & --- & --- & --- & --- & --- & --- & --- & --- & 7152.5 & 5.40 & 0.07\\
--- & --- & --- & --- & --- & --- & --- & --- & --- & --- & --- & --- & 7181.5 & 5.37 & 0.07\\
--- & --- & --- & --- & --- & --- & --- & --- & --- & --- & --- & --- & 7182.4 & 5.38 & 0.07\\
--- & --- & --- & --- & --- & --- & --- & --- & --- & --- & --- & --- & 7225.5 & 5.35 & 0.29\\
--- & --- & --- & --- & --- & --- & --- & --- & --- & --- & --- & --- & 7759.0 & 5.34 & 0.04\\
\hline
\hline
\end{tabular}
\end{table*}

\end{document}